\documentclass[fleqn,usenatbib]{mnras}

\usepackage{newtxtext,newtxmath}

\usepackage[T1]{fontenc}

\DeclareRobustCommand{\VAN}[3]{#2}
\let\VANthebibliography\thebibliography
\def\thebibliography{\DeclareRobustCommand{\VAN}[3]{##3}\VANthebibliography}

\usepackage{pdflscape}	
\usepackage{verbatim} 
\usepackage{hyperref}
\usepackage{threeparttablex}
\usepackage{longtable}
\usepackage{geometry}
\usepackage{url}

\usepackage{graphicx}	
\usepackage{amsmath}	
\usepackage{orcidlink}

\usepackage[switch]{lineno}

\title[The ambiguous AT2022rze]{The ambiguous AT2022rze: Changing-look AGN mimicking a supernova in a merging galaxy system}

\author[P.~J. Pessi et al.]{
P.~J. Pessi\orcidlink{0000-0002-8041-8559},$^{1}$\thanks{E-mail: priscila.pessi@astro.su.se}
R. Lunnan\orcidlink{0000-0001-9454-4639},$^{1}$
J. Sollerman\orcidlink{0000-0003-1546-6615},$^{1}$
L. Yan\orcidlink{0000-0003-1710-9339},$^{2}$
A. Le Reste\orcidlink{0000-0003-1767-6421},$^{3}$
Y. Yao\orcidlink{0000-0001-6747-8509},$^{4,5}$
S. Nordblom\orcidlink{0009-0003-8750-7110},$^{6}$
\newauthor
Y. Sharma\orcidlink{0000-0003-4531-1745},$^{7}$
M. Gilfanov,$^{8,9}$ 
R. Sunyaev,$^{8,9}$ 
S. Schulze\orcidlink{0000-0001-6797-1889},$^{10}$
J. Johansson\orcidlink{0000-0001-5975-290X},$^{11}$
A. Gangopadhyay\orcidlink{0000-0002-3884-5637},$^{1}$
\newauthor
C. Fremling\orcidlink{0000-0002-4223-103X},$^{2,7}$
K. Tristram\orcidlink{0000-0001-8281-5059},$^{12}$
M. J. Hayes,$^{1}$
C. Fransson,$^{1}$
Y. Hu\orcidlink{0000-0002-9744-3910},$^{1}$
S. J. Brennan\orcidlink{0000-0003-1325-6235},$^{1}$
S. Rose\orcidlink{0000-0003-4725-4481},$^{13}$
\newauthor
K. De\orcidlink{0000-0002-8989-0542},$^{14}$
K-R. Hinds\orcidlink{0000-0002-0129-806X},$^{15}$
C. Liu\orcidlink{0000-0002-7866-4531},$^{16,10}$
A. A. Miller\orcidlink{0000-0001-9515-478X},$^{16,10,17}$
Y-J. Qin,$^{13}$
P. Charalampopoulos,$^{18}$
A. Gkini\orcidlink{0009-0000-9383-2305},$^{1}$
\newauthor
M. J. Graham\orcidlink{0000-0002-3168-0139},$^{7}$
C. P. Guti\'errez\orcidlink{0000-0003-2375-2064},$^{19,20}$
S. Mattila\orcidlink{0000-0001-7497-2994},$^{18,21}$
T. Nagao\orcidlink{0000-0002-3933-7861},$^{18,22,23}$
I. P\'erez-Fournon\orcidlink{0000-0002-2807-6459},$^{24,25}$
\newauthor
F. Poidevin\orcidlink{0000-0002-5391-5568},$^{24,25}$
J. S. Bloom\orcidlink{0000-0002-7777-216X},$^{5,26}$
J. Brugger,$^{2}$
T. X. Chen\orcidlink{0000-0001-9152-6224},$^{27}$
M. M. Kasliwal\orcidlink{0000-0002-5619-4938},$^{7}$
F. J. Masci\orcidlink{000-0002-8532-9395},$^{27}$
\newauthor
and
J. N. Purdum,$^{2}$
\\
$^{1}$The Oskar Klein Centre, Department of Astronomy, Stockholm University, Albanova University Center, SE 106 91 Stockholm, Sweden\\
$^{2}$Caltech Optical Observatories, California Institute of Technology, Pasadena, CA 91125, USA\\
$^{3}$Minnesota Institute for Astrophysics, University of Minnesota, 116 Church Street SE, Minneapolis, MN 55455, USA\\
$^{4}$Miller Institute for Basic Research in Science, 468 Donner Lab, Berkeley, CA 94720, USA\\
$^{5}$Department of Astronomy, University of California, Berkeley, CA 94720, USA\\
$^{6}$Independent Researcher, Sweden\\
$^{7}$Division of Physics, Mathematics and Astronomy, California Institute of Technology, Pasadena, CA 91125, USA\\
$^{8}$Space Research Institute, Russian Academy of Sciences, Profsoyuznaya 84/32, 117997 Moscow, Russia\\
$^{9}$Max Planck Institute for Astrophysics, Karl-Schwarzschild-Str 1, Garching b. München D-85741, Germany\\
$^{10}$Center for Interdisciplinary Exploration and Research in Astrophysics (CIERA), 1800 Sherman Ave., Evanston, IL 60201, USA\\
$^{11}$The Oskar Klein Centre, Department of Physics, Stockholm University, Albanova University Center, SE 106 91 Stockholm, Sweden\\
$^{12}$European Southern Observatory, Alonso de Córdova 3107, Vitacura, Santiago, Chile\\
$^{13}$Cahill Center for Astrophysics, California Institute of Technology, Pasadena, CA 91125, USA\\
$^{14}$MIT-Kavli Institute for Astrophysics and Space Research, 77 Massachusetts Ave., Cambridge, MA 02139, USA\\
$^{15}$Astrophysics Research Institute, Liverpool John Moores University, 146 Brownlow Hill, Liverpool L3 5RF, UK\\
$^{16}$Department of Physics and Astronomy, Northwestern University, 2145 Sheridan Rd, Evanston, IL 60208, USA\\
$^{17}$ NSF-Simons AI Institute for the Sky (SkAI), 172 E. Chestnut St., Chicago, IL 60611, USA\\
$^{18}$Department of Physics and Astronomy, FI-20014 University of Turku, Finland\\
$^{19}$Institut d'Estudis Espacials de Catalunya (IEEC), Edifici RDIT,
Campus UPC, 08860 Castelldefels (Barcelona), Spain \\
$^{20}$Institute of Space Sciences (ICE, CSIC), Campus UAB, Carrer de Can
Magrans, s/n, E-08193 Barcelona, Spain \\
$^{21}$School of Sciences, European University Cyprus, Diogenes Street, Engomi, 1516 Nicosia, Cyprus\\
$^{22}$Aalto University Metsähovi Radio Observatory, Metsähovintie 114,
02540 Kylmälä, Finland\\
$^{23}$Aalto University Department of Electronics and Nanoengineering,
P.O. BOX 15500, FI-00076 AALTO, Finland\\
$^{24}$Instituto de Astrof\'isica de Canarias, C/V\'ia Láctea, s/n, E-38205 San Crist\'obal de La Laguna, Tenerife, Spain\\
$^{25}$Universidad de La Laguna, Dpto. Astrof\'isica, E-38206 San Crist\'obal de La Laguna, Tenerife, Spain\\
$^{26}$Physics Division, Lawrence Berkeley National Laboratory, 1 Cyclotron Road, MS 50B-4206, Berkeley, CA 94720, USA\\
$^{27}$IPAC, California Institute of Technology, 1200 E. California Blvd,
Pasadena, CA 91125, USA
}

\date{Accepted XXX. Received YYY; in original form ZZZ}

\pubyear{\the\year{}}

\begin{document}
\label{firstpage}
\pagerange{\pageref{firstpage}--\pageref{lastpage}}
\maketitle

\begin{abstract}
AT2022rze is a luminous, ambiguous transient located South-East of the geometric center of its host galaxy at redshift $z = 0.08$. The host appears to be formed by a merging galaxy system. The observed characteristics of AT2022rze are reminiscent of active galactic nuclei (AGN), tidal disruption events (TDEs), and superluminous supernovae (SLSNe). The transient reached a peak absolute magnitude of $-$20.2 $\pm$ 0.2 mag, showing a sharp rise (t$_{\mathrm{rise,1/e}} = 27.5 \pm 0.6$ days) followed by a slow decline (t$_{\mathrm{dec,1/e}} = 382.9 \pm 0.6$). Its bumpy light curve and narrow Balmer lines indicate the presence of gas (and dust). Its light curve shows rather red colors, indicating that the transient could be affected by significant host extinction. The spectra reveal coronal lines, indicative of high-energy (X-ray/UV) emission. Archival data reveal no prior activity at this location, disfavoring a steady-state AGN, although an optical spectrum obtained prior to the transient is consistent with an AGN classification of the host. Based on this, we conclude that the transient most likely represents a Changing-look AGN at the center of the smallest component of the merging system.  
\end{abstract}

\begin{keywords}
transients: supernovae -- transients: tidal disruption events -- galaxies: active  -- galaxies: interactions 
\end{keywords}



\section{Introduction}
\label{sec:intro}

Transient astronomical events provide critical insights into the dynamic processes of the universe. Modern high-cadence, all-sky surveys have dramatically enhanced our ability to detect and study a broad range of transient phenomena, many of which display complex and often puzzling observational characteristics. Among these, transients displaying extreme luminosity variations are particularly challenging to interpret, as the energy required to power them often pushes the boundaries of standard theoretical models. Three prominent classes of such energetic transients are superluminous supernovae (SLSNe), tidal disruption events (TDEs), and changing-look active galactic nuclei (CLAGNs).

Core collapse supernovae (CCSNe) mark the explosive deaths of massive stars (> 8M$_{\odot}$). CCSNe can be classified by spectroscopic and/or photometric features \citep[e.g:][]{2017hsn..book..195G}. In particular, events that display exceptionally high peak luminosities, typically brighter than $\sim -$20~mag in optical bands, are classified as SLSNe \citep[e.g:][]{2019ARA&A..57..305G}. When hydrogen (H) lines are visible in the spectra of these events, they are classified as Type II SLSNe or Type IIn SLSNe if the hydrogen features are narrow, which indicates interaction between the SN ejecta and a dense circumstellar material \citep[CSM, e.g:][]{2017hsn..book..403S}. 
Tidal disruption events (TDEs)  occur when a star intersects the tidal radius of a black hole (BH) and is torn apart by tidal forces, producing a luminous flare powered by the fallback and accretion of stellar debris \citep[e.g.,][]{2021ARA&A..59...21G}. 
Active galactic nuclei (AGNs) are sustained by accretion of gas onto supermassive BHs (SMBHs) at galactic centers, producing intense, persistent emission across the electromagnetic spectrum \citep[e.g.,][]{2017A&ARv..25....2P}. A subclass of AGNs have been discovered to change their aspect and/or ignite when their host galaxies transition from quiescence to active accretion into the central BH, displaying luminous, energetic light curves. These have been labeled changing-look AGN \citep[CLAGN, e.g.,][]{2023NatAs...7.1282R}. 

In most cases, distinguishing SLSNe, TDEs and AGNs is relatively straightforward by spectral matching or light curve analysis \citep[a large variety of classification tools exits, here we only list a small fraction as example:][]{2005ApJ...634.1190H,2007ApJ...666.1024B,2018ApJS..236....9N,2019ApJ...885...85M,2021AJ....161..242F,2024A&A...692A.208F}. For instance, AGN variability is generally stochastic and persistent, unlike the more defined temporal structure of SLSNe and TDEs. TDEs typically appear blue and exhibit little color evolution, whereas SLSNe begin with high temperatures and gradually redden as they cool. On the contrary, CLAGNs  are very diverse, and their characteristics can resemble those of SLSNe~IIn and TDEs \citep[e.g.,][]{2024SerAJ.209....1K}.

Recent studies have started to incorporate host galaxy properties alongside photometric behavior to aid in classification  \citep[e.g:][]{2020ApJ...904...74G,2023ApJ...954....6G}. However, overlap in host galaxy location, particularly for events located in nuclear or circumnuclear regions, makes classification even more ambiguous \citep[e.g.,][]{2020SSRv..216...32F,2021ApJ...913..102W,2021MNRAS.507..156G,2023ApJ...950..161L}. Additionally, dust and gas surrounding the transient may produce prolonged, bumpy, and luminous light curves, further obscuring classification. As a result SLSNe, TDEs, and (CL)AGNs can become contaminants in corresponding sample studies \citep[see for example the contaminants in the SLSN~II sample presented by][]{2025A&A...695A.142P}. 

There is an increasing number of energetic transients with ambiguous characteristics being discovered \citep[e.g.][]{2017NatAs...1..865K}. A large fraction of these events is found at the nuclei of their host galaxies and have been dub these events Ambiguous Nuclear Transients \citep[ANTs,][]{2025MNRAS.537.2024W}. ANTs exhibit heterogeneous spectral and photometric properties that do not really fit those of SLSNe, TDEs, or AGNs. Mid-infrared observations suggest that dust and molecular gas may shape the optical properties of ANTs, highlighting the need for multiwavelength observations to understand their nature. 
AT2022rze, exemplifies this ambiguity. It appears centrally located in a compact structure, possibly a small galaxy merging with a larger host, but exhibits properties that defy a straightforward classification.

Understanding enigmatic events is critical, not only for improving the classification framework of transient populations but also for understanding the physical mechanisms powering these phenomena. This study focuses on the analysis of AT2022rze, a puzzling event originally classified as a SN, but exhibiting characteristics similar to CLAGNs and TDEs. The observations of AT2022rze are presented in Sect.~\ref{sec:obs}. These are analyzed in Sect.~\ref{sec:analy} and the results of the analysis are discussed in Sect.~\ref{sec:disc}. We conclude in Sect.~\ref{sec:conc}.

\section{Observations}
\label{sec:obs}

\begin{figure}
    \includegraphics[width=\columnwidth]{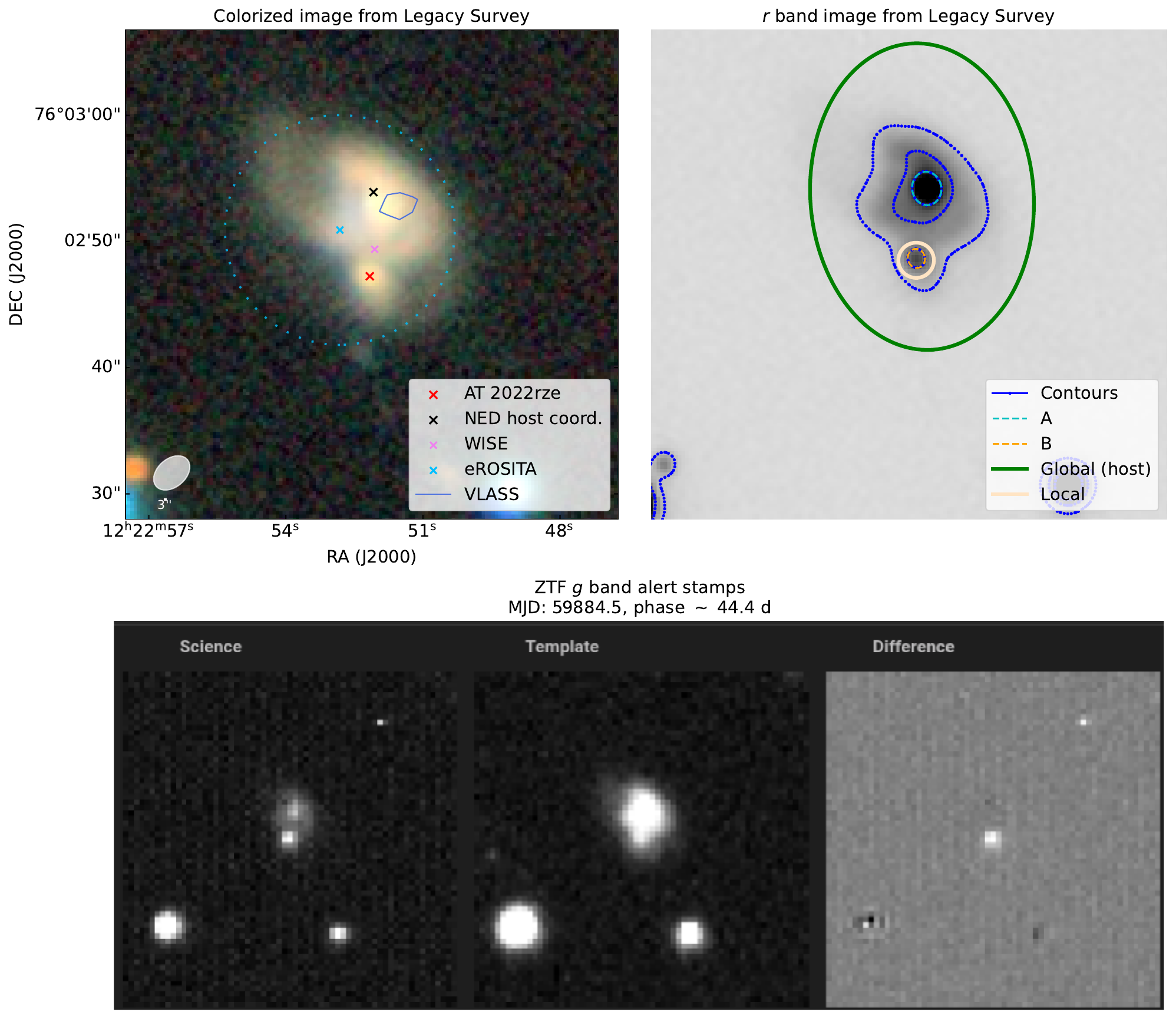}
    \caption{Host images. Top left panel: colorized Legacy Survey \citep[][]{2019AJ....157..168D} cutout centered on AT2022rze. The position of the transient is marked with a red cross. The coordinates of the host as obtained from NED are shown with a black cross. A purple cross shows the coordinates of the AllWISE observations. A light blue cross shows the coordinates of the eROSITA observations, a dotted circle of the same color with a 9.1\farcs radius shows the 98\% error radius for these observations. Tentative VLASS radio continuum emission is shown as a blue contour. The gray ellipse marks the size of the VLASS beam. 
    Top right panel: $r$ band Legacy Survey cutout centered on AT2022rze. A green and salmon circle denote the respective global and local regions considered for host characterization. Blue markers show constant value pixel contours. Cyan and orange dashed ellipses mark the respective fits to the smallest regions confined by the contours. For details see Sect.~\ref{sec:host}.
    Bottom panel: Stamps associated to a ZTF $g$ band detection alert. Science, Template and Difference of those respectively. These were obtained from the Alerce broker \citep[\url{https://alerce.online/},][]{2021AJ....161..242F}.}
    \label{fig:host}
\end{figure}

The discovery of AT2022rze (also known as Gaia22dmp, ZTF22abnfjsm and ATLAS22bmgb) was reported on August 21st 2022 by \cite{2022TNSTR2428....1H}. Its location at R.A.$=$12h22m52.16s, Dec$=+$76$^\circ$02$^\prime$47.26$^{\prime\prime}$, (J2000.0) allows the identification of the galaxy WISEAJ122252.07$+$760253.9 as the host. AT2022rze is located at the center of a small luminous ``blob'' (galaxy B, see Sect.~\ref{sec:host}) located South-East from the geometrical center of the host (see Fig.~\ref{fig:host}, see Sect.~\ref{sec:host} for further discussion of the host). AT2022rze was classified as a SN~IIn almost 4 months after its discovery, on December 1 2022, by \cite{2022TNSCR3514....1H}. Two years later, on December 16 2024, it was re-classified as an AGN by \citep[][]{2024TNSCR4932....1G}. Both classifications are based on spectroscopic features, raising questions about the true nature of this event. 

From the position of the hydrogen emission lines we derive a redshift of $z = 0.082$, consistent with the reported photometric redshift of the host \citep{2014ApJS..210....9B}. We employ the \texttt{astropy.cosmology} software adopting H$_{0} = 73$ km s$^{-1}$ Mpc$^{-1}$, $\Omega _{\mathrm{Matter}} = 0.27$, $\Omega _{\mathrm{Lambda}} = 0.73$ as cosmological parameters. With these, we obtain a distance modulus of $\mu = 37.76$~mag. We do not consider uncertainties related to host galaxy peculiar velocities as it is in the Hubble flow ($z > 0.02$) and corrections would be small. From the NASA/IPAC Extragalactic Database's (NED\footnote{The NASA/IPAC Extragalactic Database (NED) is funded by the National Aeronautics and Space Administration and operated by the California Institute of Technology.}) we obtain a Milky Way (MW) extinction of A$_{V} = 0.123$ mag. We use NED's Galactic Extinction Calculator\footnote{NED's Extinction Calculator considers the recalibration presented by \citet{2011ApJ...737..103S} to the extinction map presented by \citet{1998ApJ...500..525S}, assuming a \citet{1999PASP..111...63F} reddening law with R$_{\mathrm{v}} = 3.1$.}, accessed through the \texttt{ned\_extinction\_calc} script\footnote{\url{https://github.com/mmechtley/ned_extinction_calc}} to obtain the MW extinction in different photometric bands. The effective wavelength corresponding to each filter was obtained from the Spanish Virtual Observatory Filter Information service \citep[SVO,][]{2012ivoa.rept.1015R,2020sea..confE.182R}. We do not consider host extinction, although see Sect.~\ref{sec:host}. The available photometric and spectroscopic data are presented below.
 
\subsection{Photometry}
\label{sec:photometry}

 AT2022rze was detected by the public surveys carried out with the Gaia satellite \citep{2016A&A...595A...1G}, the Asteroid Terrestrial-impact Last Alert System \citep[ATLAS,][]{2018PASP..130f4505T,2020PASP..132h5002S} and the Zwicky Transient Facility \citep[ZTF,][]{2019PASP..131a8002B,2019PASP..131g8001G,2019PASP..131a8003M,2020PASP..132c8001D}. Thus, it has extensive optical photometric follow up. Additional photometric follow up was obtained as part of the ZTF collaboration with the Spectral Energy Distribution Machine \citep[SEDM,][]{2018PASP..130c5003B} on the 60-inch telescope (P60) at Palomar Observatory. Moreover, mid-Infrared (MIR) observations were obtained by the  Wide-field Infrared Survey Explorer (WISE) instrument.
 
\begin{figure*}
	\includegraphics[width=\textwidth]{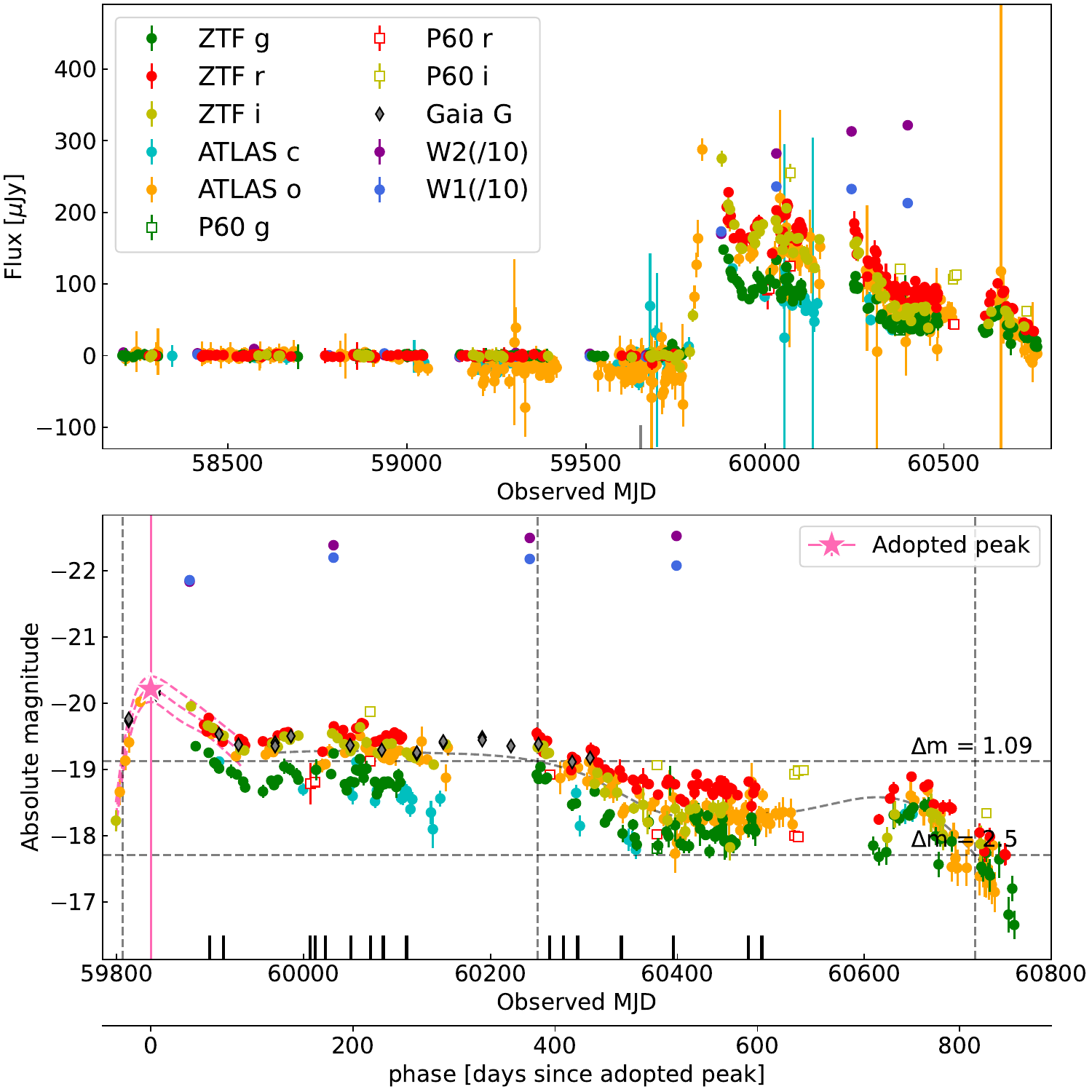}
    \caption{Top panel: Observed flux. WISE flux (but not absolute magnitude) is scaled by a factor 10 for better visibility. Bottom panel: Absolute magnitudes. Green, red and yellow circles represent the ZTF $gri$ bands respectively. Cyan and orange circles represent the ATLAS $co$ bands respectively. P60 SEDM $gri$ observations in the SDSS system are represented as green, red and yellow empty squares respectively. Black romboids represent the Gaia $G$ band. Dark magenta and dark blue circles show the WISE W2 and W1 bands, respectively. The central dashed pink line in the bottom panel indicate the Gaussian Process interpolation to the earlier portion of the light curve used to estimate the peak epoch, shown as a pink star, the top and bottom dashed pink lines indicate the associated interpolation confidence interval (see Sect.~\ref{sec:lcandgp}). The central dashed grey line shows the ALR interpolation after peak. Vertical and horizontal gray dashed lines indicate the rise and decline times. Short vertical lines on the x-axis indicate epochs of spectroscopic observations.}
    \label{fig:lcs}
\end{figure*}

The observed photometry is listed in Tables~\ref{table:ZTFandATLASphot}, \ref{table:SEDMphot} and \ref{table:WISEphot}. The corresponding light curves are shown in Fig.~\ref{fig:lcs}. Absolute magnitudes are obtained as $M_{\lambda} = m_{\lambda} - \mu - A_{\lambda} - K\_corr$, where $m_{\lambda}$ is apparent magnitude in the AB system, $\mu = 37.76$~mag is the distance modulus (see Sect.~\ref{sec:intro}), $A_{\lambda}$ is the MW extinction in the considered band and $K\_corr$ is the cosmological term for the K-correction \citep{2002astro.ph.10394H} obtained as $-2.5\times\log(1+z)$ (as presented in \citealt{2023ApJ...943...41C} and \citealt{2025A&A...695A.142P}). 

\subsubsection{Gaia}

The Gaia Science Alerts project \citep{2021A&A...652A..76H} is dedicated to search for transient events within Gaia data. Alerts for discovered transients are made public through Gaia alerts that provide information about the transient, including the average magnitude of the source across all CCD strips at each considered timestamp. 
We obtained the public light curve from the dedicated repository\footnote{\url{https://gsaweb.ast.cam.ac.uk/alerts/}}. The estimated absolute magnitudes are presented as black rhomboids in the bottom panel of Fig.~\ref{fig:lcs}. Since the public light curves do not include associated error bars, we do not include error bars here. 

\subsubsection{ATLAS}

The ATLAS survey scans the sky in two wide filters: $c$ (or ``cyan'' band, in the wavelength range 4200–6500~\AA, roughly corresponding to the typical $g + r$ range), and $o$ (or ``orange'' band, in the wavelength range of 5600–8200~\AA, roughly corresponding to the typical $r + i$ range); with a cadence of around two days \citep{2020PASP..132h5002S,2021TNSAN...7....1S}. We retrieved the ATLAS forced-photometry from the dedicated repository\footnote{\url{https://fallingstar-data.com/forcedphot/}}, and processed it utilizing the suggested pipeline developed by \cite{Young_plot_atlas_fp}, with the intra-night stacking option. The $c$ and $o$ light curves are presented in flux and absolute magnitude in the top and bottom panels of Fig.~\ref{fig:lcs}, respectively.

\subsubsection{ZTF}
\label{sec:photZTF}

ZTF is a wide-field (47-square-degree field of view), high-cadence survey (minutes to days, with a three day average for the public survey, \citealt{2019PASP..131f8003B}), that covers the whole northern sky. ZTF forced point-spread function (PSF) photometry is processed and distributed by the Science Data System at IPAC\footnote{Formerly referred to as the Infrared Processing \& Analysis Center (\url{https://www.ipac.caltech.edu}).} \citep{2019PASP..131a8003M}. We requested the available forced photometry following the steps outlined in the ZTF forced photometry guideline\footnote{\url{http://web.ipac.caltech.edu/staff/fmasci/ztf/forcedphot.pdf}}. We process the retrieved light curves in the same way as outlined by \cite{2025A&A...695A.142P}, removing points with bad quality flags and finding the zero flux baseline corresponding to each filter. The $gri$ light curves are presented in flux and absolute magnitude in the top and bottom panels of Fig.~\ref{fig:lcs}, respectively.

We obtained further P60 photometry through the ZTF collaboration. Photometry is obtained through PSF fitting, calibrated with standard stars from the Sloan Digital Sky Survey \citep[SDSS,][]{1995AAS...186.4405G}. All the photometry is processed using the \texttt{Fpipe} software \citep{2016A&A...593A..68F} using SDSS reference images. The P60 photometry is neither color-corrected nor adjusted for differences in filter response. Although we do not use these measurements in our analysis, we include them here for completeness.

\subsubsection{WISE}

The field of AT2022rze was observed with WISE in the W1 and W2 bands (with central wavelengths of 3.4 and 4.6 $\mu$m, respectively) during the NEOWISE-reactivation mission \citep[NEOWISE-R;][]{mainzer2014}. Single exposures in the W1 and W2 bands were collected from the NASA/IPAC Infrared Science Archive (IRSA) and coadded using the ICORE service \citep{masci2013}. Reference images were created by coadding data from 2015, which were subtracted from the subsequent images. 
We perform aperture photometry on the subtracted images using a standard 8.25\arcsec radius aperture (and corresponding aperture corrections), and zero magnitude fluxes for the WISE W1 and W2 bands of $F_{\nu,0}^{W1} = 306.7$ Jy and $F_{\nu,0}^{W2} = 170.7$ Jy \citep{Wright2010}.

\subsubsection{Host}
\label{sec:phot_host}

The host galaxy has been observed by the Wide-field Infrared Survey Explorer mission \citep[WISE,][]{2010AJ....140.1868W}, the Very Large Array Sky Survey \citep[VLASS,][]{2020PASP..132c5001L,2021ApJS..255...30G}, and eROSITA \citep[][]{2021A&A...647A...1P}. We note that the host is formed by two galaxies undergoing a merging event, namely galaxy A and galaxy B (see Fig.~\ref{fig:host} and Sect.~\ref{sec:host} for a detailed description). Throughout this work, we refer to the entire system as the host, specifying galaxy A or B when relevant.

MIR observations of the host are available in the AllWISE Source Catalog\footnote{\url{https://wise2.ipac.caltech.edu/docs/release/allwise/}}. AllWISE combines both the WISE cryogenic and NEOWISE \citep{2011ApJ...731...53M} post-cryogenic survey phases, providing observations at 3.4, 4.6, 12, and 22 $\mu$m, corresponding to the so called W1, W2, W3, and W4 filters, respectively. A coordinate search considering a 10$\arcsec$ cone radius retrieves two sources, one consistent with the NED coordinates of the host, and the other offset 0.4\farcs W, 4.5\farcs S from that position (see Fig.~\ref{fig:host}, see Sect.~\ref{sec:host} for further discussion). The magnitudes in each of the WISE filters are listed in Table~\ref{tab:WISE}.

\setlength{\tabcolsep}{4pt} 
\begin{table}
	\centering
	\caption{MIR observations obtained from the AllWISE Source Catalog. The magnitudes are presented in the Vega system. Error bars associated to each filter are presented in parethesis. }
	\label{tab:WISE}
	\begin{tabular}{lcccc} 
		\hline
		  & W1 & W2 & W3 & W4\\
            & [mag] & [mag] & [mag] & [mag]\\
		\hline
		Host   & 13.614 (0.039) & 13.382 (0.044) & 9.433 (0.045) & 7.542 (0.197)\\
            Other  & 13.878 (0.048) & 13.563 (0.050) & 9.895 (0.064) & 7.820 (0.251)\\		
            \hline
	\end{tabular}
\end{table}

Radio observations of the host are available in the VLASS archive\footnote{\url{http://cutouts.cirada.ca/}}.
We retrieve quick look single epoch radio continuum images from the VLASS, centered on the position of AT2022rze. These images sample the radio continuum between 2-4 GHz, and were taken on September 15 2017, September 3 2020, and March 5 2023. We tentatively detect radio continuum emission (2.5$\sigma$) close to the NED coordinates of the host in the  epoch 1 and 3 images, but do not find emission at the position of AT2022rze. In order to increase the signal-to-noise ratio of the images, we stack the three epochs together, co-adding the images weighted by their rms (respectively 0.12 mJy.beam$^{-1}$, 0.15 mJy.beam$^{-1}$, and 0.11 mJy.beam$^{-1}$). In the stacked image, we detect the 3 GHz radio continuum emission from the host with flux density $0.653\pm 0.2\,$mJy.   
We do not detect radio emission from AT2022rze, but derive a 3$\sigma$ upper limit $S_{3 GHz}<0.2\,$mJy within the beam ($\sim$3"), assuming radio continuum emission from the source is unresolved.

Observations by eROSITA performed between 2020 and 2021 (before the transient went off), detected the faint X-ray source SRGe~J122252.8$+$760251, located at $\rm R.A.=185.72006^{\circ}$, $\rm decl.=76.047474^{\circ}$. The radius of the localization region (at the 98\% confidence level) is 9.1\farcs (see Fig.~\ref{fig:host}). 

\subsection{Spectroscopy}

\begin{figure*}
	\includegraphics[width=0.84\textwidth]{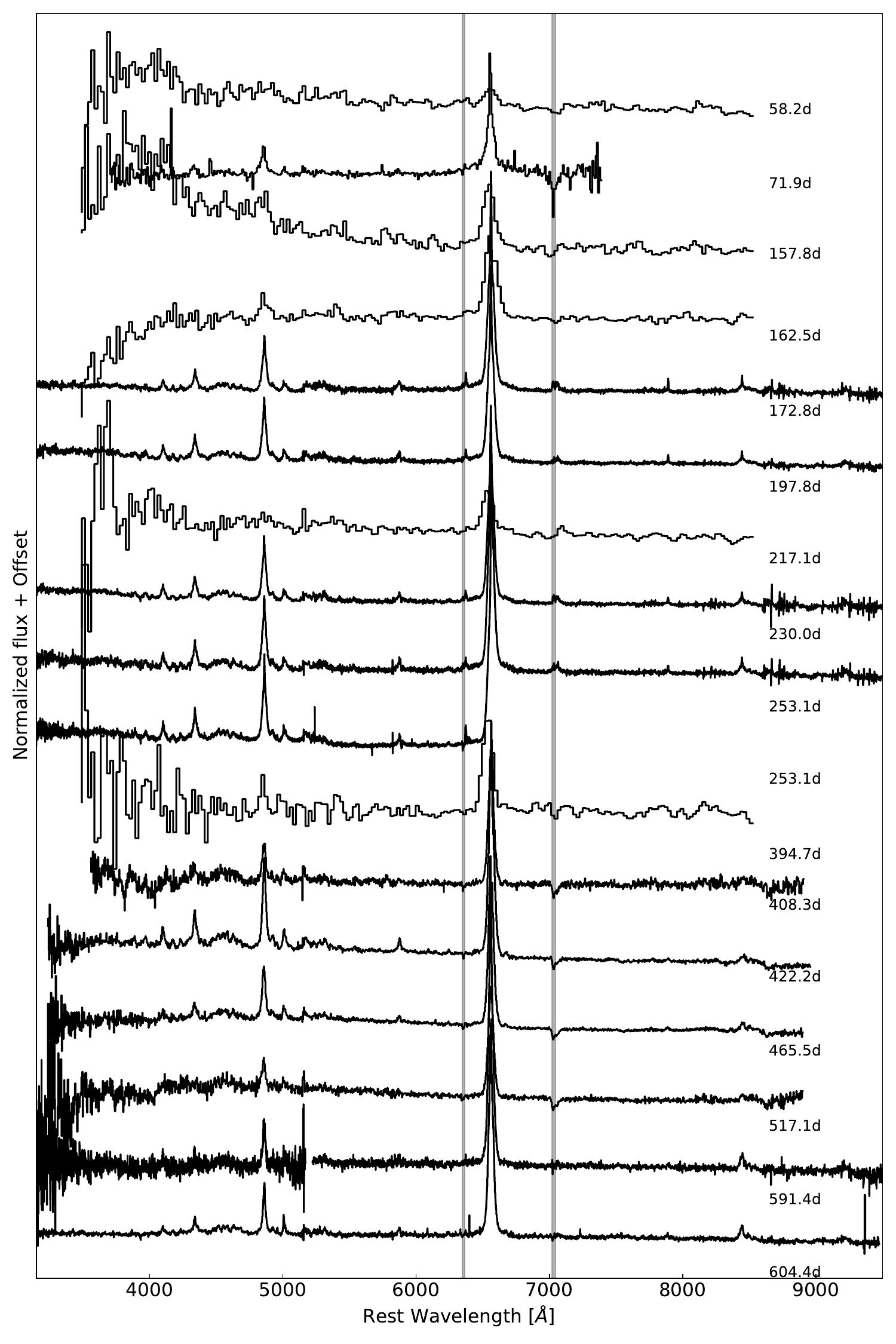}
    \caption{Spectral evolution of AT2022rze. The vertical gray stripes mark the telluric affected regions. Specific spectral lines are identified in Fig.~\ref{fig:lineids}. To the right of each spectrum we annotate the spectral rest frame phase with respect to light curve peak (see Sect.~\ref{sec:lcandgp}).}
    \label{fig:spectra}
\end{figure*}

In addition to the photometric follow up, we obtained several epochs of spectroscopic observations using the SEDM spectrograph, the Spectrograph for the Rapid Acquisition of Transients \citep[SPRAT,][]{2014SPIE.9147E..8HP} at the 2-m Liverpool Telescope, the Low-Resolution Imaging Spectrograph \citep[LRIS,][]{1995PASP..107..375O} at the 10-m Keck I telescope, the Alhambra Faint Object Spectrograph and Camera (ALFOSC) at the Nordic Optical Telescope, and the Double Spectrograph \citep[DBSP,][]{1982PASP...94..586O} at the Palomar 200-inch telescope. The SEDM spectra were reduced using the \texttt{pysedm} fully automated integral field spectrograph pipeline \citep{2019A&A...627A.115R, 2022PASP..134b4505K}. SPRAT spectra were reduced using an adaptation of the \texttt{FRODOspec} pipeline \citep{2012AN....333..101B}. LRIS spectra were reduced using the \texttt{LPipe} automated reduction pipeline \citep{2019PASP..131h4503P}. ALFOSC spectra were reduced using the \texttt{PypeIt} python spectroscopic data reduction pipeline \citep{2020JOSS....5.2308P,pypeit:joss_pub}. DBSP spectra were reduced using the \texttt{DBSP\_DRP} automated spectroscopic data reduction pipeline \citep{2022JOSS....7.3612M}. Follow-up observations were coordinated using the Fritz instance of SkyPortal \citep{2019JOSS....4.1247V, 2023ApJS..267...31C}. The log of spectroscopic observations can be found in Table~\ref{table:speclog}. The obtained spectral evolution is shown in Fig.~\ref{fig:spectra}.

Most of the spectra were obtained with the slit aligned along the parallactic angle to minimize atmospheric dispersion effects. The ALFOSC spectra obtained at 422.2 and 465.5 days were observed with a slit rotation that allowed to include both the transient (located at the center of galaxy B) and galaxy A (see Sect.~\ref{sec:host}) and thus, obtain two spectral epochs of the latter. These are shown in Fig.~\ref{fig:host_spec}.

The position of the transient (i.e. galaxy B) was also observed by the Dark Energy Spectroscopic Instrument \citep[DESI,][]{2016arXiv161100036D,2016arXiv161100037D} on March 15th 2022, $\sim$ five months before the transient was discovered. The spectrum is included in their recent data release \citep{2025arXiv250314745D} and we include it in Fig.~\ref{fig:host_spec}.

\section{Analysis}
\label{sec:analy}

AT2022rze was originally classified as a SN~IIn. Because of its peak absolute magnitude, it is included in the SLSN~II sample presented by \cite{2025A&A...695A.142P}. AT2022rze does not stand out from the rest of the SLSN~II sample. However, \cite{2025A&A...695A.142P} only include observations up to December 12th 2022, an arbitrary date that could result in missed light curve characteristics. This event was re-classified as an AGN in December 16th 2024. Thus, we go back to AT2022rze and find that its light curve continues to show activity for over 700 days after peak. Yet, there is an apparent lack of evolution of the spectral lines that resulted in the original SN classification, suggesting that such classification could still be valid. In the following, we analyze the available observations in order to decipher the true nature of this transient.

\subsection{Light curve}

Although the reported discovery date of AT2022rze is August 21st 2022 (MJD $=$ 59812.8), we can see in the bottom panel of Fig.~\ref{fig:lcs} that the transient was first detected on August 8th 2022 (MJD $=$ 59799.18) in the ZTF $i$ band at $-18.2 \pm 0.2$~mag. The top panel of Fig.~\ref{fig:lcs} shows the measured historical flux at the position of the target since March 28th 2018 (MJD $=$ 58205.3). We can see that the historical optical and NIR light curves show no flux variations until the first ZTF $i$ band detection, meaning that there was no transient event at the time, or at least, no optical activity brighter than the surveys limiting magnitude.  

\subsubsection{Light curve peak and timescales}
\label{sec:lcandgp}

The gaps in observations around the first peak prevent an accurate estimate of the peak epoch in any given filter. Given the overlap between the $rio$ and $G$ filters, we consider no color evolution around the time of peak, and include the observations in all these bands to estimate an overall approximate peak epoch. To do this, we use Gaussian Process \citep[GP, e.g:][]{2006gpml.book.....R} to interpolate the observations that cover the main (more luminous) light curve peak in the mentioned filters. GP is implemented utilizing the \texttt{GPy} Python package\footnote{\url{https://gpy.readthedocs.io/en/deploy/}.}. We consider the epoch and absolute magnitude of the peak of the GP median to be the peak epoch and peak absolute magnitude of AT2022rze. Errorbars represent the standard deviation of the estimated peak of the median of 1000 posterior samples. The interpolation together with the confidence region is shown as dashed pink lines in the bottom panel of Fig.~\ref{fig:lcs}. The adopted peak epoch is MJD $=$ 59836.5 $\pm$ 2.4, and the adopted peak absolute magnitude is $-$20.2 $\pm$ 0.2 mag, shown as a pink star in the bottom panel of Fig.~\ref{fig:lcs}. Since the interpolation considers $rio$ and $G$, this peak epoch does not correspond to a single band, but we would associate the peak with the $o$ band, as this is the filter with the wider passband and the most observations in the considered region. All the phases in this work are presented as rest frame days with respect to the adopted peak, unless stated otherwise.

\cite{2025A&A...695A.142P} follows the work of \cite{2023ApJ...943...41C} and define rise and decline timescales as flux fractions with respect to peak. The considered fractions are 10\% and 1/e, which correspond to $\Delta$mag $=$ 2.5 and $\Delta$mag $=$ 1.09, respectively. Dashed gray horizontal lines in Fig.~\ref{fig:lcs} indicate the considered $\Delta$mag's. Using the GP interpolation described above, we can estimate a 1/e rise time t$_{\mathrm{rise,1/e}} = 27.5 \pm 0.6$ days. t$_{\mathrm{rise,10\%}}$ cannot be determined due to the lack of early observations. 
We can see that the light curve shows many wiggles and bumps after peak, it also has several gaps in the observations. Thus, GP interpolation is not ideal to interpolate the whole light curve. Instead, we interpolate the post-peak $o$ band light curve using the Automated Loess Regression (\texttt{ALR}) pipeline presented by \cite{2019MNRAS.483.5459R}. The resulting interpolation is presented as a gray-dashed line in Fig.~\ref{fig:lcs}. We use this ALR interpolation to determine the decline timescales t$_{\mathrm{dec,1/e}}$ and t$_{\mathrm{dec,10\%}}$, which are $382.9 \pm 0.6$ and $815.2 \pm 0.6$ days, respectively. 

These rise times of AT2022rze are comparable to the rise times presented for SLSNe~II by \cite{2025A&A...695A.142P}. They report a median t$_{\mathrm{rise,1/e}}$ of 33.8, 37.0 and 46.9 days in the $gri$ bands, respectively. On the contrary, the decline times of AT2022rze are much longer than those presented for SLSNe~II. The median t$_{\mathrm{dec,1/e}}$ of SLSNe~II being 75.0, 89.3 and 98.3 days in the $gri$ bands, respectively; and the median t$_{\mathrm{dec,10\%}}$ being 235.8, 248.1, 246.8 days in the $gri$ bands, respectively. However, \cite{2025A&A...695A.142P} find that a fraction of slow declining SLSNe~II exist ($\sim$ 5\% of their sample), that take over a year to fade. 

\subsubsection{Color evolution}
\label{sec:color}

\begin{figure*}
	\includegraphics[width=\textwidth]{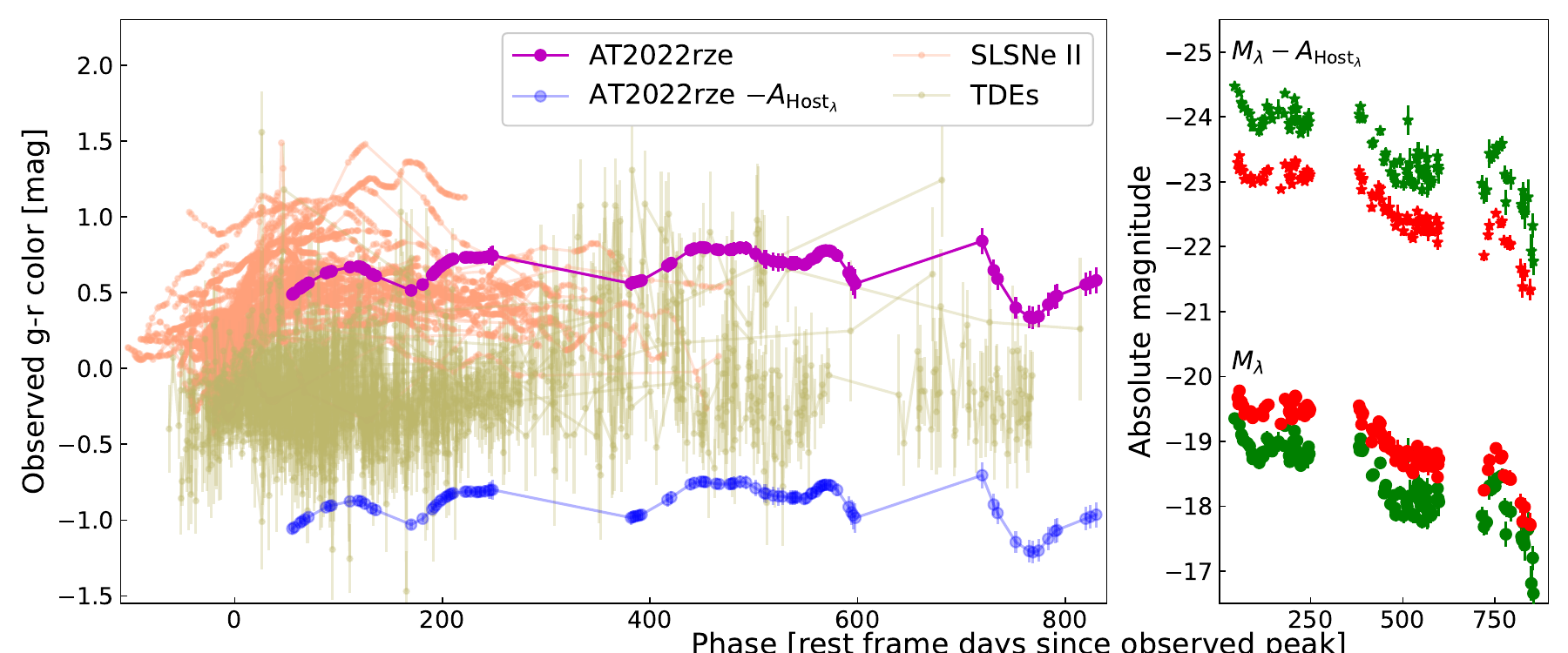}
    \caption{Colors. Left panel: $g-r$ color evolution of AT2022rze (magenta) compared to the same colors of the SLSN~II sample (salmon) presented by \protect\cite{2025A&A...695A.142P} and the sample of TDEs presented by \protect\cite{2023ApJ...955L...6Y}. In blue we show the respective color of AT2022rze when considering host extinction (see Sect.~\ref{sec:host}). Right panel: ZTF $g$ and $r$ band absolute magnitude light curves in green and red respectively. Round markers show the absolute magnitude light curves obtained as described in Sect.~\ref{sec:photometry}. Star markers show the absolute magnitude light curves obtained considering substantial host extinction (see Sect.~\ref{sec:host}).}
    \label{fig:color}
\end{figure*}

The full ZTF $gr$ band light curves of AT2022rze were interpolated with \texttt{ALR}. These interpolations were used to construct the $g-r$ color curve presented in Fig.~\ref{fig:color}. We compare the observed color of AT2022rze to those of the SLSN~II sample presented by \cite{2025A&A...695A.142P} and to those of TDEs presented by \cite{2023ApJ...955L...6Y}. We find better agreement with SLSNe~II (although see Sect.~\ref{sec:host}). We note that we do not have color information before or around peak, but the evolution at the earliest observed phases seem to follow the trend of the SLSN~II sample. 

\subsubsection{Pseudo-bolometric light curve}
\label{sec:pbol_lc}

\begin{figure}
	\includegraphics[width=\columnwidth]{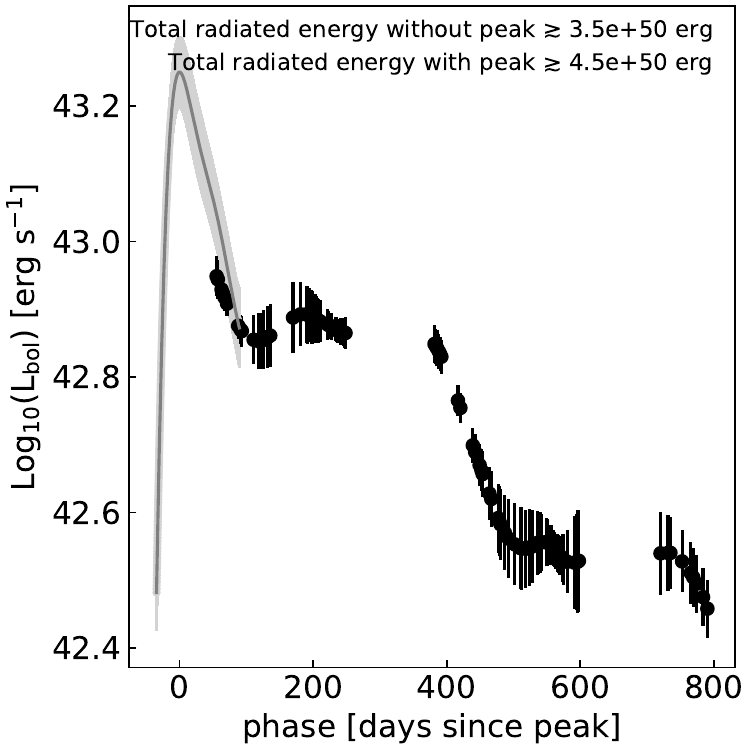}
    \caption{AT2022rze pseudo-bolometric light curve. Black markers show the light curve estimated using only $gri$ bands. The solid grey curve shows the peak estimated using $\mathrm{BC}_{o}$ on the early time GP interpolation.}
    \label{fig:pseudobol}
\end{figure}

To constrain the total radiated energy we construct the pseudo-bolometric light curve of AT2022rze. We only consider ZTF $gri$ observations. To have a consistent estimate of the spectral energy distribution (SED) during the evolution of the event, we only consider epochs in which all three bands were observed. The SED is estimated by interpolating the $gri$ bands with respect to the adopted peak epoch using \texttt{ALR}, and integrating over the resulting interpolated curve. The obtained light curve is presented in Fig.~\ref{fig:pseudobol}. This approach only provides a lower limit for the bolometric luminosity as it completely ignores UV and IR contributions to the SED, as well as the effect of host galaxy extinction. Still, it allow us to place a lower limit on the total radiated energy, which we estimate to be $E_{\mathrm{rad}} \gtrsim 3.5 \times 10^{50}$~erg. 

Although the peak of the light curve was missed in the $gri$ bands, it was observed in the $o$ band. We note that there is some overlap between the first $gri$ observations and the earlier $o$-band photometry. We decided to consider this overlapping region to estimate a bolometric correction as:

\begin{equation}
    \mathrm{BC}_{o} = -2.5 \times log(\mathrm{L_{bol}}/\mathrm{L}_{o}),
\end{equation}

\noindent where $\mathrm{BC}_{o}$ is the bolometric correction in the $o$ band, $\mathrm{L_{bol}}$ is the estimated pseudo-bolometric luminosity and $\mathrm{L}_{o}$ is the luminosity in the $o$ band. Once we obtained $\mathrm{BC}_{o}$, we consider the GP interpolation described in Sect.~\ref{sec:lcandgp} to represent the $o$ band at earlier times, and convert it to pseudo-bolometric luminosity through the estimated $\mathrm{BC}_{o}$. The solid grey curve in Fig.~\ref{fig:pseudobol} shows the additional peak. The total radiated energy when considering this light curve is $E_{\mathrm{rad}} \gtrsim 4.5 \times 10^{50}$~erg. The complete pseudo-bolometric light curve is presented in Table~\ref{table:pbolLC}.

\subsection{Spectra}
\label{sec:spectra}

\begin{figure*}
    \hspace*{-8em}
	\includegraphics[scale=0.58]{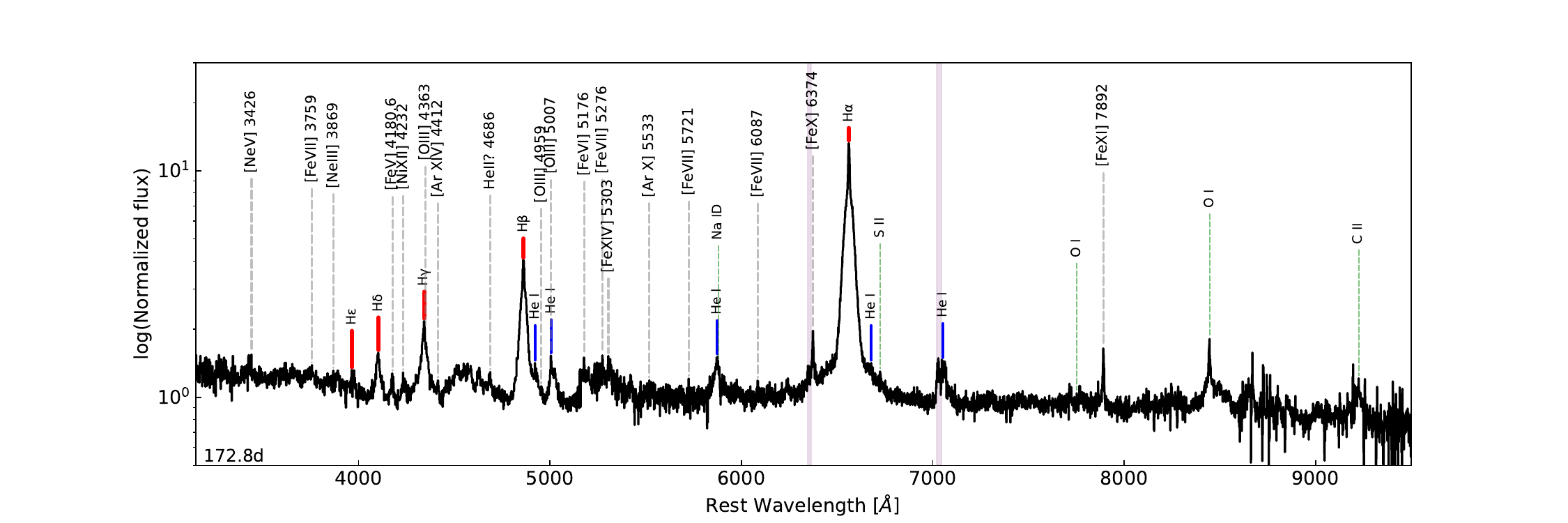}
    \caption{Line identification. The position of each spectral line is marked with vertical lines and the corresponding ion label at the top of each one. Shaded regions show areas affected by telluric lines.}
    \label{fig:lineids}
\end{figure*}

The spectral evolution of AT2022rze is presented in Fig.~\ref{fig:spectra}. 
The first available spectrum has low resolution and was observed $\sim$ 58 days after peak. The following observed spectra seem to show little to no spectral line evolution. In Fig.~\ref{fig:lineids} we show 
the identified spectral lines on the spectrum obtained at $\sim$ 173~d. Besides the hydrogen lines that originally led to the Type II SN classification, we can see several lines from \ion{He}{I}. In particular, the \ion{He}{I} $\lambda$5876 is blended with \ion{Na}{ID}. It is unclear whether \ion{Fe}{II} lines are present at $\lambda$4924 and $\lambda$5018, as they overlap with \ion{He}{I} $\lambda$4922 and $\lambda$5016 respectively. There is also weak \ion{S}{II} $\lambda$6715 and possibly \ion{C}{II} $\lambda$9234, although the redder part of the spectrum is quite noisy, making it difficult to properly identify lines. We see \ion{O}{I} $\lambda$8446 but there is barely any indication of \ion{O}{I} $\lambda$7774. The absence of \ion{O}{I} $\lambda$7774 and presence of \ion{O}{I} $\lambda$8446 could indicate the presence of Bowen resonance-fluorescence mechanisms, which may indicate the existence of intense far-UV radiation fields \citep[see][]{2006agna.book.....O}. We tentatively identify \ion{He}{II} $\lambda$4686. In addition, we see several high excitation lines, such as \ion{[Ne}{V]} $\lambda$3426, \ion{[Ne}{III]} $\lambda$3869, \ion{[Fe}{VII]} $\lambda$3759, \ion{[Fe}{V]} $\lambda$4180.6, \ion{[Ni}{XII]} $\lambda$4232, \ion{[O}{III]} $\lambda$4363, \ion{[Ar}{XIV]} $\lambda$4412, \ion{[O}{III]} $\lambda$4959, \ion{[O}{III]} $\lambda$5007, \ion{[Fe}{VI]} $\lambda$5176, \ion{[Fe}{VII]} $\lambda$5276, \ion{[Fe}{XIV]} $\lambda$5303, \ion{[Ar}{X]} $\lambda$5533, \ion{[Fe}{VII]} $\lambda$5721, \ion{[Fe}{VII]} $\lambda$6087, \ion{[Fe}{X]} $\lambda$6374, and \ion{[Fe}{XI]} $\lambda$7892. Some of these lines are typically known as coronal lines, and require intense UV/X-ray radiation or collisionally ionized hot gas to be formed \citep[see][]{2006agna.book.....O}. The spectra show little evolution and the same lines can be identified during most of the spectral sequence.

\subsection{Host}
\label{sec:host}

AT2022rze has been ingested by the \texttt{Blast} web application\footnote{\url{https://blast.scimma.org/}} \citep{2024arXiv241017322J}, which provides global and local (within 2~kpc) properties of a transients' host galaxy. The software uses UV, optical and IR archival data of the host to fit the spectral energy distribution (SED) and estimate different characteristics. \texttt{Blast} considers the (global) host galaxy of AT2022rze to be enclosed by the green ellipse shown in the top right panel of Fig.~\ref{fig:host}. Throughout this work, we have been considering that to be the host of AT2022rze, WISEAJ122252.07$+$760253.9. 

Images from many shallow surveys (including WISE) suggest that host WISEAJ122252.07$+$760253.9 has an irregular morphology. A visual inspection of deeper images from the Legacy Survey suggest that this is because it is formed by two separate galaxies undergoing a merging event (see Fig.~\ref{fig:host}). To explore this possibility and better understand the transient’s location within its host, we perform a qualitative analysis of the Legacy Survey $r$ band image shown in the top right panel of Fig.~\ref{fig:host}. We first displayed the image on the SAOImage DS9 software \citep{2000ascl.soft03002S,2003ASPC..295..489J}, considering an intensity range from $-$0.1 to 0.6. Then, we used the contours analysis tool to find the curves defined by constant value pixels, and find two different regions of maximum intensity, indicating that the host is a merging system. 

To be consistent with the above, we will still consider the host to be the overall merging system. This is, we will still call host to the area contained by the green ellipse in Fig.~\ref{fig:host}. The center of the host defined in this way is R.A.$=$12h22m52.04s, Dec$=+$76$^\circ$02$^\prime$52.03$^{\prime\prime}$, this is 1.9\farcs away from the WISEAJ122252.07$+$760253.9 coordinates reported on NED. The host is formed by galaxy A and galaxy B (for ``blob''), the nuclear regions of which are marked by dashed cyan and orange ellipses, respectively, in Fig.~\ref{fig:host}. These ellipses were obtained by fitting the contours found by the SAOImage DS9 software. The center of galaxy A is R.A.$=$12h22m51.944s, Dec$=+$76$^\circ$02$^\prime$52.643$^{\prime\prime}$. The center of galaxy B is R.A.$=$12h22m52.155s, Dec$=+$76$^\circ$02$^\prime$47.401$^{\prime\prime}$. 

The location of AT2022rze is then offset 4.8\farcs (7.1~kpc), 5.4\farcs (8.1~kpc) and 0.1\farcs (0.2~kpc) from the center of the host, the center of galaxy A and the center of galaxy B, respectively. This is a rough approximation and dedicated analysis of deeper, higher resolution images are needed to better locate the transient. Still, the host galaxy appears to be a galaxy merger with disturbed morphology and AT2022rze appears coincident with one of the nuclei of this galaxy merger (identified as galaxy B).

The \texttt{Blast} analysis concludes that the host of AT2022rze is consistent with an AGN, both in their global and local properties, even though the latter only include optical bands. Note that the global analysis includes the entirety of the host while the local analysis only includes galaxy B (see upper right panel of Fig.~\ref{fig:host}). While the \texttt{Blast} approach has some limitations \citep[see][]{2024arXiv241017322J}, it offers a useful and efficient means of rapidly assessing the host galaxy and estimating its properties. The parameters inferred for AT2022rze can be found at \url{https://blast.scimma.org/transients/2022rze/}. Most relevant to decipher the nature of AT2022rze are: the estimated median mass, which is 5.5$\times 10^{10}$~M$_{\odot}$ for the global analysis, and 4.9$\times 10^{9}$~M$_{\odot}$ for the local analysis; the estimated star formation rate (SFR), which is 4.3~M$_{\odot}$~y$^{-1}$ and 0.03~M$_{\odot}$~y$^{-1}$ for the global and local analysis, respectively; and the AGN role (or $f_{\mathrm{AGN}}$, fraction of bolometric luminosity emission due to AGN), which is 1.9\% and 18\% for the global and local analysis, respectively. These parameters indicate that the host is a massive galaxy with a moderate star formation and a very weak influence of an AGN. On the contrary, the local analysis is consistent with a low-mass galaxy (similar to a dwarf galaxy), with very low star formation and a non-negligible probability of AGN activity. Although this analysis is limited and insufficient to securely claim the presence of an AGN, and the local photometry would need to be deconvolved to accurately estimate the mass of both galaxies, the differences between the global and local analysis, together with the qualitative assessment of the morphology of the galaxy, indicates that galaxy B, at which AT2022rze is centered, is indeed a smaller galaxy with a nuclear AGN merging with a bigger, star forming galaxy.

\begin{figure}
    \includegraphics[width=\linewidth]{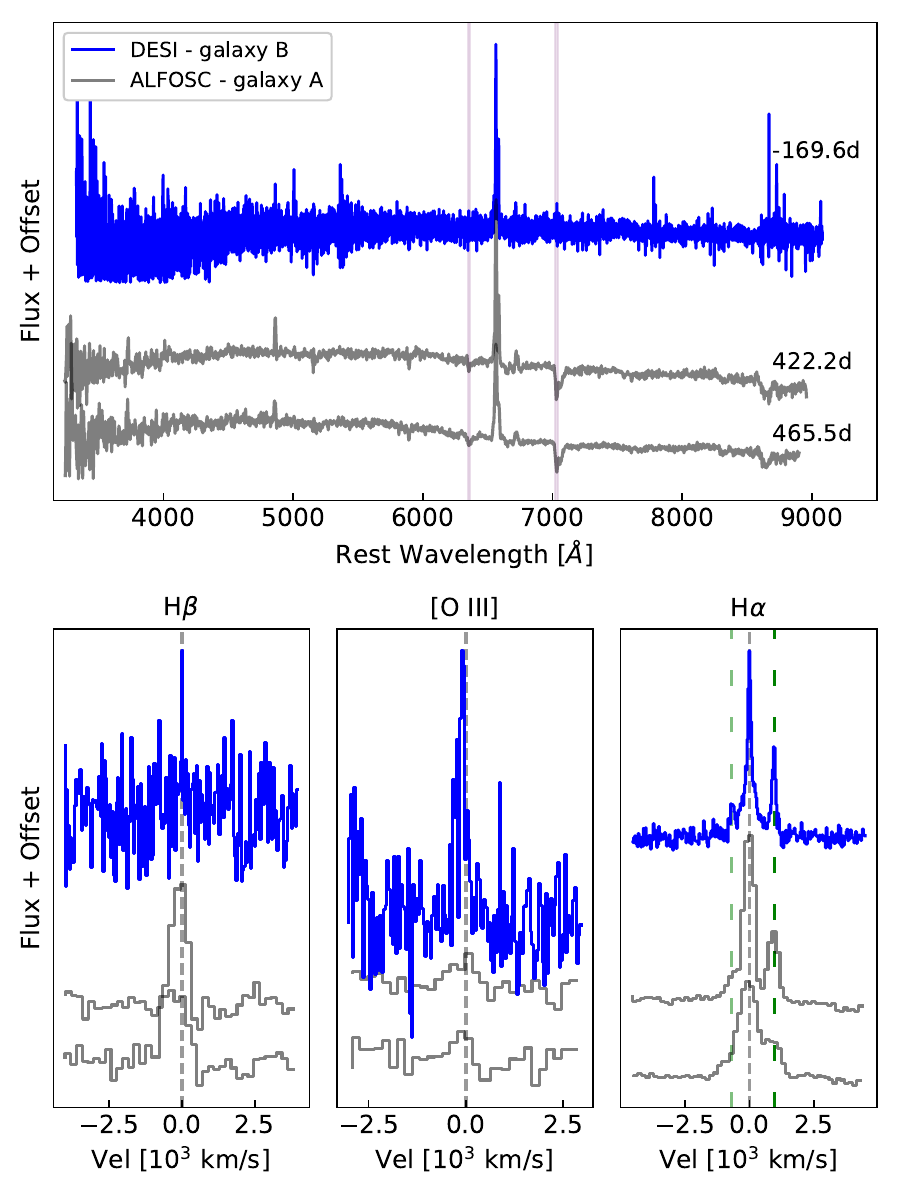}
    \caption{Host spectra normalized to the continuum. Top panel: DESI spectrum obtained at galaxy B $\sim$ 170 days prior to peak in blue. ALFOSC spectra obtained  over 400 days after peak at galaxy A in grey. Shaded areas show the telluric region. Bottom panel: zoom in to the H$\beta$, \ion{[O}{III]} $\lambda$5007 and H$\alpha$ lines (left to right, respectively). Gray vertical dashed lines show respective rest velocities. Green vertical dashed lines in the rightmost panel mark the position of the \ion{[N}{II]} $\lambda$6548 and $\lambda$6584 .}
    \label{fig:host_spec}
\end{figure}

Galaxy B was observed by DESI a few months before AT2022rze went off. The spectrum is rather noisy (see Fig.~\ref{fig:host_spec}). 
When compared to our ALFOSC spectra obtained at the center of galaxy A, we can see that the H$\beta$ feature of galaxy B is narrower and the \ion{[O}{III]} $\lambda$5007 feature is much stronger than that seen in galaxy A (bottom panel of Fig.~\ref{fig:host_spec}).

\begin{figure}
	\includegraphics[width=\columnwidth]{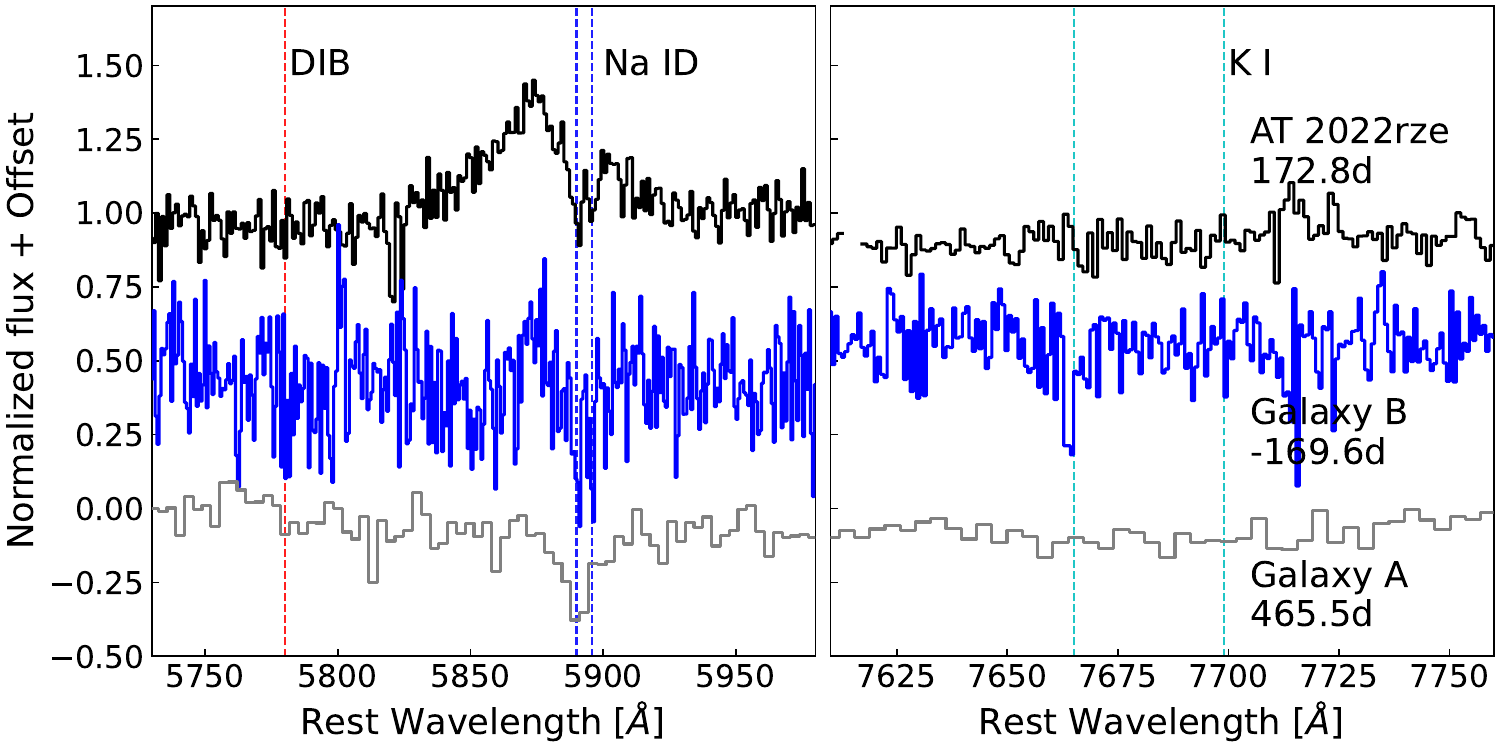}
    \caption{Spectral lines associated to host extinction. AT2022rze is presented in black, galaxy B in blue, and galaxy A in grey.}
    \label{fig:hostext}
\end{figure}

Although we do not consider host extinction for AT2022rze, we can see \ion{Na}{ID} absorption lines both in the transient and in the host (see Fig.~\ref{fig:hostext}). It has been argued that the pseudo-equivalent width (pEW) of these lines can be used to estimate the associated host extinction as:

\begin{equation}
\label{eq:host_extinction}
    log10(\mathrm{E}(B - V )) = 1.17 \times \mathrm{pEW(D1} + \mathrm{D2)} - 1.85 \pm 0.08,
\end{equation}

\noindent where $\mathrm{pEW(D1} + \mathrm{D2)}$ is the sum of the pEW of both \ion{Na}{ID} components \citep{2012MNRAS.426.1465P}. This method has several caveats and becomes insensitive at pEW(\ion{Na}{ID}) $\gtrsim$ 1\AA\ \citep{2013ApJ...779...38P}. 
From the spectrum of AT2022rze, we obtain $\mathrm{pEW(D1)} = 1.1 \pm 0.4$ \AA\ and $\mathrm{pEW(D2)} = 0.6 \pm 0.3 $ \AA\, this is consistent with the pEW of the same lines in the galaxy spectra. Thus, attempts to estimate host extinction using this method would be inadequate. \cite{2013ApJ...779...38P} also consider the diffuse interstellar band (DIB) $\lambda$5780 and the \ion{K}{I} lines $\lambda$7665 and $\lambda$7699 as proxy for independently estimating host extinction. We do not clearly see these lines neither on the transient nor on the galaxies (see Fig.~\ref{fig:hostext}). Still, we measure a Balmer decrements $f_{\mathrm{H\alpha}}/f_{\mathrm{H\beta}} > 3.1$ at all phases (see Fig.~\ref{fig:tde} and Table~\ref{table:BalmerDecandVel}), which is typically indicative of extinction.

Given that there is evidence of host extinction, we decided to consider equation~\ref{eq:host_extinction}, even if inadequate, to provide a rough estimate of the possible impact of such extinction. We obtained a host extinction of E$_{\mathrm{B-V}} = 1.38 \pm 0.08$ mag that, considering R$_{\mathrm{V}} = 3.1$, implies an A$_{V}{_\mathrm{host}} \sim 4.3$ mag. This would imply a peak absolute magnitude of $\sim -$24.5~mag ($\sim$ 4~mag brighter than without considering host extinction) and total radiated energies of $E_{\mathrm{rad}} \gtrsim 1.4 \times 10^{52}$~erg (considering the pseudo bolometric peak, see Sect.~\ref{sec:pbol_lc}), similar to the radiated energy of other ambiguous transients \citep[e.g.,][]{2017NatAs...1..865K,2018Sci...361..482M}. 

The color of AT2022rze also significantly changes when considering host extinction. The observed $g-r$ color is presented in the left panel of Fig.~\ref{fig:color}, while the absolute magnitude $g$ and $r$ light curves are presented in the right panel of the same figure. Inspecting the light curves before applying host extinction (circles), we would conclude that AT2022rze is rather red, and definitely redder than TDEs at early times. However, after applying the estimated host extinction, the transient (stars) becomes much bluer. Given our approximations, the true host galaxy extinction would likely be somewhere between these two extremes. In order to effectively determine the host extinction, we would need better resolution spectra, to be able to identify and measure the column density of the extinction indicators. Still, our rough estimation suggests that AT2022rze may be bluer than observed.

\subsection{Comparison to other events}
\label{sec:compTOothers}

Although AT2022rze was classified as a SN and shows some SN characteristics, it was later classified as an AGN. Yet, both classifications are based on spectral characteristics, which highlights the uncertainties present when performing spectral matching. When the classification between SN and AGN is dubious, one should also consider the possibility of a TDE, as these nuclear transients can also appear similar. Below we compare AT2022rze to well studied
events, including: SNe (SN~2010jl, \citealt{2014ApJ...797..118F}; iPTF13z, \citealt{2017A&A...605A...6N}; iPTF14hls, \citealt{2017Natur.551..210A,2019A&A...621A..30S}), AGNs (Mrk1044 and ESO424-12 both obtained from the BAT AGN Spectroscopic
Survey\footnote{BASS: \url{https://www.bass-survey.com/}}) and TDEs (SDSS1115+0544, originally considered a CLAGN by \citealt{2019ApJ...874...44Y} it was re-classified as a TDE by \citealt{2022ApJS..258...21W} and \citealt{2025arXiv250108812Z}; ASASSN-14li, \citealt{2016MNRAS.455.2918H,2022A&A...659A..34C}) that show similarities, either in the light curve or spectra, to AT2022rze. We also include comparisons to samples of (SL)SN~IIn \citep{2013A&A...555A..10T,2025A&A...695A..29S}, TDEs \citep{2022A&A...659A..34C} and CLAGNs \citep{2024ApJ...966..128W,2024ApJ...966...85Z}.

\begin{figure}
	\includegraphics[width=\linewidth]{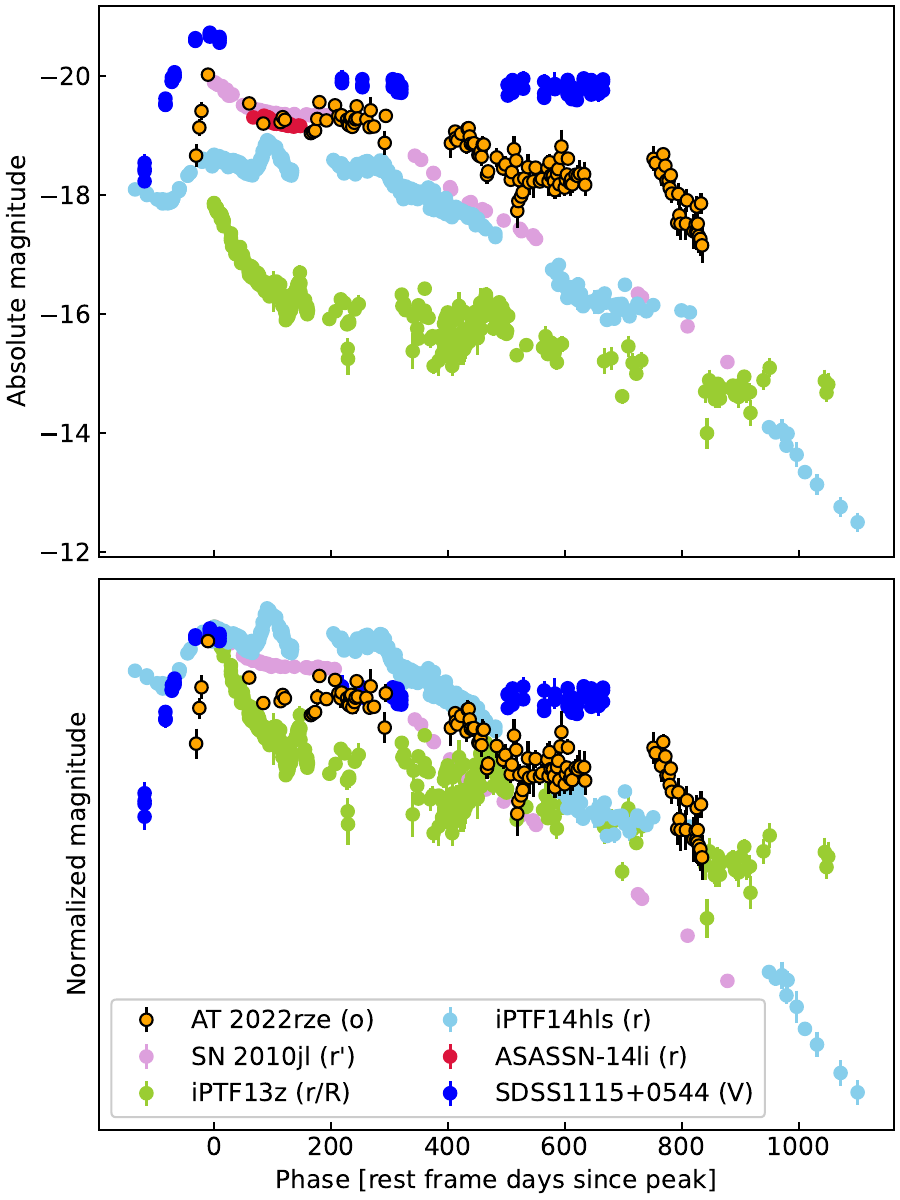}
    \caption{Light curve comparisons. Top panel: Absolute magnitude versus phase with respect to the corresponding light curve peak. Bottom panel: Same as top panel but normalized light curves with respect to peak absolute magnitude.}
    \label{fig:lc_comp}
\end{figure}

Figure~\ref{fig:lc_comp} shows the optical light curves of the considered comparison events (except Mrk1044 and ESO424-12 for which we do not have optical light curves). There is a large diversity among the light curves and a qualitative analysis is not enough to claim that AT2022rze is more similar to one event than to another. Compared to Type IIn SN~2010jl, AT2022rze shows a similar peak absolute magnitude and duration before a main steep decline, although the light curve of AT2022rze is ``bumpier''. In that sense, AT2022rze is similar to Type IIn iPTF13z, although the latter is not as luminous, and its bumps are more pronounced. The same is true when comparing to Type II iPTF14hls, although the late time decline of AT2022rze shows some resemblance to the late time decline of the former. SDSS1115+0544 is more luminous than AT2022rze and its rise to peak is smoother, its overall light curve seems not to be as bumpy as that of AT2022rze, although the light curve coverage is poorer. Unfortunately, the sampling of the light curve of ASASSN-14li is very limited. Still, we can see similar absolute magnitudes to those seen right after peak in AT2022rze. 

\begin{figure}
    \includegraphics[width=\linewidth]{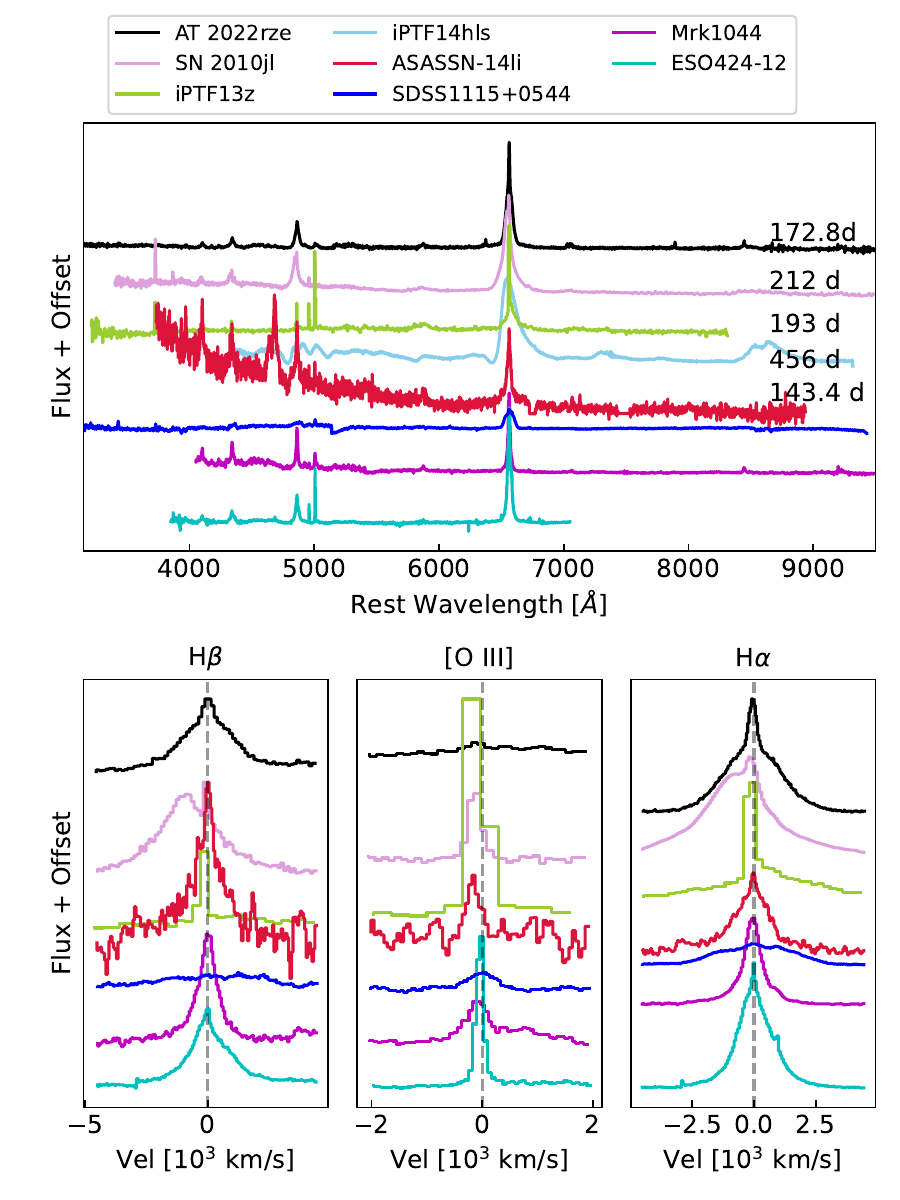}
    \caption{Spectroscopic comparisons. Spectra normalized to the respective continuum. Top panel: normalized optical spectra. Bottom panel: zoom in on the three main lines used to classify SNe and AGNs, namely H$\beta$ on the left panel, \ion{[O}{III]} $\lambda$5007 on the central panel, and H$\alpha$ on the right panel.}
    \label{fig:spec_comp}
\end{figure}

Figure~\ref{fig:spec_comp} shows spectral comparison between AT2022rze and the considered comparison events. Most objects show similar narrow lines, except iPTF14hls, that shows a typical Type II SN H$\alpha$ P-Cygni profile, and SDSS115$+$0544, which has broader H$\alpha$ and H$\beta$ emission lines. In the bottom panel of Fig.~\ref{fig:spec_comp} we show a zoom into the H$\beta$, \ion{[O}{III]} $\lambda$5007, and H$\alpha$ profiles. We exclude iPTF14hls here as its lines, typical of a regular SN~II, are much broader than for the rest of the sample. The H features of AT2022rze are more symmetrical with respect to rest velocity than the lines of the comparison SNe. In this sense, these features are more similar to those of the comparison AGNs and TDEs, although the broad component of these features is more pronounced for AT2022rze. The \ion{[O}{III]} $\lambda$5007 feature of every comparison event is stronger than that of AT2022rze.

\begin{figure}
    \includegraphics[scale=0.4]{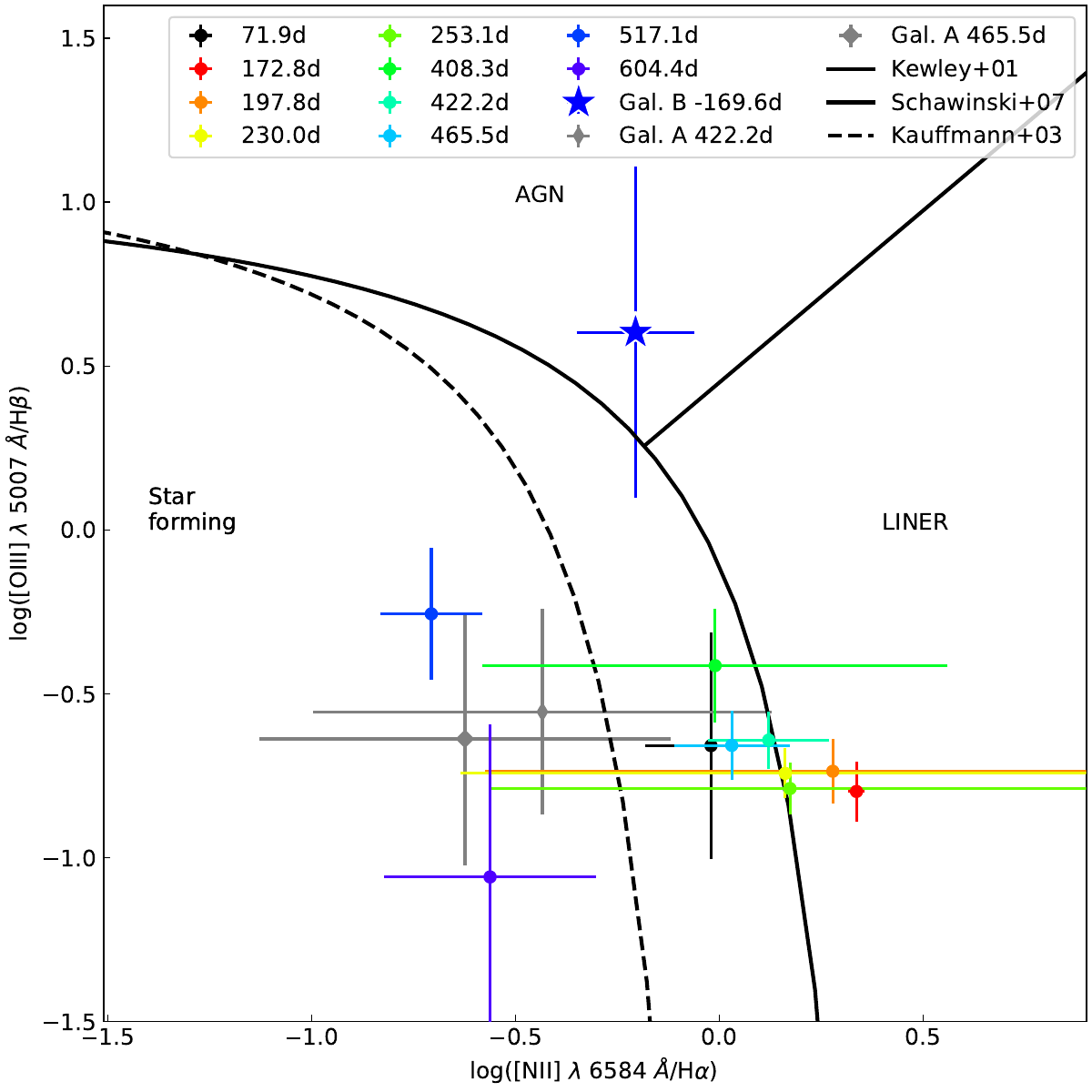}
    \caption{BPT diagram. The black lines delimiting each region are based on models by \protect\cite{2001ApJ...556..121K}, \protect\cite{2007MNRAS.382.1415S}, and \protect\cite{2003MNRAS.346.1055K}.}
    \label{fig:bpt}
\end{figure}

These three lines are of particular interest because, together with \ion{[N}{II]} $\lambda$6584, can be used as AGN diagnostic when plotting corresponding ratios in a BPT diagram \citep[][]{1981PASP...93....5B}. We caution that \ion{[N}{II]} $\lambda$6584 is not clearly visible in the spectra of AT2022rze, even though it is present in the spectra of the host (see Fig.~\ref{fig:host_spec}). We assume that the \ion{[N}{II]} lines are blended with H$\alpha$ and use the the {\sc Python} package {\sc lmfit} \citep[][]{2021zndo....598352N} to fit the H$\alpha$ using a three Gaussian model. We center one of the Gaussians at the H$\alpha$ rest wavelength and the other two at the rest wavelengths of \ion{[N}{II]} $\lambda$6548 and $\lambda$6584, respectively. We consider the sigma of \ion{[N}{II]} $\lambda$6584 to be the same of the \ion{[O}{III]} $\lambda$5007 line (as forbidden lines should have similar dispersion). The results are presented in Fig.~\ref{fig:bpt}. While BPT diagrams are known to suffer from biases related to H distribution within galaxies \citep[e.g.,][]{2015ApJ...811...26T} and their applicability to transient phenomena remains uncertain \citep[e.g.,][]{2019ApJ...883...31F}, they can nonetheless provide useful additional context when interpreting events. The measurements for AT2022rze span a broad range that includes the regions corresponding both to star forming galaxies and liners, the latter formed both by weak AGNs and galaxies that no longer form stars.
All the relevant lines are clearly visible in the host spectra, and measurements are straightforward. Galaxy A (grey markers) is consistent with a star forming galaxy, while galaxy B is consistent with an AGN. 

WISE colors can also be used as diagnostic to determine whether an event may be an (obscured) AGN \citep[see][]{2010AJ....140.1868W}. In particular, a single epoch W1$-$W2 $\gtrsim$ 0.5~mag color could suggest an AGN origin if it is accompanied by a W2$-$W3 $\gtrsim$ 2.5~mag. The AllWISE W1$-$W2 color measured near the position of AT2022rze before it went off are < 0.5~mag (see Table~\ref{tab:WISE}). In particular, the color corresponding to the position closer to the transient (labeled ``other'' in Table~\ref{tab:WISE}) is W1$-$W2 $= 0.3$~mag. This value is similar to the colors measured for CLAGNs \citep[][]{2021ApJ...920...56F}. 
That being said, the WISE light curve luminosity (see Fig.~\ref{fig:lcs} and Sect.~\ref{sec:clagn} for further discussion) is comparable to the luminosity observed in other CLAGNs and ANTs \citep[][]{2023MNRAS.524..188L,2024MNRAS.531.2603H}. ANTs are ambiguous transients associated to either CLAGNs or obscured TDEs \citep[][]{2025MNRAS.537.2024W}. Distinguishing between the two is challenging, and some previously classified CLAGNs have been reclassified as TDEs, as is the example of SDSS1115+0544 \citep{2019ApJ...874...44Y,2022ApJS..258...21W,2025arXiv250108812Z}.

\cite{2025A&A...697A.159G} present a sample of TDEs detected with eROSITA. The eROSITA dectection of AT2022rze is fainter than their TDE sample. However, their W1$-$W2 cut to remove AGNs is similar to that measured for AT2022rze. TDEs are associated to massive black holes (MBHs) with masses $M_{\rm BH} \lesssim 10^{8}$~M$_{\odot}$ \citep[although higher masses are possible for highly-spinning black holes and low-density stars][]{2012PhRvD..85b4037K,2016NatAs...1E...2L}. \cite{2023ApJ...955L...6Y} present a relation between the host galaxy stellar mass and the mass of the MBH producing the TDE. Considering the local galaxy stellar mass presented in Sect.~\ref{sec:host}, we infer a BH mass $M_{\rm BH}\sim 7.9 \times 10^{5}$~M$_{\odot}$. Considering the relation between BH mass and galaxy stellar mass presented by \cite{2015ApJ...813...82R}, we find $M_{\rm BH}\sim 1.2 \times 10^{6}$~M$_{\odot}$. These values indicate that a TDE is possible. While not conclusive, these values are much lower than those presented by \cite{2015ApJ...813...82R} for their AGN sample. However, the population of obscured AGN associated to lower-mass BHs is increasing \citep[e.g][]{2025MNRAS.538.2116P}.

\begin{figure}
    \includegraphics[width=\columnwidth]{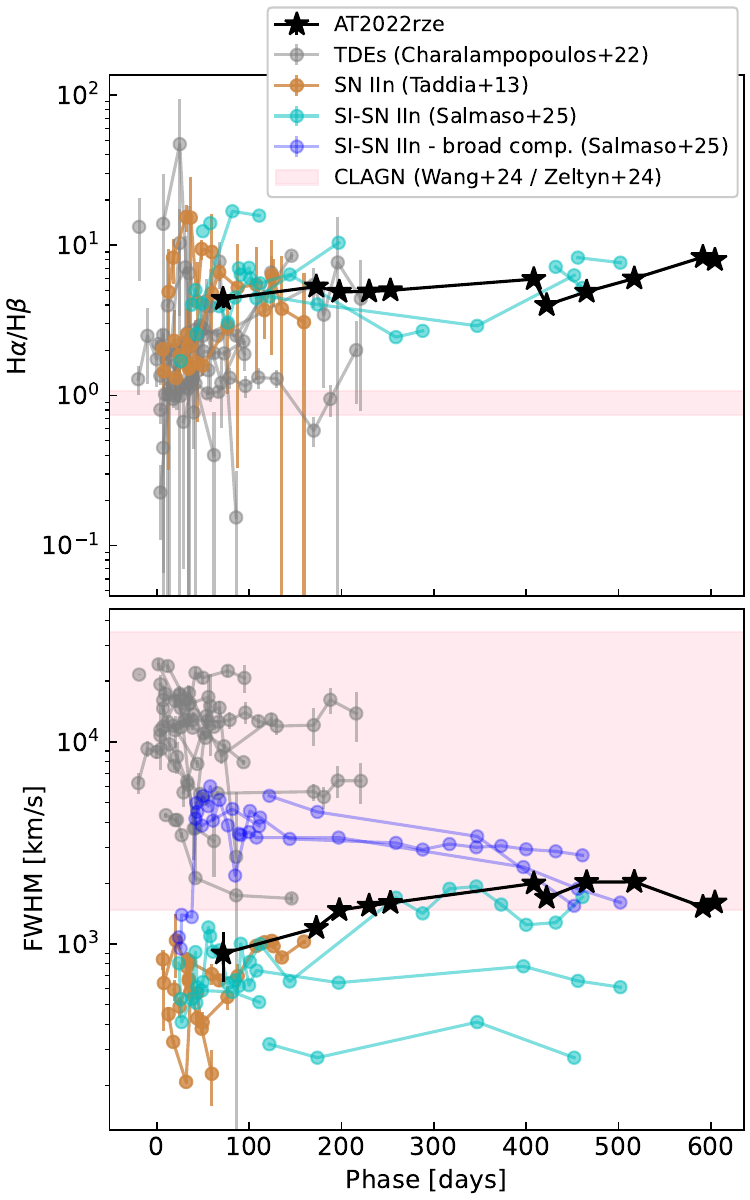}
    \caption{Comparison of AT2022rze to samples of TDEs, SNe~IIn and CLAGNs. Top panel: H$\alpha$ to H$\beta$ ratio. Bottom panel: H$\alpha$ FWHM velocity. AT2022rze is presented as black stars. The sample of TDEs of \protect\cite{2022A&A...659A..34C} is presented in grey. The sample of SN~IIn of \protect\cite{2013A&A...555A..10T} is presented in brown. The parameters for the narrow component features of the sample of  \protect\cite{2025A&A...695A..29S} are presented in cyan and the broad components in blue. Pink regions mark the higher and lower values of the H$\alpha$ to H$\beta$ ratio and H$\alpha$ velocity of the samples of \protect\cite{2024ApJ...966..128W} and \protect\cite{2024ApJ...966...85Z}, respectively. Note that these pink regions include measurements that do not consider light curve phases.}
    \label{fig:tde}
\end{figure}

Fig.~\ref{fig:tde} shows a comparison of the H$\alpha$ to H$\beta$ ratio (top panel) and H$\alpha$ FWHM velocity (bottom panel) of AT2022rze to samples of TDEs, (SL)SNe~IIn and CLAGNs. The H$\alpha$ FWHM velocity of AT2022rze was obtained by fitting Gaussian models with {\sc lmfit}. The associated error bars were calculated by arbitrarily modifying the limits of the line continua and repeating the Gaussian fit. The corresponding measured parameters for AT2022rze are listed in Table~\ref{table:BalmerDecandVel}. In this case, we only focus on comparison events studied in bulk, as they are analyzed using consistent methods and are likely to involve fewer biases than a collection of results obtained from individual object analyses. We selected the TDE sample of \cite{2022A&A...659A..34C}, the sample of SN~IIn of \cite{2013A&A...555A..10T}, the sample of strongly interacting (SI-) SN~IIn of \cite{2025A&A...695A..29S} that includes two SLSN~IIn\footnote{We consider SLSN~II any Type II event whose light curve reach a peak magnitude brighter than $-20$~mag \citep{2025A&A...695A.142P}.}, and the CLAGN samples of \cite{2024ApJ...966..128W} for H$\alpha$ to H$\beta$ ratio and \cite{2024ApJ...966...85Z} for H$\alpha$ FWHM velocity. The relevant parameters of CLAGNs are taken from two separate studies, as we did not find a single work that provides both. Furthermore, we show only the highest and lowest measurements of each parameter as a shaded region, since the considered studies do not account for different light-curve phases. Note that the considered phases also vary from study to study, we present rest frame days since peak similar to \cite{2022A&A...659A..34C}, although they consider peaks at different bands. \cite{2013A&A...555A..10T} consider days since discovery and \cite{2025A&A...695A..29S} days since explosion. Thus a phase shift of up to 90~days (largest rise time upper limit in \citealt{2025A&A...695A..29S}) may need to be consider for the SN~IIn comparison samples.

In the top panel of Fig.~\ref{fig:tde} we can see that the H$\alpha$ to H$\beta$ ratios of AT2022rze are among the highest values measured for TDEs and SN~IIn, and are consistent with the values measured for SI-SN~IIn at later times. The comparison to CLAGN is somewhat unfair as our measurements consider the entire observed spectrum and the CLAGN measurements of \cite{2024ApJ...966..128W} subtract the galaxy component, so they represent only a lower limit in this case. In the bottom panel of Fig.~\ref{fig:tde} we can see that the measured H$\alpha$ velocity for AT2022rze is much lower than the velocities measured for the considered sample of TDEs. At early times, the velocity of AT2022rze is consistent with the velocity of the narrow components of SN~IIn and SI-SN~IIn. Both \cite{2013A&A...555A..10T} and \cite{2025A&A...695A..29S} present measurements for the broad and narrow components of the H$\alpha$ features, if these are distinguishable. We only consider the velocity of the narrow component of the sample of \cite{2013A&A...555A..10T} as it provides a good enough fit to the features they present. The sample of \cite{2025A&A...695A..29S} show significantly broader features so we include both the narrow and broad components separately. At late times the broad H$\alpha$ component velocity of SI-SN~IIn is comparable to that of AT2022rze, while the rest of the events show slower narrow component velocities, except for SN~2021acya. The CLAGNs' lower H$\alpha$ velocities are also consistent with the late time velocities of AT2022rze.

\section{Discussion}
\label{sec:disc}

AT2022rze is a puzzling transient event that share characteristics with SLSNe~II, AGNs, and TDEs. Below, we explore the implications of our analysis within the context of each of these possible scenarios.

\subsection{SLSNe~II}
\label{sec:SLSN}

AT2022rze was originally classified as a H-rich SN based on the presence of narrow Balmer lines. We show in Sect.~\ref{sec:host} that the position of the transient (galaxy B, see Fig.~\ref{fig:host}) may be associated to an AGN. This association does not exclude the possibility of a SN. On the contrary, it has been proposed that the rates of CCSNe are the same or even higher in AGNs or star-forming galaxies \citep[][]{1990A&A...239...63P,2005AJ....129.1369P}.

In SNe, the presence of narrow lines indicate CSM interaction. Its bumpy light curve is also consistent with the presence of CSM, and it is reminiscent of other long-lived H-rich events (see Sect.~\ref{sec:compTOothers}). It has been argued that these bumpy, long-lived light curves result from the interaction between the SN ejecta and a CSM with large density variations. Such a CSM structure can be produced by episodic mass-loss events. These are often associated with Luminous Blue Variable-like progenitors \citep[e.g.,][]{2011MNRAS.415..773S} or Pulsational Pair Instability SNe \citep[PPISNe, e.g:][]{2017ApJ...836..244W, 2018ApJ...863..105W}. 

At early times, the observed colors of AT2022rze agree more with those observed for SLSNe~II than for TDEs (see Fig.~\ref{fig:color}). The estimated total radiated energy (see Sect.~\ref{sec:pbol_lc}) can be explained by CSM interaction. However, a rough analysis of the impact of the potential host extinction indicates that the transient may be rather blue (see Sect.~\ref{sec:host}) and much more energetic, reaching energies of the order of $10^{52}$~erg. This would present a challenge to a supernova interpretation, as the radiated energy would exceed what is available in a neutrino-driven core-collapse explosion \citep{2012ARNPS..62..407J}. In this case, the rotational energy of a central engine would be required, in addition to CSM interaction to produce the spectral signatures and the structure in the light curve \citep{2017ApJ...836..244W}.  

The spectra of AT2022rze show several coronal lines (see Sect.~\ref{sec:spectra}), with \ion{[Fe}{VII]} $\lambda$6087, \ion{[Fe}{X]} $\lambda$6374, and \ion{[Fe}{XI]} $\lambda$7892 being the strongest ones. Coronal lines are uncommon in SN although not previously unseen in SN~IIn \citep[e.g.,][]{2002ApJ...572..350F,2009ApJ...695.1334S,2009ApJ...707.1560I}, where X-rays from the SN excite the CSM. Yet, it has been argued that coronal line emitting transients are associated to TDEs \citep[e.g.,][]{2024MNRAS.528.4775H} and AGN \citep[e.g.,][]{2015MNRAS.448.2900R}.

\subsection{TDE}

TDEs are characterized for their blue colors and broad Balmer lines without a blueshifted absorption. AT2022rze show red colors unless we consider high host extinction (see Fig.\ref{fig:color}). In addition, The H$\alpha$ line are much narrower than those in other TDEs (see Fig.~\ref{fig:tde}). We calculate the time to and from half-peak (t$_{\mathrm{rise,1/2}} = 23.7 \pm 0.6$ and t$_{\mathrm{dec,1/2}} = 73.6 \pm 0.6$, respectively), and compare them to the same values presented for TDEs by \cite{2023ApJ...955L...6Y}. We find that the half-peak rise and decline time of AT2022rze is similar to those of TDEs, with AT2022rze being among the TDEs with the longest rest-frame duration above half-peak. We note, however, that it is possible that galaxy B hosts an AGN, and it is unclear what the characteristics of the TDE associated to an AGN should look like \citep[][]{2021SSRv..217...54Z}. 

The host galaxy of AT2022rze is experiencing a merger. Although a few TDEs have been suggested to be associated to merging galaxies \citep[e.g.,][]{2018Sci...361..482M,2020MNRAS.498.2167K,2022A&A...664A.158R}, there are only two confirmed TDEs in optical wavelenghts in such systems \citep[][]{2024A&A...689A.350C,2025MNRAS.540..498O}. These events exhibit much smoother light curves and spectroscopic features that differ significantly from those of AT2022rze, making direct comparison unwarranted.

\subsection{(CL)AGN}
\label{sec:clagn}

It is often difficult to differentiate between CLAGNs and TDEs due to their overlapping observational features. It has even been proposed that CLAGN are triggered by TDEs \citep[][]{2015MNRAS.452...69M,2024arXiv240612096W}. 
The analysis of the DESI spectrum of galaxy B, together with the \texttt{Blast} analysis, suggest that galaxy B is associated with an AGN. 

\begin{figure}
    \includegraphics[width=\columnwidth]{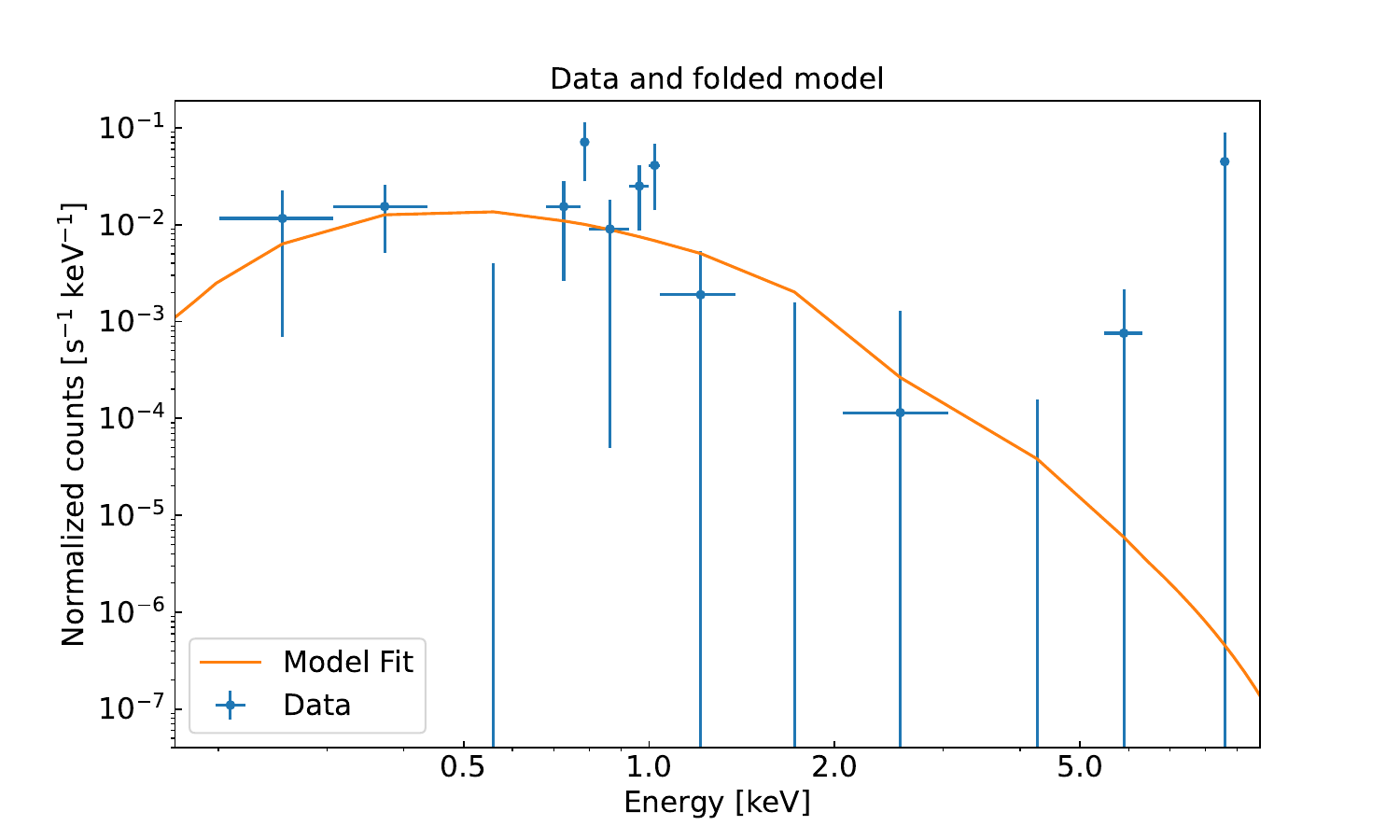}
    \caption{X-ray spectrum fit.}
    \label{fig:xrayspec}
\end{figure}

Using C-statistics \citep{Cash1979} to model the eROSITA spectrum with an absorbed power-law (\texttt{tbabs*cflux*powerlaw} in \textsc{xspec}, see Fig.~\ref{fig:xrayspec}), with the Galactic column density fixed at $N_{\rm H} = 3.25\times 10^{20}\,{\rm cm^{-2}}$ \citep{HI4PI2016}, we find a decent fit with \texttt{cstat/dof=18.1/11}, a photon index of $\Gamma = 2.5\pm0.7$ and an unabsorbed 0.3--2.3\,keV flux of ${\rm log}[f_{\rm X} / ({\rm erg\,s^{-1}\,cm^{-2}})] = -13.53 \pm 0.13$. The latter corresponds to an X-ray luminosity of ${\rm log}[L_{\rm X}/({\rm erg\,s^{-1}})] = 41.67\pm0.13$, 
which is on the fainter end of AGN, consistent with low-luminosity or obscured AGN \citep{Hasinger2005}. 

There are several sub-classifications of AGN, in particular, they can be divided in radio-loud and radio-quiet \citep{1989AJ.....98.1195K}. VLASS archival pre-transient images put sub-mJy upper limits near the center of the host (see Sect.~\ref{sec:obs}), but there is no radio detection of galaxy B. This suggests a radio-quiet AGN \citep[e.g:][]{2009ApJ...694..235P}. Radio-quiet AGN are often associated to spiral galaxies, and share many characteristics with star forming galaxies \citep[][]{2013MNRAS.436.3759B}. 

Given the evidence of the presence of an AGN at the center of galaxy B, it is possible that AT2022rze is associated to changes on the environment or accretion rate of said AGN. Considering the transition from the AGN (galaxy B) to the LINER region (earlier spectra of AT2022rze, see Fig.~\ref{fig:bpt}), AT2022rze could be considered a Changing-look AGN (CLAGN). 
CLAGN is a fairly recent and slowly growing class of transients, but the number of events is still low. Their characteristics are very diverse, and their searches can be considered as an anomaly detection task \citep[e.g.,][]{2021AJ....162..206S}. Therefore, systematic analysis of their time-scales and overall light curve and spectral characteristics is scarce. \cite{2023MNRAS.524..188L} developed a method to select CLAGN using multi-wavelength observations. Unfortunately, we do not have multi-wavelength coverage of AT2022rze itself. Still, the host X-ray flux measured before the transient went off, is consistent with the pre-activity X-ray fluxes that \cite{2023MNRAS.524..188L} measure for their sample of CLAGNs. In addition, the evolution of W1$-$W2 color at the position of AT2022rze that goes from < 0.5~mag  before it is detected (see Table~\ref{tab:WISE}) to $\sim$ 1.1~mag (see last column of Table~\ref{table:WISEphot}) for the last observed epoch, is in agreement with the interpretation presented by \cite{2023MNRAS.524..188L}, who argue that this effect is produced by a stronger contribution of hot dust produced by the increase in brightness of the AGN. 

We argue in Sect.~\ref{sec:host} that the steep Balmer decrement values are associated to significant host extinction. However, there is evidence that the Balmer decrement can depend on the accretion-rate for some AGNs and CLAGNs, in such cases the steep Balmer decrement could be related to ionizing photons with low flux \citep[][]{2023ApJ...950..106W}. The fact that the Balmer decrement evolves with time (see Fig.~\ref{fig:tde}), suggests that other mechanisms are in place in addition to extinction. Still, the fact that coronal lines can be seen during the whole spectra evolution, suggest the presence of high energies being in place. 
 
\section{Conclusions}
\label{sec:conc}

We have studied AT2022rze to try to determine its 
nature. The host of AT2022rze seems to be formed of two galaxies (A and B) undergoing a merging event. The transient is located in the center of galaxy B, located South-East of the plane-of-sky geometrical center of the system. AT2022rze exhibits minimal spectral evolution, consistently displaying narrow Balmer lines alongside various high-excitation features, indicative of a persistent high-energy source. The X-ray luminosity observed near the position of the transient before it went off is much lower than that observed in AGNs \citep[e.g:][]{1978MNRAS.183..129E,1990ApJ...361..459E,1996ASPC..103...70S}, although we note that the observations are off-set from the position of the transient. Coronal lines present in the spectra of AT2022rze suggest the presence of soft X-ray and UV emission. As coronal lines are found in AGNs, TDEs, and SNe, their presence alone do not serve as a diagnostic. The overall observed spectroscopic and light curve features of the transient are reminiscent of SLSNe~II, TDEs and AGNs. 

An analysis of an archival spectrum of galaxy B obtained before AT2022rze occurred, suggests that the galaxy is associated with an AGN (see Fig.~\ref{fig:bpt}). This is consistent with the environmental analysis performed by \texttt{Blast} (see Sect.~\ref{sec:host}). We roughly estimate the mass of the BH to be between $\sim 1.2 \times 10^{6}$~M$_{\odot}$ and $\sim 7.9 \times 10^{5}$~M$_{\odot}$ (see Sect.~\ref{sec:host}). This does not discard the possibility of a SLSN~II or a TDE. 

It is worth noting that the sustained presence of coronal lines in the spectra of some transients has led to the proposal of a new class, referred to as coronal line emitters (ECLs). Although this class groups together events that appear to exhibit similar phenomenology, it is not defined by a shared underlying physical mechanism, as the SLSN, TDE, or AGN classes are. Although ECLs have been mostly associated with TDEs \citep[][]{2024MNRAS.528.4775H}, it is still not clear what is their association to CLAGNs \citep[][]{2024MNRAS.528.7076C,2025arXiv250204080C}. 

We find possible evidence of Bowen resonance-fluorescence (see Sect.~\ref{sec:spectra}). The number of events displaying Bowen fluorescence features has been steadily increasing and a new class of transients dubbed Bowen fluorescence flares has been proposed \citep[BFFs][]{2019NatAs...3..242T}. Again, this class is defined by observational characteristics rather than by an underlying physical mechanism, and the nature of these transients is still debated \citep[e.g.,][]{2025ApJ...989..173S}.

AT2022rze is a luminous ambiguous transient in the nucleus of a merging galaxy. Many such ambiguous events have recently been found and have been dubbed ambiguous nuclear transients \citep[ANTs,][]{2025MNRAS.537.2024W}, which is another phenomenological class. \cite{2025MNRAS.537.2024W} propose that ANTs could be TDEs produced by different mechanisms than that of standard TDEs. 

The estimated total radiated energy of AT2022rze would be difficult to explain in a SLSN~II scenario, in particular if we consider significant host extinction (see Sect.~\ref{sec:host}). Although we cannot discard the possibility of AT2022rze being an unusual TDE, given the closeness of AT2022rze to the nucleus of galaxy B ($\sim$ 0.2 kpc), the low H$\alpha$ FWHM velocities (see Fig.~\ref{fig:tde}), the overall red colors and long duration of the light curve, the increase in W1$-$W2 color, and the steep Balmer decrement (see Fig.~\ref{fig:bpt}),make us conclude that AT2022rze is most likely a CLAGN, triggered by the ongoing merger in its host galaxy.

Panoramic surveys continue to uncover an ever increasing number of ambiguous transients. To disentangle and accurately classify the most luminous events, those that lie at the intersection of SLSNe~II, AGN, and TDEs, comprehensive multiwavelength follow-up and systematic comparative analysis are essential. Classification is not merely a matter of labeling; it provides critical insight into the underlying physical mechanisms powering these phenomena. The difficulty in confidently assigning these events to a specific class underscores the current limitations in our understanding of the energetic processes linked to MBHs.

\section*{Acknowledgements}

P.J.P thanks Stephen Thorp, Laureano Martinez, and Alberto Saldana-Lopez for useful discussion.
We thank the referee for their constructive comments.
Funded by the European Union (ERC, project number 101042299, TransPIre). Views and opinions expressed are however those of the author(s) only and do not necessarily reflect those of the European Union or the European Research Council Executive Agency. Neither the European Union nor the granting authority can be held responsible for them.
C.L. is supported by DoE award \#DE-SC0025599.
CPG acknowledges financial support from the Secretary of Universities
and Research (Government of Catalonia) and by the Horizon 2020 Research
and Innovation Programme of the European Union under the Marie
Sk\l{}odowska-Curie and the Beatriu de Pin\'os 2021 BP 00168 programme,
from the Spanish Ministerio de Ciencia e Innovaci\'on (MCIN) and the
Agencia Estatal de Investigaci\'on (AEI) 10.13039/501100011033 under the
PID2023-151307NB-I00 SNNEXT project, from Centro Superior de
Investigaciones Cient\'ificas (CSIC) under the PIE project 20215AT016
and the program Unidad de Excelencia Mar\'ia de Maeztu CEX2020-001058-M,
and from the Departament de Recerca i Universitats de la Generalitat de
Catalunya through the 2021-SGR-01270 grant.
SM acknowledges support from the Research Council of Finland project 350458.
FP acknowledges support from the Spanish Ministerio de Ciencia, Innovación y Universidades (MICINN) under grant numbers PID2022-141915NB-C21.
AAM and CL are partially supported by DoE award \#\,DE-SC0025599 and Cottrell Scholar Award \#\,CS-CSA-2025-059 from Research Corporation for Science Advancement.
This work is based on observations obtained with the Samuel Oschin Telescope 48-inch and the 60-inch Telescope at the Palomar Observatory as part of the Zwicky Transient Facility project. ZTF is supported by the National Science Foundation under Grants No. AST-1440341 and AST-2034437 and a collaboration including current partners Caltech, IPAC, the Weizmann Institute of Science, the Oskar Klein Center at Stockholm University, the University of Maryland, Deutsches Elektronen-Synchrotron and Humboldt University, the TANGO Consortium of Taiwan, the University of Wisconsin at Milwaukee, Trinity College Dublin, Lawrence Livermore National Laboratories, IN2P3, University of Warwick, Ruhr University Bochum, Northwestern University and former partners the University of Washington, Los Alamos National Laboratories, and Lawrence Berkeley National Laboratories. Operations are conducted by COO, IPAC, and UW.
The ZTF forced-photometry service was funded under the Heising-Simons Foundation grant No. 12540303 (PI: Graham).
SED Machine is based upon work supported by the National Science Foundation under Grant No. 1106171.
The Gordon and Betty Moore Foundation, through both the Data-Driven Investigator Program and a dedicated grant, provided critical funding for SkyPortal. 
W. M. Keck Observatory access was supported by Northwestern University and the Center for Interdisciplinary Exploration and Research in Astrophysics (CIERA).
This work has made use of data from the Asteroid Terrestrial-impact Last Alert System (ATLAS) project. The Asteroid Terrestrial-impact Last Alert System (ATLAS) project is primarily funded to search for near earth asteroids through NASA grants NN12AR55G, 80NSSC18K0284, and 80NSSC18K1575; byproducts of the NEO search include images and catalogs from the survey area. This work was partially funded by Kepler/K2 grant J1944/80NSSC19K0112 and HST GO-15889, and STFC grants ST/T000198/1 and ST/S006109/1. The ATLAS science products have been made possible through the contributions of the University of Hawaii Institute for Astronomy, the Queen’s University Belfast, the Space Telescope Science Institute, the South African Astronomical Observatory, and The Millennium Institute of Astrophysics (MAS), Chile.
We acknowledge ESA Gaia, DPAC and the Photometric Science Alerts Team (\url{http://gsaweb.ast.cam.ac.uk/alerts}).
This research has made use of the NASA/IPAC Extragalactic Database, which is funded by the National Aeronautics and Space Administration and operated by the California Institute of Technology.
This research has made use of the SVO Filter Profile Service "Carlos Rodrigo", funded by MCIN/AEI/10.13039/501100011033/ through grant PID2020-112949GB-I00
This research has made use of the SVO Filter Profile Service ``Carlos Rodrigo'', funded by MCIN/AEI/10.13039/501100011033/ through grant PID2020-112949GB-I00
The National Radio Astronomy Observatory is a facility of the National Science Foundation operated under cooperative agreement by Associated Universities, Inc. 
This research has made use of the CIRADA cutout service at \url{cutouts.cirada.ca}, operated by the Canadian Initiative for Radio Astronomy Data Analysis (CIRADA). CIRADA is funded by a grant from the Canada Foundation for Innovation 2017 Innovation Fund (Project 35999), as well as by the Provinces of Ontario, British Columbia, Alberta, Manitoba and Quebec, in collaboration with the National Research Council of Canada, the US National Radio Astronomy Observatory and Australia’s Commonwealth Scientific and Industrial Research Organisation.
The Legacy Surveys consist of three individual and complementary projects: the Dark Energy Camera Legacy Survey (DECaLS; Proposal ID 2014B-0404; PIs: David Schlegel and Arjun Dey), the Beijing-Arizona Sky Survey (BASS; NOAO Prop. ID 2015A-0801; PIs: Zhou Xu and Xiaohui Fan), and the Mayall z-band Legacy Survey (MzLS; Prop. ID 2016A-0453; PI: Arjun Dey). DECaLS, BASS and MzLS together include data obtained, respectively, at the Blanco telescope, Cerro Tololo Inter-American Observatory, NSF’s NOIRLab; the Bok telescope, Steward Observatory, University of Arizona; and the Mayall telescope, Kitt Peak National Observatory, NOIRLab. Pipeline processing and analyses of the data were supported by NOIRLab and the Lawrence Berkeley National Laboratory (LBNL). The Legacy Surveys project is honored to be permitted to conduct astronomical research on Iolkam Du’ag (Kitt Peak), a mountain with particular significance to the Tohono O’odham Nation.

NOIRLab is operated by the Association of Universities for Research in Astronomy (AURA) under a cooperative agreement with the National Science Foundation. LBNL is managed by the Regents of the University of California under contract to the U.S. Department of Energy.

This project used data obtained with the Dark Energy Camera (DECam), which was constructed by the Dark Energy Survey (DES) collaboration. Funding for the DES Projects has been provided by the U.S. Department of Energy, the U.S. National Science Foundation, the Ministry of Science and Education of Spain, the Science and Technology Facilities Council of the United Kingdom, the Higher Education Funding Council for England, the National Center for Supercomputing Applications at the University of Illinois at Urbana-Champaign, the Kavli Institute of Cosmological Physics at the University of Chicago, Center for Cosmology and Astro-Particle Physics at the Ohio State University, the Mitchell Institute for Fundamental Physics and Astronomy at Texas A\&M University, Financiadora de Estudos e Projetos, Fundacao Carlos Chagas Filho de Amparo, Financiadora de Estudos e Projetos, Fundacao Carlos Chagas Filho de Amparo a Pesquisa do Estado do Rio de Janeiro, Conselho Nacional de Desenvolvimento Cientifico e Tecnologico and the Ministerio da Ciencia, Tecnologia e Inovacao, the Deutsche Forschungsgemeinschaft and the Collaborating Institutions in the Dark Energy Survey. The Collaborating Institutions are Argonne National Laboratory, the University of California at Santa Cruz, the University of Cambridge, Centro de Investigaciones Energeticas, Medioambientales y Tecnologicas-Madrid, the University of Chicago, University College London, the DES-Brazil Consortium, the University of Edinburgh, the Eidgenossische Technische Hochschule (ETH) Zurich, Fermi National Accelerator Laboratory, the University of Illinois at Urbana-Champaign, the Institut de Ciencies de l’Espai (IEEC/CSIC), the Institut de Fisica d’Altes Energies, Lawrence Berkeley National Laboratory, the Ludwig Maximilians Universitat Munchen and the associated Excellence Cluster Universe, the University of Michigan, NSF’s NOIRLab, the University of Nottingham, the Ohio State University, the University of Pennsylvania, the University of Portsmouth, SLAC National Accelerator Laboratory, Stanford University, the University of Sussex, and Texas A\&M University.

BASS is a key project of the Telescope Access Program (TAP), which has been funded by the National Astronomical Observatories of China, the Chinese Academy of Sciences (the Strategic Priority Research Program “The Emergence of Cosmological Structures” Grant XDB09000000), and the Special Fund for Astronomy from the Ministry of Finance. The BASS is also supported by the External Cooperation Program of Chinese Academy of Sciences (Grant 114A11KYSB20160057), and Chinese National Natural Science Foundation (Grant 12120101003, 11433005).
This research used data obtained with the Dark Energy Spectroscopic Instrument (DESI). DESI construction and operations is managed by the Lawrence Berkeley National Laboratory. This material is based upon work supported by the U.S. Department of Energy, Office of Science, Office of High-Energy Physics, under Contract No. DE–AC02–05CH11231, and by the National Energy Research Scientific Computing Center, a DOE Office of Science User Facility under the same contract. Additional support for DESI was provided by the U.S. National Science Foundation (NSF), Division of Astronomical Sciences under Contract No. AST-0950945 to the NSF’s National Optical-Infrared Astronomy Research Laboratory; the Science and Technology Facilities Council of the United Kingdom; the Gordon and Betty Moore Foundation; the Heising-Simons Foundation; the French Alternative Energies and Atomic Energy Commission (CEA); the National Council of Humanities, Science and Technology of Mexico (CONAHCYT); the Ministry of Science and Innovation of Spain (MICINN), and by the DESI Member Institutions: www.desi.lbl.gov/collaborating-institutions. The DESI collaboration is honored to be permitted to conduct scientific research on I’oligam Du’ag (Kitt Peak), a mountain with particular significance to the Tohono O’odham Nation. Any opinions, findings, and conclusions or recommendations expressed in this material are those of the author(s) and do not necessarily reflect the views of the U.S. National Science Foundation, the U.S. Department of Energy, or any of the listed funding agencies.
This work used data of eROSITA telescope onboard SRG observatory. The SRG observatory was built by Roskosmos in the interests of the Russian Academy of Sciences represented by its Space Research Institute (IKI) in the framework of the Russian Federal Space Program, with the participation of the Deutsches Zentrum für Luft- und Raumfahrt (DLR). The SRG/eROSITA X-ray telescope was built by a consortium of German Institutes led by MPE, and supported by DLR.  The SRG spacecraft was designed, built, launched and is operated by the Lavochkin Association and its subcontractors. The science data are downlinked via the Deep Space Network Antennae in Bear Lakes, Ussurijsk, and Baykonur, funded by Roskosmos. The eROSITA data used in this work were processed using the eSASS software system developed by the German eROSITA consortium and proprietary data reduction and analysis software developed by the Russian eROSITA Consortium.

\section*{Data Availability}

The photometric data used in this paper are presented in Appendix \ref{app:lcs}, with except of Gaia photometry that can be found at the dedicated repository at \url{https://gsaweb.ast.cam.ac.uk/alerts/}. The spectra can be found on WISeREP at \url{https://www.wiserep.org}.



\bibliographystyle{mnras}
\bibliography{2022rze} 

@ARTICLE{mainzer2014,
       author = {{Mainzer}, A. and {Bauer}, J. and {Cutri}, R.~M. and {Grav}, T. and {Masiero}, J. and {Beck}, R. and {Clarkson}, P. and {Conrow}, T. and {Dailey}, J. and {Eisenhardt}, P. and {Fabinsky}, B. and {Fajardo-Acosta}, S. and {Fowler}, J. and {Gelino}, C. and {Grillmair}, C. and {Heinrichsen}, I. and {Kendall}, M. and {Kirkpatrick}, J. Davy and {Liu}, F. and {Masci}, F. and {McCallon}, H. and {Nugent}, C.~R. and {Papin}, M. and {Rice}, E. and {Royer}, D. and {Ryan}, T. and {Sevilla}, P. and {Sonnett}, S. and {Stevenson}, R. and {Thompson}, D.~B. and {Wheelock}, S. and {Wiemer}, D. and {Wittman}, M. and {Wright}, E. and {Yan}, L.},
        title = "{Initial Performance of the NEOWISE Reactivation Mission}",
      journal = {\apj},
         year = 2014,
        month = sep,
       volume = {792},
       number = {1},
          eid = {30},
        pages = {30},
          doi = {10.1088/0004-637X/792/1/30},
archivePrefix = {arXiv},
       eprint = {1406.6025},
 primaryClass = {astro-ph.EP},
       adsurl = {https://ui.adsabs.harvard.edu/abs/2014ApJ...792...30M}
}

@ARTICLE{masci2013,
       author = {{Masci}, Frank},
        title = "{ICORE: Image Co-addition with Optional Resolution Enhancement}",
      journal = {arXiv e-prints},
         year = 2013,
        month = jan,
          eid = {arXiv:1301.2718},
        pages = {arXiv:1301.2718},
          doi = {10.48550/arXiv.1301.2718},
archivePrefix = {arXiv},
       eprint = {1301.2718},
 primaryClass = {astro-ph.IM},
       adsurl = {https://ui.adsabs.harvard.edu/abs/2013arXiv1301.2718M}
}

@ARTICLE{Wright2010,
       author = {{Wright}, Edward L. and {Eisenhardt}, Peter R.~M. and {Mainzer}, Amy K. and {Ressler}, Michael E. and {Cutri}, Roc M. and {Jarrett}, Thomas and {Kirkpatrick}, J. Davy and {Padgett}, Deborah and {McMillan}, Robert S. and {Skrutskie}, Michael and {Stanford}, S.~A. and {Cohen}, Martin and {Walker}, Russell G. and {Mather}, John C. and {Leisawitz}, David and {Gautier}, Thomas N., III and {McLean}, Ian and {Benford}, Dominic and {Lonsdale}, Carol J. and {Blain}, Andrew and {Mendez}, Bryan and {Irace}, William R. and {Duval}, Valerie and {Liu}, Fengchuan and {Royer}, Don and {Heinrichsen}, Ingolf and {Howard}, Joan and {Shannon}, Mark and {Kendall}, Martha and {Walsh}, Amy L. and {Larsen}, Mark and {Cardon}, Joel G. and {Schick}, Scott and {Schwalm}, Mark and {Abid}, Mohamed and {Fabinsky}, Beth and {Naes}, Larry and {Tsai}, Chao-Wei},
        title = "{The Wide-field Infrared Survey Explorer (WISE): Mission Description and Initial On-orbit Performance}",
      journal = {\aj},
         year = 2010,
        month = dec,
       volume = {140},
       number = {6},
        pages = {1868-1881},
          doi = {10.1088/0004-6256/140/6/1868},
archivePrefix = {arXiv},
       eprint = {1008.0031},
 primaryClass = {astro-ph.IM},
       adsurl = {https://ui.adsabs.harvard.edu/abs/2010AJ....140.1868W}
}

@ARTICLE{2019ApJ...874...44Y,
       author = {{Yan}, Lin and {Wang}, Tinggui and {Jiang}, Ning and {Stern}, Daniel and {Dou}, Liming and {Fremling}, C. and {Graham}, M.~J. and {Drake}, A.~J. and {Yang}, Chenwei and {Burdge}, K. and {Kasliwal}, M.~M.},
        title = "{Rapid {\textquotedblleft}Turn-on{\textquotedblright} of Type-1 AGN in a Quiescent Early-type Galaxy SDSS1115+0544}",
      journal = {\apj},
         year = 2019,
        month = mar,
       volume = {874},
       number = {1},
          eid = {44},
        pages = {44},
          doi = {10.3847/1538-4357/ab074b},
archivePrefix = {arXiv},
       eprint = {1902.04163},
 primaryClass = {astro-ph.HE},
       adsurl = {https://ui.adsabs.harvard.edu/abs/2019ApJ...874...44Y}
}

@ARTICLE{Hasinger2005,
       author = {{Hasinger}, G. and {Miyaji}, T. and {Schmidt}, M.},
        title = "{Luminosity-dependent evolution of soft X-ray selected AGN. New Chandra and XMM-Newton surveys}",
      journal = {\aap},
         year = 2005,
        month = oct,
       volume = {441},
       number = {2},
        pages = {417-434},
          doi = {10.1051/0004-6361:20042134},
archivePrefix = {arXiv},
       eprint = {astro-ph/0506118},
 primaryClass = {astro-ph},
       adsurl = {https://ui.adsabs.harvard.edu/abs/2005A&A...441..417H}
}

@ARTICLE{Cash1979,
       author = {{Cash}, W.},
        title = "{Parameter estimation in astronomy through application of the likelihood ratio.}",
      journal = {\apj},
         year = 1979,
        month = mar,
       volume = {228},
        pages = {939-947},
          doi = {10.1086/156922},
       adsurl = {https://ui.adsabs.harvard.edu/abs/1979ApJ...228..939C}
}

@ARTICLE{HI4PI2016,
       author = {{HI4PI Collaboration} and {Ben Bekhti}, N. and {Fl{\"o}er}, L. and {Keller}, R. and {Kerp}, J. and {Lenz}, D. and {Winkel}, B. and {Bailin}, J. and {Calabretta}, M.~R. and {Dedes}, L. and {Ford}, H.~A. and {Gibson}, B.~K. and {Haud}, U. and {Janowiecki}, S. and {Kalberla}, P.~M.~W. and {Lockman}, F.~J. and {McClure-Griffiths}, N.~M. and {Murphy}, T. and {Nakanishi}, H. and {Pisano}, D.~J. and {Staveley-Smith}, L.},
        title = "{HI4PI: A full-sky H I survey based on EBHIS and GASS}",
      journal = {\aap},
         year = 2016,
        month = oct,
       volume = {594},
          eid = {A116},
        pages = {A116},
          doi = {10.1051/0004-6361/201629178},
archivePrefix = {arXiv},
       eprint = {1610.06175},
 primaryClass = {astro-ph.GA},
       adsurl = {https://ui.adsabs.harvard.edu/abs/2016A&A...594A.116H}
}

@ARTICLE{2022TNSTR2428....1H,
       author = {{Hodgkin}, S.~T. and {Breedt}, E. and {Delgado}, A. and {Harrison}, D.~L. and {Leeuwen}, M.~V. and {Rixon}, G. and {Wevers}, T. and {Yoldas}, A. and {Ihanec}, N. and {Kruszy{\'n}ska}, K. and {Rybicki}, K.~A. and {Wyrzykowski}, {\L}. and {Kostrzewa-Rutkowska}, Z. and {Eappachen}, D. and {Marton}, G.},
        title = "{GaiaAlerts Transient Discovery Report for 2022-08-23}",
      journal = {Transient Name Server Discovery Report},
         year = 2022,
        month = aug,
       volume = {2022-2428},
        pages = {1},
       adsurl = {https://ui.adsabs.harvard.edu/abs/2022TNSTR2428....1H}
}

@ARTICLE{2022TNSCR3514....1H,
       author = {{Hinds}, K. and {Perley}, D. and {Chu}, M. and {Dahiwale}, A. and {Fremling}, C.},
        title = "{ZTF Transient Classification Report for 2022-12-01}",
      journal = {Transient Name Server Classification Report},
         year = 2022,
        month = dec,
       volume = {2022-3514},
        pages = {1},
       adsurl = {https://ui.adsabs.harvard.edu/abs/2022TNSCR3514....1H}
}

@ARTICLE{2016A&A...595A...1G,
       author = {{Gaia Collaboration} and {Prusti}, T. and {de Bruijne}, J.~H.~J. and {Brown}, A.~G.~A. and {Vallenari}, A. and {Babusiaux}, C. and {Bailer-Jones}, C.~A.~L. and {Bastian}, U. and {Biermann}, M. and {Evans}, D.~W. and {Eyer}, L. and {Jansen}, F. and {Jordi}, C. and {Klioner}, S.~A. and {Lammers}, U. and {Lindegren}, L. and {Luri}, X. and {Mignard}, F. and {Milligan}, D.~J. and {Panem}, C. and {Poinsignon}, V. and {Pourbaix}, D. and {Randich}, S. and {Sarri}, G. and {Sartoretti}, P. and {Siddiqui}, H.~I. and {Soubiran}, C. and {Valette}, V. and {van Leeuwen}, F. and {Walton}, N.~A. and {Aerts}, C. and {Arenou}, F. and {Cropper}, M. and {Drimmel}, R. and {H{\o}g}, E. and {Katz}, D. and {Lattanzi}, M.~G. and {O'Mullane}, W. and {Grebel}, E.~K. and {Holland}, A.~D. and {Huc}, C. and {Passot}, X. and {Bramante}, L. and {Cacciari}, C. and {Casta{\~n}eda}, J. and {Chaoul}, L. and {Cheek}, N. and {De Angeli}, F. and {Fabricius}, C. and {Guerra}, R. and {Hern{\'a}ndez}, J. and {Jean-Antoine-Piccolo}, A. and {Masana}, E. and {Messineo}, R. and {Mowlavi}, N. and {Nienartowicz}, K. and {Ord{\'o}{\~n}ez-Blanco}, D. and {Panuzzo}, P. and {Portell}, J. and {Richards}, P.~J. and {Riello}, M. and {Seabroke}, G.~M. and {Tanga}, P. and {Th{\'e}venin}, F. and {Torra}, J. and {Els}, S.~G. and {Gracia-Abril}, G. and {Comoretto}, G. and {Garcia-Reinaldos}, M. and {Lock}, T. and {Mercier}, E. and {Altmann}, M. and {Andrae}, R. and {Astraatmadja}, T.~L. and {Bellas-Velidis}, I. and {Benson}, K. and {Berthier}, J. and {Blomme}, R. and {Busso}, G. and {Carry}, B. and {Cellino}, A. and {Clementini}, G. and {Cowell}, S. and {Creevey}, O. and {Cuypers}, J. and {Davidson}, M. and {De Ridder}, J. and {de Torres}, A. and {Delchambre}, L. and {Dell'Oro}, A. and {Ducourant}, C. and {Fr{\'e}mat}, Y. and {Garc{\'\i}a-Torres}, M. and {Gosset}, E. and {Halbwachs}, J. -L. and {Hambly}, N.~C. and {Harrison}, D.~L. and {Hauser}, M. and {Hestroffer}, D. and {Hodgkin}, S.~T. and {Huckle}, H.~E. and {Hutton}, A. and {Jasniewicz}, G. and {Jordan}, S. and {Kontizas}, M. and {Korn}, A.~J. and {Lanzafame}, A.~C. and {Manteiga}, M. and {Moitinho}, A. and {Muinonen}, K. and {Osinde}, J. and {Pancino}, E. and {Pauwels}, T. and {Petit}, J. -M. and {Recio-Blanco}, A. and {Robin}, A.~C. and {Sarro}, L.~M. and {Siopis}, C. and {Smith}, M. and {Smith}, K.~W. and {Sozzetti}, A. and {Thuillot}, W. and {van Reeven}, W. and {Viala}, Y. and {Abbas}, U. and {Abreu Aramburu}, A. and {Accart}, S. and {Aguado}, J.~J. and {Allan}, P.~M. and {Allasia}, W. and {Altavilla}, G. and {{\'A}lvarez}, M.~A. and {Alves}, J. and {Anderson}, R.~I. and {Andrei}, A.~H. and {Anglada Varela}, E. and {Antiche}, E. and {Antoja}, T. and {Ant{\'o}n}, S. and {Arcay}, B. and {Atzei}, A. and {Ayache}, L. and {Bach}, N. and {Baker}, S.~G. and {Balaguer-N{\'u}{\~n}ez}, L. and {Barache}, C. and {Barata}, C. and {Barbier}, A. and {Barblan}, F. and {Baroni}, M. and {Barrado y Navascu{\'e}s}, D. and {Barros}, M. and {Barstow}, M.~A. and {Becciani}, U. and {Bellazzini}, M. and {Bellei}, G. and {Bello Garc{\'\i}a}, A. and {Belokurov}, V. and {Bendjoya}, P. and {Berihuete}, A. and {Bianchi}, L. and {Bienaym{\'e}}, O. and {Billebaud}, F. and {Blagorodnova}, N. and {Blanco-Cuaresma}, S. and {Boch}, T. and {Bombrun}, A. and {Borrachero}, R. and {Bouquillon}, S. and {Bourda}, G. and {Bouy}, H. and {Bragaglia}, A. and {Breddels}, M.~A. and {Brouillet}, N. and {Br{\"u}semeister}, T. and {Bucciarelli}, B. and {Budnik}, F. and {Burgess}, P. and {Burgon}, R. and {Burlacu}, A. and {Busonero}, D. and {Buzzi}, R. and {Caffau}, E. and {Cambras}, J. and {Campbell}, H. and {Cancelliere}, R. and {Cantat-Gaudin}, T. and {Carlucci}, T. and {Carrasco}, J.~M. and {Castellani}, M. and {Charlot}, P. and {Charnas}, J. and {Charvet}, P. and {Chassat}, F. and {Chiavassa}, A. and {Clotet}, M. and {Cocozza}, G. and {Collins}, R.~S. and {Collins}, P. and {Costigan}, G. and {Crifo}, F. and {Cross}, N.~J.~G. and {Crosta}, M. and {Crowley}, C. and {Dafonte}, C. and {Damerdji}, Y. and {Dapergolas}, A. and {David}, P. and {David}, M. and {De Cat}, P. and {de Felice}, F. and {de Laverny}, P. and {De Luise}, F. and {De March}, R. and {de Martino}, D. and {de Souza}, R. and {Debosscher}, J. and {del Pozo}, E. and {Delbo}, M. and {Delgado}, A. and {Delgado}, H.~E. and {di Marco}, F. and {Di Matteo}, P. and {Diakite}, S. and {Distefano}, E. and {Dolding}, C. and {Dos Anjos}, S. and {Drazinos}, P. and {Dur{\'a}n}, J. and {Dzigan}, Y. and {Ecale}, E. and {Edvardsson}, B. and {Enke}, H. and {Erdmann}, M. and {Escolar}, D. and {Espina}, M. and {Evans}, N.~W. and {Eynard Bontemps}, G. and {Fabre}, C. and {Fabrizio}, M. and {Faigler}, S. and {Falc{\~a}o}, A.~J. and {Farr{\`a}s Casas}, M. and {Faye}, F. and {Federici}, L. and {Fedorets}, G. and {Fern{\'a}ndez-Hern{\'a}ndez}, J. and {Fernique}, P. and {Fienga}, A. and {Figueras}, F. and {Filippi}, F. and {Findeisen}, K. and {Fonti}, A. and {Fouesneau}, M. and {Fraile}, E. and {Fraser}, M. and {Fuchs}, J. and {Furnell}, R. and {Gai}, M. and {Galleti}, S. and {Galluccio}, L. and {Garabato}, D. and {Garc{\'\i}a-Sedano}, F. and {Gar{\'e}}, P. and {Garofalo}, A. and {Garralda}, N. and {Gavras}, P. and {Gerssen}, J. and {Geyer}, R. and {Gilmore}, G. and {Girona}, S. and {Giuffrida}, G. and {Gomes}, M. and {Gonz{\'a}lez-Marcos}, A. and {Gonz{\'a}lez-N{\'u}{\~n}ez}, J. and {Gonz{\'a}lez-Vidal}, J.~J. and {Granvik}, M. and {Guerrier}, A. and {Guillout}, P. and {Guiraud}, J. and {G{\'u}rpide}, A. and {Guti{\'e}rrez-S{\'a}nchez}, R. and {Guy}, L.~P. and {Haigron}, R. and {Hatzidimitriou}, D. and {Haywood}, M. and {Heiter}, U. and {Helmi}, A. and {Hobbs}, D. and {Hofmann}, W. and {Holl}, B. and {Holland}, G. and {Hunt}, J.~A.~S. and {Hypki}, A. and {Icardi}, V. and {Irwin}, M. and {Jevardat de Fombelle}, G. and {Jofr{\'e}}, P. and {Jonker}, P.~G. and {Jorissen}, A. and {Julbe}, F. and {Karampelas}, A. and {Kochoska}, A. and {Kohley}, R. and {Kolenberg}, K. and {Kontizas}, E. and {Koposov}, S.~E. and {Kordopatis}, G. and {Koubsky}, P. and {Kowalczyk}, A. and {Krone-Martins}, A. and {Kudryashova}, M. and {Kull}, I. and {Bachchan}, R.~K. and {Lacoste-Seris}, F. and {Lanza}, A.~F. and {Lavigne}, J. -B. and {Le Poncin-Lafitte}, C. and {Lebreton}, Y. and {Lebzelter}, T. and {Leccia}, S. and {Leclerc}, N. and {Lecoeur-Taibi}, I. and {Lemaitre}, V. and {Lenhardt}, H. and {Leroux}, F. and {Liao}, S. and {Licata}, E. and {Lindstr{\o}m}, H.~E.~P. and {Lister}, T.~A. and {Livanou}, E. and {Lobel}, A. and {L{\"o}ffler}, W. and {L{\'o}pez}, M. and {Lopez-Lozano}, A. and {Lorenz}, D. and {Loureiro}, T. and {MacDonald}, I. and {Magalh{\~a}es Fernandes}, T. and {Managau}, S. and {Mann}, R.~G. and {Mantelet}, G. and {Marchal}, O. and {Marchant}, J.~M. and {Marconi}, M. and {Marie}, J. and {Marinoni}, S. and {Marrese}, P.~M. and {Marschalk{\'o}}, G. and {Marshall}, D.~J. and {Mart{\'\i}n-Fleitas}, J.~M. and {Martino}, M. and {Mary}, N. and {Matijevi{\v{c}}}, G. and {Mazeh}, T. and {McMillan}, P.~J. and {Messina}, S. and {Mestre}, A. and {Michalik}, D. and {Millar}, N.~R. and {Miranda}, B.~M.~H. and {Molina}, D. and {Molinaro}, R. and {Molinaro}, M. and {Moln{\'a}r}, L. and {Moniez}, M. and {Montegriffo}, P. and {Monteiro}, D. and {Mor}, R. and {Mora}, A. and {Morbidelli}, R. and {Morel}, T. and {Morgenthaler}, S. and {Morley}, T. and {Morris}, D. and {Mulone}, A.~F. and {Muraveva}, T. and {Musella}, I. and {Narbonne}, J. and {Nelemans}, G. and {Nicastro}, L. and {Noval}, L. and {Ord{\'e}novic}, C. and {Ordieres-Mer{\'e}}, J. and {Osborne}, P. and {Pagani}, C. and {Pagano}, I. and {Pailler}, F. and {Palacin}, H. and {Palaversa}, L. and {Parsons}, P. and {Paulsen}, T. and {Pecoraro}, M. and {Pedrosa}, R. and {Pentik{\"a}inen}, H. and {Pereira}, J. and {Pichon}, B. and {Piersimoni}, A.~M. and {Pineau}, F. -X. and {Plachy}, E. and {Plum}, G. and {Poujoulet}, E. and {Pr{\v{s}}a}, A. and {Pulone}, L. and {Ragaini}, S. and {Rago}, S. and {Rambaux}, N. and {Ramos-Lerate}, M. and {Ranalli}, P. and {Rauw}, G. and {Read}, A. and {Regibo}, S. and {Renk}, F. and {Reyl{\'e}}, C. and {Ribeiro}, R.~A. and {Rimoldini}, L. and {Ripepi}, V. and {Riva}, A. and {Rixon}, G. and {Roelens}, M. and {Romero-G{\'o}mez}, M. and {Rowell}, N. and {Royer}, F. and {Rudolph}, A. and {Ruiz-Dern}, L. and {Sadowski}, G. and {Sagrist{\`a} Sell{\'e}s}, T. and {Sahlmann}, J. and {Salgado}, J. and {Salguero}, E. and {Sarasso}, M. and {Savietto}, H. and {Schnorhk}, A. and {Schultheis}, M. and {Sciacca}, E. and {Segol}, M. and {Segovia}, J.~C. and {Segransan}, D. and {Serpell}, E. and {Shih}, I. -C. and {Smareglia}, R. and {Smart}, R.~L. and {Smith}, C. and {Solano}, E. and {Solitro}, F. and {Sordo}, R. and {Soria Nieto}, S. and {Souchay}, J. and {Spagna}, A. and {Spoto}, F. and {Stampa}, U. and {Steele}, I.~A. and {Steidelm{\"u}ller}, H. and {Stephenson}, C.~A. and {Stoev}, H. and {Suess}, F.~F. and {S{\"u}veges}, M. and {Surdej}, J. and {Szabados}, L. and {Szegedi-Elek}, E. and {Tapiador}, D. and {Taris}, F. and {Tauran}, G. and {Taylor}, M.~B. and {Teixeira}, R. and {Terrett}, D. and {Tingley}, B. and {Trager}, S.~C. and {Turon}, C. and {Ulla}, A. and {Utrilla}, E. and {Valentini}, G. and {van Elteren}, A. and {Van Hemelryck}, E. and {van Leeuwen}, M. and {Varadi}, M. and {Vecchiato}, A. and {Veljanoski}, J. and {Via}, T. and {Vicente}, D. and {Vogt}, S. and {Voss}, H. and {Votruba}, V. and {Voutsinas}, S. and {Walmsley}, G. and {Weiler}, M. and {Weingrill}, K. and {Werner}, D. and {Wevers}, T. and {Whitehead}, G. and {Wyrzykowski}, {\L}. and {Yoldas}, A. and {{\v{Z}}erjal}, M. and {Zucker}, S. and {Zurbach}, C. and {Zwitter}, T. and {Alecu}, A. and {Allen}, M. and {Allende Prieto}, C. and {Amorim}, A. and {Anglada-Escud{\'e}}, G. and {Arsenijevic}, V. and {Azaz}, S. and {Balm}, P. and {Beck}, M. and {Bernstein}, H. -H. and {Bigot}, L. and {Bijaoui}, A. and {Blasco}, C. and {Bonfigli}, M. and {Bono}, G. and {Boudreault}, S. and {Bressan}, A. and {Brown}, S. and {Brunet}, P. -M. and {Bunclark}, P. and {Buonanno}, R. and {Butkevich}, A.~G. and {Carret}, C. and {Carrion}, C. and {Chemin}, L. and {Ch{\'e}reau}, F. and {Corcione}, L. and {Darmigny}, E. and {de Boer}, K.~S. and {de Teodoro}, P. and {de Zeeuw}, P.~T. and {Delle Luche}, C. and {Domingues}, C.~D. and {Dubath}, P. and {Fodor}, F. and {Fr{\'e}zouls}, B. and {Fries}, A. and {Fustes}, D. and {Fyfe}, D. and {Gallardo}, E. and {Gallegos}, J. and {Gardiol}, D. and {Gebran}, M. and {Gomboc}, A. and {G{\'o}mez}, A. and {Grux}, E. and {Gueguen}, A. and {Heyrovsky}, A. and {Hoar}, J. and {Iannicola}, G. and {Isasi Parache}, Y. and {Janotto}, A. -M. and {Joliet}, E. and {Jonckheere}, A. and {Keil}, R. and {Kim}, D. -W. and {Klagyivik}, P. and {Klar}, J. and {Knude}, J. and {Kochukhov}, O. and {Kolka}, I. and {Kos}, J. and {Kutka}, A. and {Lainey}, V. and {LeBouquin}, D. and {Liu}, C. and {Loreggia}, D. and {Makarov}, V.~V. and {Marseille}, M.~G. and {Martayan}, C. and {Martinez-Rubi}, O. and {Massart}, B. and {Meynadier}, F. and {Mignot}, S. and {Munari}, U. and {Nguyen}, A. -T. and {Nordlander}, T. and {Ocvirk}, P. and {O'Flaherty}, K.~S. and {Olias Sanz}, A. and {Ortiz}, P. and {Osorio}, J. and {Oszkiewicz}, D. and {Ouzounis}, A. and {Palmer}, M. and {Park}, P. and {Pasquato}, E. and {Peltzer}, C. and {Peralta}, J. and {P{\'e}turaud}, F. and {Pieniluoma}, T. and {Pigozzi}, E. and {Poels}, J. and {Prat}, G. and {Prod'homme}, T. and {Raison}, F. and {Rebordao}, J.~M. and {Risquez}, D. and {Rocca-Volmerange}, B. and {Rosen}, S. and {Ruiz-Fuertes}, M.~I. and {Russo}, F. and {Sembay}, S. and {Serraller Vizcaino}, I. and {Short}, A. and {Siebert}, A. and {Silva}, H. and {Sinachopoulos}, D. and {Slezak}, E. and {Soffel}, M. and {Sosnowska}, D. and {Strai{\v{z}}ys}, V. and {ter Linden}, M. and {Terrell}, D. and {Theil}, S. and {Tiede}, C. and {Troisi}, L. and {Tsalmantza}, P. and {Tur}, D. and {Vaccari}, M. and {Vachier}, F. and {Valles}, P. and {Van Hamme}, W. and {Veltz}, L. and {Virtanen}, J. and {Wallut}, J. -M. and {Wichmann}, R. and {Wilkinson}, M.~I. and {Ziaeepour}, H. and {Zschocke}, S.},
        title = "{The Gaia mission}",
      journal = {\aap},
         year = 2016,
        month = nov,
       volume = {595},
          eid = {A1},
        pages = {A1},
          doi = {10.1051/0004-6361/201629272},
archivePrefix = {arXiv},
       eprint = {1609.04153},
 primaryClass = {astro-ph.IM},
       adsurl = {https://ui.adsabs.harvard.edu/abs/2016A&A...595A...1G}
}

@ARTICLE{2018PASP..130f4505T,
       author = {{Tonry}, J.~L. and {Denneau}, L. and {Heinze}, A.~N. and {Stalder}, B. and {Smith}, K.~W. and {Smartt}, S.~J. and {Stubbs}, C.~W. and {Weiland}, H.~J. and {Rest}, A.},
        title = "{ATLAS: A High-cadence All-sky Survey System}",
      journal = {\pasp},
         year = 2018,
        month = jun,
       volume = {130},
       number = {988},
        pages = {064505},
          doi = {10.1088/1538-3873/aabadf},
archivePrefix = {arXiv},
       eprint = {1802.00879},
 primaryClass = {astro-ph.IM},
       adsurl = {https://ui.adsabs.harvard.edu/abs/2018PASP..130f4505T}
}

@ARTICLE{2020PASP..132h5002S,
       author = {{Smith}, K.~W. and {Smartt}, S.~J. and {Young}, D.~R. and {Tonry}, J.~L. and {Denneau}, L. and {Flewelling}, H. and {Heinze}, A.~N. and {Weiland}, H.~J. and {Stalder}, B. and {Rest}, A. and {Stubbs}, C.~W. and {Anderson}, J.~P. and {Chen}, T. -W. and {Clark}, P. and {Do}, A. and {F{\"o}rster}, F. and {Fulton}, M. and {Gillanders}, J. and {McBrien}, O.~R. and {O'Neill}, D. and {Srivastav}, S. and {Wright}, D.~E.},
        title = "{Design and Operation of the ATLAS Transient Science Server}",
      journal = {\pasp},
         year = 2020,
        month = aug,
       volume = {132},
       number = {1014},
          eid = {085002},
        pages = {085002},
          doi = {10.1088/1538-3873/ab936e},
archivePrefix = {arXiv},
       eprint = {2003.09052},
 primaryClass = {astro-ph.IM},
       adsurl = {https://ui.adsabs.harvard.edu/abs/2020PASP..132h5002S}
}

@ARTICLE{2019PASP..131a8002B,
       author = {{Bellm}, Eric C. and {Kulkarni}, Shrinivas R. and {Graham}, Matthew J. and {Dekany}, Richard and {Smith}, Roger M. and {Riddle}, Reed and {Masci}, Frank J. and {Helou}, George and {Prince}, Thomas A. and {Adams}, Scott M. and {Barbarino}, C. and {Barlow}, Tom and {Bauer}, James and {Beck}, Ron and {Belicki}, Justin and {Biswas}, Rahul and {Blagorodnova}, Nadejda and {Bodewits}, Dennis and {Bolin}, Bryce and {Brinnel}, Valery and {Brooke}, Tim and {Bue}, Brian and {Bulla}, Mattia and {Burruss}, Rick and {Cenko}, S. Bradley and {Chang}, Chan-Kao and {Connolly}, Andrew and {Coughlin}, Michael and {Cromer}, John and {Cunningham}, Virginia and {De}, Kishalay and {Delacroix}, Alex and {Desai}, Vandana and {Duev}, Dmitry A. and {Eadie}, Gwendolyn and {Farnham}, Tony L. and {Feeney}, Michael and {Feindt}, Ulrich and {Flynn}, David and {Franckowiak}, Anna and {Frederick}, S. and {Fremling}, C. and {Gal-Yam}, Avishay and {Gezari}, Suvi and {Giomi}, Matteo and {Goldstein}, Daniel A. and {Golkhou}, V. Zach and {Goobar}, Ariel and {Groom}, Steven and {Hacopians}, Eugean and {Hale}, David and {Henning}, John and {Ho}, Anna Y.~Q. and {Hover}, David and {Howell}, Justin and {Hung}, Tiara and {Huppenkothen}, Daniela and {Imel}, David and {Ip}, Wing-Huen and {Ivezi{\'c}}, {\v{Z}}eljko and {Jackson}, Edward and {Jones}, Lynne and {Juric}, Mario and {Kasliwal}, Mansi M. and {Kaspi}, S. and {Kaye}, Stephen and {Kelley}, Michael S.~P. and {Kowalski}, Marek and {Kramer}, Emily and {Kupfer}, Thomas and {Landry}, Walter and {Laher}, Russ R. and {Lee}, Chien-De and {Lin}, Hsing Wen and {Lin}, Zhong-Yi and {Lunnan}, Ragnhild and {Giomi}, Matteo and {Mahabal}, Ashish and {Mao}, Peter and {Miller}, Adam A. and {Monkewitz}, Serge and {Murphy}, Patrick and {Ngeow}, Chow-Choong and {Nordin}, Jakob and {Nugent}, Peter and {Ofek}, Eran and {Patterson}, Maria T. and {Penprase}, Bryan and {Porter}, Michael and {Rauch}, Ludwig and {Rebbapragada}, Umaa and {Reiley}, Dan and {Rigault}, Mickael and {Rodriguez}, Hector and {van Roestel}, Jan and {Rusholme}, Ben and {van Santen}, Jakob and {Schulze}, S. and {Shupe}, David L. and {Singer}, Leo P. and {Soumagnac}, Maayane T. and {Stein}, Robert and {Surace}, Jason and {Sollerman}, Jesper and {Szkody}, Paula and {Taddia}, F. and {Terek}, Scott and {Van Sistine}, Angela and {van Velzen}, Sjoert and {Vestrand}, W. Thomas and {Walters}, Richard and {Ward}, Charlotte and {Ye}, Quan-Zhi and {Yu}, Po-Chieh and {Yan}, Lin and {Zolkower}, Jeffry},
        title = "{The Zwicky Transient Facility: System Overview, Performance, and First Results}",
      journal = {\pasp},
         year = 2019,
        month = jan,
       volume = {131},
       number = {995},
        pages = {018002},
          doi = {10.1088/1538-3873/aaecbe},
archivePrefix = {arXiv},
       eprint = {1902.01932},
 primaryClass = {astro-ph.IM},
       adsurl = {https://ui.adsabs.harvard.edu/abs/2019PASP..131a8002B}
}

@ARTICLE{2019PASP..131g8001G,
       author = {{Graham}, Matthew J. and {Kulkarni}, S.~R. and {Bellm}, Eric C. and {Adams}, Scott M. and {Barbarino}, Cristina and {Blagorodnova}, Nadejda and {Bodewits}, Dennis and {Bolin}, Bryce and {Brady}, Patrick R. and {Cenko}, S. Bradley and {Chang}, Chan-Kao and {Coughlin}, Michael W. and {De}, Kishalay and {Eadie}, Gwendolyn and {Farnham}, Tony L. and {Feindt}, Ulrich and {Franckowiak}, Anna and {Fremling}, Christoffer and {Gezari}, Suvi and {Ghosh}, Shaon and {Goldstein}, Daniel A. and {Golkhou}, V. Zach and {Goobar}, Ariel and {Ho}, Anna Y.~Q. and {Huppenkothen}, Daniela and {Ivezi{\'c}}, {\v{Z}}eljko and {Jones}, R. Lynne and {Juric}, Mario and {Kaplan}, David L. and {Kasliwal}, Mansi M. and {Kelley}, Michael S.~P. and {Kupfer}, Thomas and {Lee}, Chien-De and {Lin}, Hsing Wen and {Lunnan}, Ragnhild and {Mahabal}, Ashish A. and {Miller}, Adam A. and {Ngeow}, Chow-Choong and {Nugent}, Peter and {Ofek}, Eran O. and {Prince}, Thomas A. and {Rauch}, Ludwig and {van Roestel}, Jan and {Schulze}, Steve and {Singer}, Leo P. and {Sollerman}, Jesper and {Taddia}, Francesco and {Yan}, Lin and {Ye}, Quan-Zhi and {Yu}, Po-Chieh and {Barlow}, Tom and {Bauer}, James and {Beck}, Ron and {Belicki}, Justin and {Biswas}, Rahul and {Brinnel}, Valery and {Brooke}, Tim and {Bue}, Brian and {Bulla}, Mattia and {Burruss}, Rick and {Connolly}, Andrew and {Cromer}, John and {Cunningham}, Virginia and {Dekany}, Richard and {Delacroix}, Alex and {Desai}, Vandana and {Duev}, Dmitry A. and {Feeney}, Michael and {Flynn}, David and {Frederick}, Sara and {Gal-Yam}, Avishay and {Giomi}, Matteo and {Groom}, Steven and {Hacopians}, Eugean and {Hale}, David and {Helou}, George and {Henning}, John and {Hover}, David and {Hillenbrand}, Lynne A. and {Howell}, Justin and {Hung}, Tiara and {Imel}, David and {Ip}, Wing-Huen and {Jackson}, Edward and {Kaspi}, Shai and {Kaye}, Stephen and {Kowalski}, Marek and {Kramer}, Emily and {Kuhn}, Michael and {Landry}, Walter and {Laher}, Russ R. and {Mao}, Peter and {Masci}, Frank J. and {Monkewitz}, Serge and {Murphy}, Patrick and {Nordin}, Jakob and {Patterson}, Maria T. and {Penprase}, Bryan and {Porter}, Michael and {Rebbapragada}, Umaa and {Reiley}, Dan and {Riddle}, Reed and {Rigault}, Mickael and {Rodriguez}, Hector and {Rusholme}, Ben and {van Santen}, Jakob and {Shupe}, David L. and {Smith}, Roger M. and {Soumagnac}, Maayane T. and {Stein}, Robert and {Surace}, Jason and {Szkody}, Paula and {Terek}, Scott and {Van Sistine}, Angela and {van Velzen}, Sjoert and {Vestrand}, W. Thomas and {Walters}, Richard and {Ward}, Charlotte and {Zhang}, Chaoran and {Zolkower}, Jeffry},
        title = "{The Zwicky Transient Facility: Science Objectives}",
      journal = {\pasp},
         year = 2019,
        month = jul,
       volume = {131},
       number = {1001},
        pages = {078001},
          doi = {10.1088/1538-3873/ab006c},
archivePrefix = {arXiv},
       eprint = {1902.01945},
 primaryClass = {astro-ph.IM},
       adsurl = {https://ui.adsabs.harvard.edu/abs/2019PASP..131g8001G}
}

@ARTICLE{2019PASP..131a8003M,
       author = {{Masci}, Frank J. and {Laher}, Russ R. and {Rusholme}, Ben and {Shupe}, David L. and {Groom}, Steven and {Surace}, Jason and {Jackson}, Edward and {Monkewitz}, Serge and {Beck}, Ron and {Flynn}, David and {Terek}, Scott and {Landry}, Walter and {Hacopians}, Eugean and {Desai}, Vandana and {Howell}, Justin and {Brooke}, Tim and {Imel}, David and {Wachter}, Stefanie and {Ye}, Quan-Zhi and {Lin}, Hsing-Wen and {Cenko}, S. Bradley and {Cunningham}, Virginia and {Rebbapragada}, Umaa and {Bue}, Brian and {Miller}, Adam A. and {Mahabal}, Ashish and {Bellm}, Eric C. and {Patterson}, Maria T. and {Juri{\'c}}, Mario and {Golkhou}, V. Zach and {Ofek}, Eran O. and {Walters}, Richard and {Graham}, Matthew and {Kasliwal}, Mansi M. and {Dekany}, Richard G. and {Kupfer}, Thomas and {Burdge}, Kevin and {Cannella}, Christopher B. and {Barlow}, Tom and {Van Sistine}, Angela and {Giomi}, Matteo and {Fremling}, Christoffer and {Blagorodnova}, Nadejda and {Levitan}, David and {Riddle}, Reed and {Smith}, Roger M. and {Helou}, George and {Prince}, Thomas A. and {Kulkarni}, Shrinivas R.},
        title = "{The Zwicky Transient Facility: Data Processing, Products, and Archive}",
      journal = {\pasp},
         year = 2019,
        month = jan,
       volume = {131},
       number = {995},
        pages = {018003},
          doi = {10.1088/1538-3873/aae8ac},
archivePrefix = {arXiv},
       eprint = {1902.01872},
 primaryClass = {astro-ph.IM},
       adsurl = {https://ui.adsabs.harvard.edu/abs/2019PASP..131a8003M}
}

@ARTICLE{2020PASP..132c8001D,
       author = {{Dekany}, Richard and {Smith}, Roger M. and {Riddle}, Reed and {Feeney}, Michael and {Porter}, Michael and {Hale}, David and {Zolkower}, Jeffry and {Belicki}, Justin and {Kaye}, Stephen and {Henning}, John and {Walters}, Richard and {Cromer}, John and {Delacroix}, Alex and {Rodriguez}, Hector and {Reiley}, Daniel J. and {Mao}, Peter and {Hover}, David and {Murphy}, Patrick and {Burruss}, Rick and {Baker}, John and {Kowalski}, Marek and {Reif}, Klaus and {Mueller}, Phillip and {Bellm}, Eric and {Graham}, Matthew and {Kulkarni}, Shrinivas R.},
        title = "{The Zwicky Transient Facility: Observing System}",
      journal = {\pasp},
         year = 2020,
        month = mar,
       volume = {132},
       number = {1009},
          eid = {038001},
        pages = {038001},
          doi = {10.1088/1538-3873/ab4ca2},
archivePrefix = {arXiv},
       eprint = {2008.04923},
 primaryClass = {astro-ph.IM},
       adsurl = {https://ui.adsabs.harvard.edu/abs/2020PASP..132c8001D}
}

@ARTICLE{2018PASP..130c5003B,
       author = {{Blagorodnova}, Nadejda and {Neill}, James D. and {Walters}, Richard and {Kulkarni}, Shrinivas R. and {Fremling}, Christoffer and {Ben-Ami}, Sagi and {Dekany}, Richard G. and {Fucik}, Jason R. and {Konidaris}, Nick and {Nash}, Reston and {Ngeow}, Chow-Choong and {Ofek}, Eran O. and {O' Sullivan}, Donal and {Quimby}, Robert and {Ritter}, Andreas and {Vyhmeister}, Karl E.},
        title = "{The SED Machine: A Robotic Spectrograph for Fast Transient Classification}",
      journal = {\pasp},
         year = 2018,
        month = mar,
       volume = {130},
       number = {985},
        pages = {035003},
          doi = {10.1088/1538-3873/aaa53f},
archivePrefix = {arXiv},
       eprint = {1710.02917},
 primaryClass = {astro-ph.IM},
       adsurl = {https://ui.adsabs.harvard.edu/abs/2018PASP..130c5003B}
}

@ARTICLE{2019A&A...627A.115R,
       author = {{Rigault}, M. and {Neill}, J.~D. and {Blagorodnova}, N. and {Dugas}, A. and {Feeney}, M. and {Walters}, R. and {Brinnel}, V. and {Copin}, Y. and {Fremling}, C. and {Nordin}, J. and {Sollerman}, J.},
        title = "{Fully automated integral field spectrograph pipeline for the SEDMachine: pysedm}",
      journal = {\aap},
         year = 2019,
        month = jul,
       volume = {627},
          eid = {A115},
        pages = {A115},
          doi = {10.1051/0004-6361/201935344},
archivePrefix = {arXiv},
       eprint = {1902.08526},
 primaryClass = {astro-ph.IM},
       adsurl = {https://ui.adsabs.harvard.edu/abs/2019A&A...627A.115R}
}

@ARTICLE{2022PASP..134b4505K,
       author = {{Kim}, Y. -L. and {Rigault}, M. and {Neill}, J.~D. and {Briday}, M. and {Copin}, Y. and {Lezmy}, J. and {Nicolas}, N. and {Riddle}, R. and {Sharma}, Y. and {Smith}, M. and {Sollerman}, J. and {Walters}, R.},
        title = "{New Modules for the SEDMachine to Remove Contaminations from Cosmic Rays and Non-target Light: BYECR and CONTSEP}",
      journal = {\pasp},
         year = 2022,
        month = feb,
       volume = {134},
       number = {1032},
          eid = {024505},
        pages = {024505},
          doi = {10.1088/1538-3873/ac50a0},
archivePrefix = {arXiv},
       eprint = {2203.01346},
 primaryClass = {astro-ph.IM},
       adsurl = {https://ui.adsabs.harvard.edu/abs/2022PASP..134b4505K}
}

@ARTICLE{2019JOSS....4.1247V,
       author = {{van der Walt}, St{\'e}fan and {Crellin-Quick}, Arien and {Bloom}, Joshua},
        title = "{SkyPortal: An Astronomical Data Platform}",
      journal = {The Journal of Open Source Software},
         year = 2019,
        month = may,
       volume = {4},
       number = {37},
          eid = {1247},
        pages = {1247},
          doi = {10.21105/joss.01247},
       adsurl = {https://ui.adsabs.harvard.edu/abs/2019JOSS....4.1247V}
}

@ARTICLE{2012ARNPS..62..407J,
       author = {{Janka}, Hans-Thomas},
        title = "{Explosion Mechanisms of Core-Collapse Supernovae}",
      journal = {Annual Review of Nuclear and Particle Science},
         year = 2012,
        month = nov,
       volume = {62},
       number = {1},
        pages = {407-451},
          doi = {10.1146/annurev-nucl-102711-094901},
archivePrefix = {arXiv},
       eprint = {1206.2503},
 primaryClass = {astro-ph.SR},
       adsurl = {https://ui.adsabs.harvard.edu/abs/2012ARNPS..62..407J}
}

@ARTICLE{2023ApJS..267...31C,
       author = {{Coughlin}, Michael W. and {Bloom}, Joshua S. and {Nir}, Guy and {Antier}, Sarah and {du Laz}, Theophile Jegou and {van der Walt}, St{\'e}fan and {Crellin-Quick}, Arien and {Culino}, Thomas and {Duev}, Dmitry A. and {Goldstein}, Daniel A. and {Healy}, Brian F. and {Karambelkar}, Viraj and {Lilleboe}, Jada and {Shin}, Kyung Min and {Singer}, Leo P. and {Ahumada}, Tom{\'a}s and {Anand}, Shreya and {Bellm}, Eric C. and {Dekany}, Richard and {Graham}, Matthew J. and {Kasliwal}, Mansi M. and {Kostadinova}, Ivona and {Kiendrebeogo}, R. Weizmann and {Kulkarni}, Shrinivas R. and {Jenkins}, Sydney and {LeBaron}, Natalie and {Mahabal}, Ashish A. and {Neill}, James D. and {Parazin}, B. and {Peloton}, Julien and {Perley}, Daniel A. and {Riddle}, Reed and {Rusholme}, Ben and {van Santen}, Jakob and {Sollerman}, Jesper and {Stein}, Robert and {Turpin}, D. and {Wold}, Avery and {Amat}, Carla and {Bonnefon}, Adrien and {Bonnefoy}, Adrien and {Flament}, Manon and {Kerkow}, Frank and {Kishore}, Sulekha and {Jani}, Shloke and {Mahanty}, Stephen K. and {Liu}, C{\'e}line and {Llinares}, Laura and {Makarison}, Jolyane and {Olli{\'e}ric}, Alix and {Perez}, In{\`e}s and {Pont}, Lydie and {Sharma}, Vyom},
        title = "{A Data Science Platform to Enable Time-domain Astronomy}",
      journal = {\apjs},
         year = 2023,
        month = aug,
       volume = {267},
       number = {2},
          eid = {31},
        pages = {31},
          doi = {10.3847/1538-4365/acdee1},
archivePrefix = {arXiv},
       eprint = {2305.00108},
 primaryClass = {astro-ph.IM},
       adsurl = {https://ui.adsabs.harvard.edu/abs/2023ApJS..267...31C}
}

@ARTICLE{2010AJ....140.1868W,
       author = {{Wright}, Edward L. and {Eisenhardt}, Peter R.~M. and {Mainzer}, Amy K. and {Ressler}, Michael E. and {Cutri}, Roc M. and {Jarrett}, Thomas and {Kirkpatrick}, J. Davy and {Padgett}, Deborah and {McMillan}, Robert S. and {Skrutskie}, Michael and {Stanford}, S.~A. and {Cohen}, Martin and {Walker}, Russell G. and {Mather}, John C. and {Leisawitz}, David and {Gautier}, Thomas N., III and {McLean}, Ian and {Benford}, Dominic and {Lonsdale}, Carol J. and {Blain}, Andrew and {Mendez}, Bryan and {Irace}, William R. and {Duval}, Valerie and {Liu}, Fengchuan and {Royer}, Don and {Heinrichsen}, Ingolf and {Howard}, Joan and {Shannon}, Mark and {Kendall}, Martha and {Walsh}, Amy L. and {Larsen}, Mark and {Cardon}, Joel G. and {Schick}, Scott and {Schwalm}, Mark and {Abid}, Mohamed and {Fabinsky}, Beth and {Naes}, Larry and {Tsai}, Chao-Wei},
        title = "{The Wide-field Infrared Survey Explorer (WISE): Mission Description and Initial On-orbit Performance}",
      journal = {\aj},
         year = 2010,
        month = dec,
       volume = {140},
       number = {6},
        pages = {1868-1881},
          doi = {10.1088/0004-6256/140/6/1868},
archivePrefix = {arXiv},
       eprint = {1008.0031},
 primaryClass = {astro-ph.IM},
       adsurl = {https://ui.adsabs.harvard.edu/abs/2010AJ....140.1868W}
}

@ARTICLE{2020PASP..132c5001L,
       author = {{Lacy}, M. and {Baum}, S.~A. and {Chandler}, C.~J. and {Chatterjee}, S. and {Clarke}, T.~E. and {Deustua}, S. and {English}, J. and {Farnes}, J. and {Gaensler}, B.~M. and {Gugliucci}, N. and {Hallinan}, G. and {Kent}, B.~R. and {Kimball}, A. and {Law}, C.~J. and {Lazio}, T.~J.~W. and {Marvil}, J. and {Mao}, S.~A. and {Medlin}, D. and {Mooley}, K. and {Murphy}, E.~J. and {Myers}, S. and {Osten}, R. and {Richards}, G.~T. and {Rosolowsky}, E. and {Rudnick}, L. and {Schinzel}, F. and {Sivakoff}, G.~R. and {Sjouwerman}, L.~O. and {Taylor}, R. and {White}, R.~L. and {Wrobel}, J. and {Andernach}, H. and {Beasley}, A.~J. and {Berger}, E. and {Bhatnager}, S. and {Birkinshaw}, M. and {Bower}, G.~C. and {Brandt}, W.~N. and {Brown}, S. and {Burke-Spolaor}, S. and {Butler}, B.~J. and {Comerford}, J. and {Demorest}, P.~B. and {Fu}, H. and {Giacintucci}, S. and {Golap}, K. and {G{\"u}th}, T. and {Hales}, C.~A. and {Hiriart}, R. and {Hodge}, J. and {Horesh}, A. and {Ivezi{\'c}}, {\v{Z}}. and {Jarvis}, M.~J. and {Kamble}, A. and {Kassim}, N. and {Liu}, X. and {Loinard}, L. and {Lyons}, D.~K. and {Masters}, J. and {Mezcua}, M. and {Moellenbrock}, G.~A. and {Mroczkowski}, T. and {Nyland}, K. and {O'Dea}, C.~P. and {O'Sullivan}, S.~P. and {Peters}, W.~M. and {Radford}, K. and {Rao}, U. and {Robnett}, J. and {Salcido}, J. and {Shen}, Y. and {Sobotka}, A. and {Witz}, S. and {Vaccari}, M. and {van Weeren}, R.~J. and {Vargas}, A. and {Williams}, P.~K.~G. and {Yoon}, I.},
        title = "{The Karl G. Jansky Very Large Array Sky Survey (VLASS). Science Case and Survey Design}",
      journal = {\pasp},
         year = 2020,
        month = mar,
       volume = {132},
       number = {1009},
          eid = {035001},
        pages = {035001},
          doi = {10.1088/1538-3873/ab63eb},
archivePrefix = {arXiv},
       eprint = {1907.01981},
 primaryClass = {astro-ph.IM},
       adsurl = {https://ui.adsabs.harvard.edu/abs/2020PASP..132c5001L}
}

@ARTICLE{2021ApJS..255...30G,
       author = {{Gordon}, Yjan A. and {Boyce}, Michelle M. and {O'Dea}, Christopher P. and {Rudnick}, Lawrence and {Andernach}, Heinz and {Vantyghem}, Adrian N. and {Baum}, Stefi A. and {Bui}, Jean-Paul and {Dionyssiou}, Mathew and {Safi-Harb}, Samar and {Sander}, Isabel},
        title = "{A Quick Look at the 3 GHz Radio Sky. I. Source Statistics from the Very Large Array Sky Survey}",
      journal = {\apjs},
         year = 2021,
        month = aug,
       volume = {255},
       number = {2},
          eid = {30},
        pages = {30},
          doi = {10.3847/1538-4365/ac05c0},
archivePrefix = {arXiv},
       eprint = {2102.11753},
 primaryClass = {astro-ph.GA},
       adsurl = {https://ui.adsabs.harvard.edu/abs/2021ApJS..255...30G}
}

@ARTICLE{2021A&A...652A..76H,
       author = {{Hodgkin}, S.~T. and {Harrison}, D.~L. and {Breedt}, E. and {Wevers}, T. and {Rixon}, G. and {Delgado}, A. and {Yoldas}, A. and {Kostrzewa-Rutkowska}, Z. and {Wyrzykowski}, {\L}. and {van Leeuwen}, M. and {Blagorodnova}, N. and {Campbell}, H. and {Eappachen}, D. and {Fraser}, M. and {Ihanec}, N. and {Koposov}, S.~E. and {Kruszy{\'n}ska}, K. and {Marton}, G. and {Rybicki}, K.~A. and {Brown}, A.~G.~A. and {Burgess}, P.~W. and {Busso}, G. and {Cowell}, S. and {De Angeli}, F. and {Diener}, C. and {Evans}, D.~W. and {Gilmore}, G. and {Holland}, G. and {Jonker}, P.~G. and {van Leeuwen}, F. and {Mignard}, F. and {Osborne}, P.~J. and {Portell}, J. and {Prusti}, T. and {Richards}, P.~J. and {Riello}, M. and {Seabroke}, G.~M. and {Walton}, N.~A. and {{\'A}brah{\'a}m}, P. and {Altavilla}, G. and {Baker}, S.~G. and {Bastian}, U. and {O'Brien}, P. and {de Bruijne}, J. and {Butterley}, T. and {Carrasco}, J.~M. and {Casta{\~n}eda}, J. and {Clark}, J.~S. and {Clementini}, G. and {Copperwheat}, C.~M. and {Cropper}, M. and {Damljanovic}, G. and {Davidson}, M. and {Davis}, C.~J. and {Dennefeld}, M. and {Dhillon}, V.~S. and {Dolding}, C. and {Dominik}, M. and {Esquej}, P. and {Eyer}, L. and {Fabricius}, C. and {Fridman}, M. and {Froebrich}, D. and {Garralda}, N. and {Gomboc}, A. and {Gonz{\'a}lez-Vidal}, J.~J. and {Guerra}, R. and {Hambly}, N.~C. and {Hardy}, L.~K. and {Holl}, B. and {Hourihane}, A. and {Japelj}, J. and {Kann}, D.~A. and {Kiss}, C. and {Knigge}, C. and {Kolb}, U. and {Komossa}, S. and {K{\'o}sp{\'a}l}, {\'A}. and {Kov{\'a}cs}, G. and {Kun}, M. and {Leto}, G. and {Lewis}, F. and {Littlefair}, S.~P. and {Mahabal}, A.~A. and {Mundell}, C.~G. and {Nagy}, Z. and {Padeletti}, D. and {Palaversa}, L. and {Pigulski}, A. and {Pretorius}, M.~L. and {van Reeven}, W. and {Ribeiro}, V.~A.~R.~M. and {Roelens}, M. and {Rowell}, N. and {Schartel}, N. and {Scholz}, A. and {Schwope}, A. and {Sip{\H{o}}cz}, B.~M. and {Smartt}, S.~J. and {Smith}, M.~D. and {Serraller}, I. and {Steeghs}, D. and {Sullivan}, M. and {Szabados}, L. and {Szegedi-Elek}, E. and {Tisserand}, P. and {Tomasella}, L. and {van Velzen}, S. and {Whitelock}, P.~A. and {Wilson}, R.~W. and {Young}, D.~R.},
        title = "{Gaia Early Data Release 3. Gaia photometric science alerts}",
      journal = {\aap},
         year = 2021,
        month = aug,
       volume = {652},
          eid = {A76},
        pages = {A76},
          doi = {10.1051/0004-6361/202140735},
archivePrefix = {arXiv},
       eprint = {2106.01394},
 primaryClass = {astro-ph.IM},
       adsurl = {https://ui.adsabs.harvard.edu/abs/2021A&A...652A..76H}
}

@ARTICLE{2002astro.ph.10394H,
       author = {{Hogg}, David W. and {Baldry}, Ivan K. and {Blanton}, Michael R. and {Eisenstein}, Daniel J.},
        title = "{The K correction}",
      journal = {arXiv e-prints},
         year = 2002,
        month = oct,
          eid = {astro-ph/0210394},
        pages = {astro-ph/0210394},
          doi = {10.48550/arXiv.astro-ph/0210394},
archivePrefix = {arXiv},
       eprint = {astro-ph/0210394},
 primaryClass = {astro-ph},
       adsurl = {https://ui.adsabs.harvard.edu/abs/2002astro.ph.10394H}
}

@ARTICLE{2023ApJ...943...41C,
       author = {{Chen}, Z.~H. and {Yan}, Lin and {Kangas}, T. and {Lunnan}, R. and {Schulze}, S. and {Sollerman}, J. and {Perley}, D.~A. and {Chen}, T. -W. and {Taggart}, K. and {Hinds}, K.~R. and {Gal-Yam}, A. and {Wang}, X.~F. and {Andreoni}, I. and {Bellm}, E. and {Bloom}, J.~S. and {Burdge}, K. and {Burgos}, A. and {Cook}, D. and {Dahiwale}, A. and {De}, K. and {Dekany}, R. and {Dugas}, A. and {Frederik}, S. and {Fremling}, C. and {Graham}, M. and {Hankins}, M. and {Ho}, A. and {Jencson}, J. and {Karambelkar}, V. and {Kasliwal}, M. and {Kulkarni}, S. and {Laher}, R. and {Rusholme}, B. and {Sharma}, Y. and {Taddia}, F. and {Tartaglia}, L. and {Thomas}, B.~P. and {Tzanidakis}, A. and {Van Roestel}, J. and {Walter}, R. and {Yang}, Y. and {Yao}, Y.~H. and {Yaron}, O.},
        title = "{The Hydrogen-poor Superluminous Supernovae from the Zwicky Transient Facility Phase I Survey. I. Light Curves and Measurements}",
      journal = {\apj},
         year = 2023,
        month = jan,
       volume = {943},
       number = {1},
          eid = {41},
        pages = {41},
          doi = {10.3847/1538-4357/aca161},
archivePrefix = {arXiv},
       eprint = {2202.02059},
 primaryClass = {astro-ph.HE},
       adsurl = {https://ui.adsabs.harvard.edu/abs/2023ApJ...943...41C}
}

@ARTICLE{2025A&A...695A.142P,
       author = {{Pessi}, P.~J. and {Lunnan}, R. and {Sollerman}, J. and {Schulze}, S. and {Gkini}, A. and {Gangopadhyay}, A. and {Yan}, L. and {Gal-Yam}, A. and {Perley}, D.~A. and {Chen}, T. -W. and {Hinds}, K.~R. and {Brennan}, S.~J. and {Hu}, Y. and {Singh}, A. and {Andreoni}, I. and {Cook}, D.~O. and {Fremling}, C. and {Ho}, A.~Y.~Q. and {Sharma}, Y. and {van Velzen}, S. and {Kangas}, T. and {Wold}, A. and {Bellm}, E.~C. and {Bloom}, J.~S. and {Graham}, M.~J. and {Kasliwal}, M.~M. and {Kulkarni}, S.~R. and {Riddle}, R. and {Rusholme}, B.},
        title = "{Sample of hydrogen-rich superluminous supernovae from the Zwicky Transient Facility}",
      journal = {\aap},
         year = 2025,
        month = mar,
       volume = {695},
          eid = {A142},
        pages = {A142},
          doi = {10.1051/0004-6361/202452014},
archivePrefix = {arXiv},
       eprint = {2408.15086},
 primaryClass = {astro-ph.HE},
       adsurl = {https://ui.adsabs.harvard.edu/abs/2025A&A...695A.142P}
}

@MISC{2012ivoa.rept.1015R,
       author = {{Rodrigo}, Carlos and {Solano}, Enrique and {Bayo}, Amelia},
        title = "{SVO Filter Profile Service Version 1.0}",
 howpublished = {IVOA Working Draft 15 October 2012},
         year = 2012,
        month = oct,
        pages = {1015},
          doi = {10.5479/ADS/bib/2012ivoa.rept.1015R},
       adsurl = {https://ui.adsabs.harvard.edu/abs/2012ivoa.rept.1015R}
}

@INPROCEEDINGS{2020sea..confE.182R,
       author = {{Rodrigo}, C. and {Solano}, E.},
        title = "{The SVO Filter Profile Service}",
    booktitle = {XIV.0 Scientific Meeting (virtual) of the Spanish Astronomical Society},
         year = 2020,
        month = jul,
          eid = {182},
        pages = {182},
       adsurl = {https://ui.adsabs.harvard.edu/abs/2020sea..confE.182R}
}

@software{Young_plot_atlas_fp,
    author = {Young, David R.},
    doi = {10.5281/zenodo.10978968},
    license = {GPL-3.0-only},
    title = {{plot\_atlas\_fp.py}},
    url = {https://zenodo.org/doi/10.5281/zenodo.10978968},
    year = {2020}
}

@ARTICLE{2016A&A...593A..68F,
       author = {{Fremling}, C. and {Sollerman}, J. and {Taddia}, F. and {Ergon}, M. and {Fraser}, M. and {Karamehmetoglu}, E. and {Valenti}, S. and {Jerkstrand}, A. and {Arcavi}, I. and {Bufano}, F. and {Elias Rosa}, N. and {Filippenko}, A.~V. and {Fox}, D. and {Gal-Yam}, A. and {Howell}, D.~A. and {Kotak}, R. and {Mazzali}, P. and {Milisavljevic}, D. and {Nugent}, P.~E. and {Nyholm}, A. and {Pian}, E. and {Smartt}, S.},
        title = "{PTF12os and iPTF13bvn. Two stripped-envelope supernovae from low-mass progenitors in NGC 5806}",
      journal = {\aap},
         year = 2016,
        month = sep,
       volume = {593},
          eid = {A68},
        pages = {A68},
          doi = {10.1051/0004-6361/201628275},
archivePrefix = {arXiv},
       eprint = {1606.03074},
 primaryClass = {astro-ph.HE},
       adsurl = {https://ui.adsabs.harvard.edu/abs/2016A&A...593A..68F}
}

@INPROCEEDINGS{1995AAS...186.4405G,
       author = {{Gunn}, James E.},
        title = "{The Sloan Digital Sky Survey}",
    booktitle = {American Astronomical Society Meeting Abstracts \#186},
         year = 1995,
       series = {American Astronomical Society Meeting Abstracts},
       volume = {186},
        month = may,
          eid = {44.05},
        pages = {44.05},
       adsurl = {https://ui.adsabs.harvard.edu/abs/1995AAS...186.4405G}
}

@ARTICLE{2011ApJ...731...53M,
       author = {{Mainzer}, A. and {Bauer}, J. and {Grav}, T. and {Masiero}, J. and {Cutri}, R.~M. and {Dailey}, J. and {Eisenhardt}, P. and {McMillan}, R.~S. and {Wright}, E. and {Walker}, R. and {Jedicke}, R. and {Spahr}, T. and {Tholen}, D. and {Alles}, R. and {Beck}, R. and {Brandenburg}, H. and {Conrow}, T. and {Evans}, T. and {Fowler}, J. and {Jarrett}, T. and {Marsh}, K. and {Masci}, F. and {McCallon}, H. and {Wheelock}, S. and {Wittman}, M. and {Wyatt}, P. and {DeBaun}, E. and {Elliott}, G. and {Elsbury}, D. and {Gautier}, T., IV and {Gomillion}, S. and {Leisawitz}, D. and {Maleszewski}, C. and {Micheli}, M. and {Wilkins}, A.},
        title = "{Preliminary Results from NEOWISE: An Enhancement to the Wide-field Infrared Survey Explorer for Solar System Science}",
      journal = {\apj},
         year = 2011,
        month = apr,
       volume = {731},
       number = {1},
          eid = {53},
        pages = {53},
          doi = {10.1088/0004-637X/731/1/53},
archivePrefix = {arXiv},
       eprint = {1102.1996},
 primaryClass = {astro-ph.EP},
       adsurl = {https://ui.adsabs.harvard.edu/abs/2011ApJ...731...53M}
}

@INPROCEEDINGS{2014SPIE.9147E..8HP,
       author = {{Piascik}, A.~S. and {Steele}, Iain A. and {Bates}, Stuart D. and {Mottram}, Christopher J. and {Smith}, R.~J. and {Barnsley}, R.~M. and {Bolton}, B.},
        title = "{SPRAT: Spectrograph for the Rapid Acquisition of Transients}",
    booktitle = {Ground-based and Airborne Instrumentation for Astronomy V},
         year = 2014,
       editor = {{Ramsay}, Suzanne K. and {McLean}, Ian S. and {Takami}, Hideki},
       series = {Society of Photo-Optical Instrumentation Engineers (SPIE) Conference Series},
       volume = {9147},
        month = jul,
          eid = {91478H},
        pages = {91478H},
          doi = {10.1117/12.2055117},
       adsurl = {https://ui.adsabs.harvard.edu/abs/2014SPIE.9147E..8HP}
}

@ARTICLE{1995PASP..107..375O,
       author = {{Oke}, J.~B. and {Cohen}, J.~G. and {Carr}, M. and {Cromer}, J. and {Dingizian}, A. and {Harris}, F.~H. and {Labrecque}, S. and {Lucinio}, R. and {Schaal}, W. and {Epps}, H. and {Miller}, J.},
        title = "{The Keck Low-Resolution Imaging Spectrometer}",
      journal = {\pasp},
         year = 1995,
        month = apr,
       volume = {107},
        pages = {375},
          doi = {10.1086/133562},
       adsurl = {https://ui.adsabs.harvard.edu/abs/1995PASP..107..375O}
}

@ARTICLE{1982PASP...94..586O,
       author = {{Oke}, J.~B. and {Gunn}, J.~E.},
        title = "{An Efficient Low Resolution and Moderate Resolution Spectrograph for the Hale Telescope}",
      journal = {\pasp},
         year = 1982,
        month = jun,
       volume = {94},
        pages = {586},
          doi = {10.1086/131027},
       adsurl = {https://ui.adsabs.harvard.edu/abs/1982PASP...94..586O}
}

@ARTICLE{2012AN....333..101B,
       author = {{Barnsley}, R.~M. and {Smith}, R.~J. and {Steele}, I.~A.},
        title = "{A fully automated data reduction pipeline for the FRODOSpec integral field spectrograph}",
      journal = {Astronomische Nachrichten},
         year = 2012,
        month = feb,
       volume = {333},
       number = {2},
        pages = {101-117},
          doi = {10.1002/asna.201111634},
archivePrefix = {arXiv},
       eprint = {1112.2574},
 primaryClass = {astro-ph.IM},
       adsurl = {https://ui.adsabs.harvard.edu/abs/2012AN....333..101B}
}

@ARTICLE{2019PASP..131h4503P,
       author = {{Perley}, Daniel A.},
        title = "{Fully Automated Reduction of Longslit Spectroscopy with the Low Resolution Imaging Spectrometer at the Keck Observatory}",
      journal = {\pasp},
         year = 2019,
        month = aug,
       volume = {131},
       number = {1002},
        pages = {084503},
          doi = {10.1088/1538-3873/ab215d},
archivePrefix = {arXiv},
       eprint = {1903.07629},
 primaryClass = {astro-ph.IM},
       adsurl = {https://ui.adsabs.harvard.edu/abs/2019PASP..131h4503P}
}

@ARTICLE{2020JOSS....5.2308P,
       author = {{Prochaska}, J. and {Hennawi}, Joseph and {Westfall}, Kyle and {Cooke}, Ryan and {Wang}, Feige and {Hsyu}, Tiffany and {Davies}, Frederick and {Farina}, Emanuele and {Pelliccia}, Debora},
        title = "{PypeIt: The Python Spectroscopic Data Reduction Pipeline}",
      journal = {The Journal of Open Source Software},
         year = 2020,
        month = dec,
       volume = {5},
       number = {56},
          eid = {2308},
        pages = {2308},
          doi = {10.21105/joss.02308},
archivePrefix = {arXiv},
       eprint = {2005.06505},
 primaryClass = {astro-ph.IM},
       adsurl = {https://ui.adsabs.harvard.edu/abs/2020JOSS....5.2308P}
}

@article{pypeit:joss_pub,
    doi = {10.21105/joss.02308},
    url = {https://doi.org/10.21105/joss.02308},
    year = {2020},
    publisher = {The Open Journal},
    volume = {5},
    number = {56},
    pages = {2308},
    author = {J. Xavier Prochaska and Joseph F. Hennawi and Kyle B. Westfall and Ryan J. Cooke and Feige Wang and Tiffany Hsyu and Frederick B. Davies and Emanuele Paolo Farina and Debora Pelliccia},
    title = {PypeIt: The Python Spectroscopic Data Reduction Pipeline},
    journal = {Journal of Open Source Software}
}

@ARTICLE{2022JOSS....7.3612M,
       author = {{Mandigo-Stoba}, Milan Sharma and {Fremling}, Christoffer and {Kasliwal}, Mansi},
        title = "{DBSP\_DRP: A Python package for automated spectroscopic data reduction of DBSP data}",
      journal = {The Journal of Open Source Software},
         year = 2022,
        month = feb,
       volume = {7},
       number = {70},
          eid = {3612},
        pages = {3612},
          doi = {10.21105/joss.03612},
archivePrefix = {arXiv},
       eprint = {2107.12339},
 primaryClass = {astro-ph.IM},
       adsurl = {https://ui.adsabs.harvard.edu/abs/2022JOSS....7.3612M}
}

@ARTICLE{2011ApJ...737..103S,
       author = {{Schlafly}, Edward F. and {Finkbeiner}, Douglas P.},
        title = "{Measuring Reddening with Sloan Digital Sky Survey Stellar Spectra and Recalibrating SFD}",
      journal = {\apj},
         year = 2011,
        month = aug,
       volume = {737},
       number = {2},
          eid = {103},
        pages = {103},
          doi = {10.1088/0004-637X/737/2/103},
archivePrefix = {arXiv},
       eprint = {1012.4804},
 primaryClass = {astro-ph.GA},
       adsurl = {https://ui.adsabs.harvard.edu/abs/2011ApJ...737..103S}
}

@ARTICLE{1998ApJ...500..525S,
       author = {{Schlegel}, David J. and {Finkbeiner}, Douglas P. and {Davis}, Marc},
        title = "{Maps of Dust Infrared Emission for Use in Estimation of Reddening and Cosmic Microwave Background Radiation Foregrounds}",
      journal = {\apj},
         year = 1998,
        month = jun,
       volume = {500},
       number = {2},
        pages = {525-553},
          doi = {10.1086/305772},
archivePrefix = {arXiv},
       eprint = {astro-ph/9710327},
 primaryClass = {astro-ph},
       adsurl = {https://ui.adsabs.harvard.edu/abs/1998ApJ...500..525S}
}

@ARTICLE{1999PASP..111...63F,
       author = {{Fitzpatrick}, Edward L.},
        title = "{Correcting for the Effects of Interstellar Extinction}",
      journal = {\pasp},
         year = 1999,
        month = jan,
       volume = {111},
       number = {755},
        pages = {63-75},
          doi = {10.1086/316293},
archivePrefix = {arXiv},
       eprint = {astro-ph/9809387},
 primaryClass = {astro-ph},
       adsurl = {https://ui.adsabs.harvard.edu/abs/1999PASP..111...63F}
}

@ARTICLE{2021TNSAN...7....1S,
       author = {{Shingles}, L. and {Smith}, K.~W. and {Young}, D.~R. and {Smartt}, S.~J. and {Tonry}, J. and {Denneau}, L. and {Heinze}, A. and {Weiland}, H. and {Flewelling}, H. and {Stalder}, B. and {Clocchiatti}, A. and {F{\"o}rster}, F. and {Pignata}, G. and {Rest}, A. and {Anderson}, J. and {Stubbs}, C. and {Erasmus}, N.},
        title = "{Release of the ATLAS Forced Photometry server for public use}",
      journal = {Transient Name Server AstroNote},
         year = 2021,
        month = jan,
       volume = {7},
        pages = {1-7},
       adsurl = {https://ui.adsabs.harvard.edu/abs/2021TNSAN...7....1S}
}

@INCOLLECTION{2017hsn..book..195G,
       author = {{Gal-Yam}, Avishay},
        title = "{Observational and Physical Classification of Supernovae}",
    booktitle = {Handbook of Supernovae},
         year = 2017,
       editor = {{Alsabti}, Athem W. and {Murdin}, Paul},
        pages = {195},
          doi = {10.1007/978-3-319-21846-5_35},
       adsurl = {https://ui.adsabs.harvard.edu/abs/2017hsn..book..195G}
}

@ARTICLE{2025A&A...695A..29S,
       author = {{Salmaso}, I. and {Cappellaro}, E. and {Tartaglia}, L. and {Anderson}, J.~P. and {Benetti}, S. and {Bronikowski}, M. and {Cai}, Y. -Z. and {Charalampopoulos}, P. and {Chen}, T. -W. and {Concepcion}, E. and {Elias-Rosa}, N. and {Galbany}, L. and {Gromadzki}, M. and {Guti{\'e}rrez}, C.~P. and {Kankare}, E. and {Lundqvist}, P. and {Matilainen}, K. and {Mazzali}, P.~A. and {Moran}, S. and {M{\"u}ller-Bravo}, T.~E. and {Nicholl}, M. and {Pastorello}, A. and {Pessi}, P.~J. and {Pessi}, T. and {Petrushevska}, T. and {Pignata}, G. and {Reguitti}, A. and {Sollerman}, J. and {Srivastav}, S. and {Stritzinger}, M. and {Tomasella}, L. and {Valerin}, G.},
        title = "{The diversity of strongly interacting Type IIn supernovae}",
      journal = {\aap},
         year = 2025,
        month = mar,
       volume = {695},
          eid = {A29},
        pages = {A29},
          doi = {10.1051/0004-6361/202451764},
archivePrefix = {arXiv},
       eprint = {2410.06111},
 primaryClass = {astro-ph.HE},
       adsurl = {https://ui.adsabs.harvard.edu/abs/2025A&A...695A..29S}
}

@ARTICLE{2021ARA&A..59...21G,
       author = {{Gezari}, Suvi},
        title = "{Tidal Disruption Events}",
      journal = {\araa},
         year = 2021,
        month = sep,
       volume = {59},
        pages = {21-58},
          doi = {10.1146/annurev-astro-111720-030029},
archivePrefix = {arXiv},
       eprint = {2104.14580},
 primaryClass = {astro-ph.HE},
       adsurl = {https://ui.adsabs.harvard.edu/abs/2021ARA&A..59...21G}
}

@ARTICLE{2017A&ARv..25....2P,
       author = {{Padovani}, P. and {Alexander}, D.~M. and {Assef}, R.~J. and {De Marco}, B. and {Giommi}, P. and {Hickox}, R.~C. and {Richards}, G.~T. and {Smol{\v{c}}i{\'c}}, V. and {Hatziminaoglou}, E. and {Mainieri}, V. and {Salvato}, M.},
        title = "{Active galactic nuclei: what's in a name?}",
      journal = {\aapr},
         year = 2017,
        month = aug,
       volume = {25},
       number = {1},
          eid = {2},
        pages = {2},
          doi = {10.1007/s00159-017-0102-9},
archivePrefix = {arXiv},
       eprint = {1707.07134},
 primaryClass = {astro-ph.GA},
       adsurl = {https://ui.adsabs.harvard.edu/abs/2017A&ARv..25....2P}
}

@ARTICLE{2021ApJ...913..102W,
       author = {{Ward}, Charlotte and {Gezari}, Suvi and {Frederick}, Sara and {Hammerstein}, Erica and {Nugent}, Peter and {van Velzen}, Sjoert and {Drake}, Andrew and {Garc{\'\i}a-P{\'e}rez}, Abigail and {Oyoo}, Immaculate and {Bellm}, Eric C. and {Duev}, Dmitry A. and {Graham}, Matthew J. and {Kasliwal}, Mansi M. and {Kaye}, Stephen and {Mahabal}, Ashish A. and {Masci}, Frank J. and {Rusholme}, Ben and {Soumagnac}, Maayane T. and {Yan}, Lin},
        title = "{AGNs on the Move: A Search for Off-nuclear AGNs from Recoiling Supermassive Black Holes and Ongoing Galaxy Mergers with the Zwicky Transient Facility}",
      journal = {\apj},
         year = 2021,
        month = jun,
       volume = {913},
       number = {2},
          eid = {102},
        pages = {102},
          doi = {10.3847/1538-4357/abf246},
archivePrefix = {arXiv},
       eprint = {2011.11656},
 primaryClass = {astro-ph.GA},
       adsurl = {https://ui.adsabs.harvard.edu/abs/2021ApJ...913..102W}
}

@ARTICLE{2021MNRAS.507..156G,
       author = {{Grishin}, Evgeni and {Bobrick}, Alexey and {Hirai}, Ryosuke and {Mandel}, Ilya and {Perets}, Hagai B.},
        title = "{Supernova explosions in active galactic nuclear discs}",
      journal = {\mnras},
         year = 2021,
        month = oct,
       volume = {507},
       number = {1},
        pages = {156-174},
          doi = {10.1093/mnras/stab1957},
archivePrefix = {arXiv},
       eprint = {2105.09953},
 primaryClass = {astro-ph.HE},
       adsurl = {https://ui.adsabs.harvard.edu/abs/2021MNRAS.507..156G}
}

@ARTICLE{2023ApJ...950..161L,
       author = {{Li}, Fu-Lin and {Liu}, Yu and {Fan}, Xiao and {Hu}, Mao-Kai and {Yang}, Xuan and {Geng}, Jin-Jun and {Wu}, Xue-Feng},
        title = "{Core-collapse Supernova Explosions in Active Galactic Nucleus Accretion Disks}",
      journal = {\apj},
         year = 2023,
        month = jun,
       volume = {950},
       number = {2},
          eid = {161},
        pages = {161},
          doi = {10.3847/1538-4357/acd2d1},
archivePrefix = {arXiv},
       eprint = {2305.04010},
 primaryClass = {astro-ph.HE},
       adsurl = {https://ui.adsabs.harvard.edu/abs/2023ApJ...950..161L}
}

@ARTICLE{2020SSRv..216...32F,
       author = {{French}, K. Decker and {Wevers}, Thomas and {Law-Smith}, Jamie and {Graur}, Or and {Zabludoff}, Ann I.},
        title = "{The Host Galaxies of Tidal Disruption Events}",
      journal = {\ssr},
         year = 2020,
        month = mar,
       volume = {216},
       number = {3},
          eid = {32},
        pages = {32},
          doi = {10.1007/s11214-020-00657-y},
archivePrefix = {arXiv},
       eprint = {2003.02863},
 primaryClass = {astro-ph.HE},
       adsurl = {https://ui.adsabs.harvard.edu/abs/2020SSRv..216...32F}
}

@ARTICLE{2023NatAs...7.1282R,
       author = {{Ricci}, Claudio and {Trakhtenbrot}, Benny},
        title = "{Changing-look active galactic nuclei}",
      journal = {Nature Astronomy},
         year = 2023,
        month = nov,
       volume = {7},
        pages = {1282-1294},
          doi = {10.1038/s41550-023-02108-4},
archivePrefix = {arXiv},
       eprint = {2211.05132},
 primaryClass = {astro-ph.GA},
       adsurl = {https://ui.adsabs.harvard.edu/abs/2023NatAs...7.1282R}
}

@ARTICLE{2007ApJ...666.1024B,
       author = {{Blondin}, St{\'e}phane and {Tonry}, John L.},
        title = "{Determining the Type, Redshift, and Age of a Supernova Spectrum}",
      journal = {\apj},
         year = 2007,
        month = sep,
       volume = {666},
       number = {2},
        pages = {1024-1047},
          doi = {10.1086/520494},
archivePrefix = {arXiv},
       eprint = {0709.4488},
 primaryClass = {astro-ph},
       adsurl = {https://ui.adsabs.harvard.edu/abs/2007ApJ...666.1024B}
}

@ARTICLE{2005ApJ...634.1190H,
       author = {{Howell}, D.~A. and {Sullivan}, M. and {Perrett}, K. and {Bronder}, T.~J. and {Hook}, I.~M. and {Astier}, P. and {Aubourg}, E. and {Balam}, D. and {Basa}, S. and {Carlberg}, R.~G. and {Fabbro}, S. and {Fouchez}, D. and {Guy}, J. and {Lafoux}, H. and {Neill}, J.~D. and {Pain}, R. and {Palanque-Delabrouille}, N. and {Pritchet}, C.~J. and {Regnault}, N. and {Rich}, J. and {Taillet}, R. and {Knop}, R. and {McMahon}, R.~G. and {Perlmutter}, S. and {Walton}, N.~A.},
        title = "{Gemini Spectroscopy of Supernovae from the Supernova Legacy Survey: Improving High-Redshift Supernova Selection and Classification}",
      journal = {\apj},
         year = 2005,
        month = dec,
       volume = {634},
       number = {2},
        pages = {1190-1201},
          doi = {10.1086/497119},
archivePrefix = {arXiv},
       eprint = {astro-ph/0509195},
 primaryClass = {astro-ph},
       adsurl = {https://ui.adsabs.harvard.edu/abs/2005ApJ...634.1190H}
}

@ARTICLE{2019ApJ...885...85M,
       author = {{Muthukrishna}, Daniel and {Parkinson}, David and {Tucker}, Brad E.},
        title = "{DASH: Deep Learning for the Automated Spectral Classification of Supernovae and Their Hosts}",
      journal = {\apj},
         year = 2019,
        month = nov,
       volume = {885},
       number = {1},
          eid = {85},
        pages = {85},
          doi = {10.3847/1538-4357/ab48f4},
archivePrefix = {arXiv},
       eprint = {1903.02557},
 primaryClass = {astro-ph.IM},
       adsurl = {https://ui.adsabs.harvard.edu/abs/2019ApJ...885...85M}
}

@ARTICLE{2021AJ....161..242F,
       author = {{F{\"o}rster}, F. and {Cabrera-Vives}, G. and {Castillo-Navarrete}, E. and {Est{\'e}vez}, P.~A. and {S{\'a}nchez-S{\'a}ez}, P. and {Arredondo}, J. and {Bauer}, F.~E. and {Carrasco-Davis}, R. and {Catelan}, M. and {Elorrieta}, F. and {Eyheramendy}, S. and {Huijse}, P. and {Pignata}, G. and {Reyes}, E. and {Reyes}, I. and {Rodr{\'\i}guez-Mancini}, D. and {Ruz-Mieres}, D. and {Valenzuela}, C. and {{\'A}lvarez-Maldonado}, I. and {Astorga}, N. and {Borissova}, J. and {Clocchiatti}, A. and {De Cicco}, D. and {Donoso-Oliva}, C. and {Hern{\'a}ndez-Garc{\'\i}a}, L. and {Graham}, M.~J. and {Jord{\'a}n}, A. and {Kurtev}, R. and {Mahabal}, A. and {Maureira}, J.~C. and {Mu{\~n}oz-Arancibia}, A. and {Molina-Ferreiro}, R. and {Moya}, A. and {Palma}, W. and {P{\'e}rez-Carrasco}, M. and {Protopapas}, P. and {Romero}, M. and {Sabatini-Gacitua}, L. and {S{\'a}nchez}, A. and {San Mart{\'\i}n}, J. and {Sep{\'u}lveda-Cobo}, C. and {Vera}, E. and {Vergara}, J.~R.},
        title = "{The Automatic Learning for the Rapid Classification of Events (ALeRCE) Alert Broker}",
      journal = {\aj},
         year = 2021,
        month = may,
       volume = {161},
       number = {5},
          eid = {242},
        pages = {242},
          doi = {10.3847/1538-3881/abe9bc},
archivePrefix = {arXiv},
       eprint = {2008.03303},
 primaryClass = {astro-ph.IM},
       adsurl = {https://ui.adsabs.harvard.edu/abs/2021AJ....161..242F}
}

@ARTICLE{2024A&A...692A.208F,
       author = {{Fraga}, B.~M.~O. and {Bom}, C.~R. and {Santos}, A. and {Russeil}, E. and {Leoni}, M. and {Peloton}, J. and {Ishida}, E.~E.~O. and {M{\"o}ller}, A. and {Blondin}, S.},
        title = "{Transient classifiers for Fink: Benchmarks for LSST}",
      journal = {\aap},
         year = 2024,
        month = dec,
       volume = {692},
          eid = {A208},
        pages = {A208},
          doi = {10.1051/0004-6361/202450370},
archivePrefix = {arXiv},
       eprint = {2404.08798},
 primaryClass = {astro-ph.IM},
       adsurl = {https://ui.adsabs.harvard.edu/abs/2024A&A...692A.208F}
}

@ARTICLE{2018ApJS..236....9N,
       author = {{Narayan}, Gautham and {Zaidi}, Tayeb and {Soraisam}, Monika D. and {Wang}, Zhe and {Lochner}, Michelle and {Matheson}, Thomas and {Saha}, Abhijit and {Yang}, Shuo and {Zhao}, Zhenge and {Kececioglu}, John and {Scheidegger}, Carlos and {Snodgrass}, Richard T. and {Axelrod}, Tim and {Jenness}, Tim and {Maier}, Robert S. and {Ridgway}, Stephen T. and {Seaman}, Robert L. and {Evans}, Eric Michael and {Singh}, Navdeep and {Taylor}, Clark and {Toeniskoetter}, Jackson and {Welch}, Eric and {Zhu}, Songzhe and {ANTARES Collaboration}},
        title = "{Machine-learning-based Brokers for Real-time Classification of the LSST Alert Stream}",
      journal = {\apjs},
         year = 2018,
        month = may,
       volume = {236},
       number = {1},
          eid = {9},
        pages = {9},
          doi = {10.3847/1538-4365/aab781},
archivePrefix = {arXiv},
       eprint = {1801.07323},
 primaryClass = {astro-ph.IM},
       adsurl = {https://ui.adsabs.harvard.edu/abs/2018ApJS..236....9N}
}

@ARTICLE{2014ApJS..210....9B,
       author = {{Bilicki}, Maciej and {Jarrett}, Thomas H. and {Peacock}, John A. and {Cluver}, Michelle E. and {Steward}, Louise},
        title = "{Two Micron All Sky Survey Photometric Redshift Catalog: A Comprehensive Three-dimensional Census of the Whole Sky}",
      journal = {\apjs},
         year = 2014,
        month = jan,
       volume = {210},
       number = {1},
          eid = {9},
        pages = {9},
          doi = {10.1088/0067-0049/210/1/9},
archivePrefix = {arXiv},
       eprint = {1311.5246},
 primaryClass = {astro-ph.CO},
       adsurl = {https://ui.adsabs.harvard.edu/abs/2014ApJS..210....9B}
}

@ARTICLE{2019PASP..131f8003B,
       author = {{Bellm}, Eric C. and {Kulkarni}, Shrinivas R. and {Barlow}, Tom and {Feindt}, Ulrich and {Graham}, Matthew J. and {Goobar}, Ariel and {Kupfer}, Thomas and {Ngeow}, Chow-Choong and {Nugent}, Peter and {Ofek}, Eran and {Prince}, Thomas A. and {Riddle}, Reed and {Walters}, Richard and {Ye}, Quan-Zhi},
        title = "{The Zwicky Transient Facility: Surveys and Scheduler}",
      journal = {\pasp},
         year = 2019,
        month = jun,
       volume = {131},
       number = {1000},
        pages = {068003},
          doi = {10.1088/1538-3873/ab0c2a},
archivePrefix = {arXiv},
       eprint = {1905.02209},
 primaryClass = {astro-ph.IM},
       adsurl = {https://ui.adsabs.harvard.edu/abs/2019PASP..131f8003B}
}

@ARTICLE{2019MNRAS.483.5459R,
       author = {{Rodr{\'\i}guez}, {\'O}. and {Pignata}, G. and {Hamuy}, M. and {Clocchiatti}, A. and {Phillips}, M.~M. and {Krisciunas}, K. and {Morrell}, N.~I. and {Folatelli}, G. and {Roth}, M. and {Castell{\'o}n}, S. and {Jang}, I.~S. and {Apostolovski}, Y. and {L{\'o}pez}, P. and {Marchi}, S. and {Ram{\'\i}rez}, R. and {S{\'a}nchez}, P.},
        title = "{Type II supernovae as distance indicators at near-IR wavelengths}",
      journal = {\mnras},
         year = 2019,
        month = mar,
       volume = {483},
       number = {4},
        pages = {5459-5479},
          doi = {10.1093/mnras/sty3396},
archivePrefix = {arXiv},
       eprint = {1812.04982},
 primaryClass = {astro-ph.CO},
       adsurl = {https://ui.adsabs.harvard.edu/abs/2019MNRAS.483.5459R}
}

@ARTICLE{2012MNRAS.426.1465P,
       author = {{Poznanski}, Dovi and {Prochaska}, J. Xavier and {Bloom}, Joshua S.},
        title = "{An empirical relation between sodium absorption and dust extinction}",
      journal = {\mnras},
         year = 2012,
        month = oct,
       volume = {426},
       number = {2},
        pages = {1465-1474},
          doi = {10.1111/j.1365-2966.2012.21796.x},
archivePrefix = {arXiv},
       eprint = {1206.6107},
 primaryClass = {astro-ph.IM},
       adsurl = {https://ui.adsabs.harvard.edu/abs/2012MNRAS.426.1465P}
}

@ARTICLE{2013ApJ...779...38P,
       author = {{Phillips}, M.~M. and {Simon}, Joshua D. and {Morrell}, Nidia and {Burns}, Christopher R. and {Cox}, Nick L.~J. and {Foley}, Ryan J. and {Karakas}, Amanda I. and {Patat}, F. and {Sternberg}, A. and {Williams}, R.~E. and {Gal-Yam}, A. and {Hsiao}, E.~Y. and {Leonard}, D.~C. and {Persson}, Sven E. and {Stritzinger}, Maximilian and {Thompson}, I.~B. and {Campillay}, Abdo and {Contreras}, Carlos and {Folatelli}, Gast{\'o}n and {Freedman}, Wendy L. and {Hamuy}, Mario and {Roth}, Miguel and {Shields}, Gregory A. and {Suntzeff}, Nicholas B. and {Chomiuk}, Laura and {Ivans}, Inese I. and {Madore}, Barry F. and {Penprase}, B.~E. and {Perley}, Daniel and {Pignata}, G. and {Preston}, G. and {Soderberg}, Alicia M.},
        title = "{On the Source of the Dust Extinction in Type Ia Supernovae and the Discovery of Anomalously Strong Na I Absorption}",
      journal = {\apj},
         year = 2013,
        month = dec,
       volume = {779},
       number = {1},
          eid = {38},
        pages = {38},
          doi = {10.1088/0004-637X/779/1/38},
archivePrefix = {arXiv},
       eprint = {1311.0147},
 primaryClass = {astro-ph.CO},
       adsurl = {https://ui.adsabs.harvard.edu/abs/2013ApJ...779...38P}
}

@ARTICLE{2019A&A...621A..30S,
       author = {{Sollerman}, J. and {Taddia}, F. and {Arcavi}, I. and {Fremling}, C. and {Fransson}, C. and {Burke}, J. and {Cenko}, S.~B. and {Andersen}, O. and {Andreoni}, I. and {Barbarino}, C. and {Blagorodova}, N. and {Brink}, T.~G. and {Filippenko}, A.~V. and {Gal-Yam}, A. and {Hiramatsu}, D. and {Hosseinzadeh}, G. and {Howell}, D.~A. and {de Jaeger}, T. and {Lunnan}, R. and {McCully}, C. and {Perley}, D.~A. and {Tartaglia}, L. and {Terreran}, G. and {Valenti}, S. and {Wang}, X.},
        title = "{Late-time observations of the extraordinary Type II supernova iPTF14hls}",
      journal = {\aap},
         year = 2019,
        month = jan,
       volume = {621},
          eid = {A30},
        pages = {A30},
          doi = {10.1051/0004-6361/201833689},
archivePrefix = {arXiv},
       eprint = {1806.10001},
 primaryClass = {astro-ph.HE},
       adsurl = {https://ui.adsabs.harvard.edu/abs/2019A&A...621A..30S}
}

@ARTICLE{2017Natur.551..210A,
       author = {{Arcavi}, Iair and {Howell}, D. Andrew and {Kasen}, Daniel and {Bildsten}, Lars and {Hosseinzadeh}, Griffin and {McCully}, Curtis and {Wong}, Zheng Chuen and {Katz}, Sarah Rebekah and {Gal-Yam}, Avishay and {Sollerman}, Jesper and {Taddia}, Francesco and {Leloudas}, Giorgos and {Fremling}, Christoffer and {Nugent}, Peter E. and {Horesh}, Assaf and {Mooley}, Kunal and {Rumsey}, Clare and {Cenko}, S. Bradley and {Graham}, Melissa L. and {Perley}, Daniel A. and {Nakar}, Ehud and {Shaviv}, Nir J. and {Bromberg}, Omer and {Shen}, Ken J. and {Ofek}, Eran O. and {Cao}, Yi and {Wang}, Xiaofeng and {Huang}, Fang and {Rui}, Liming and {Zhang}, Tianmeng and {Li}, Wenxiong and {Li}, Zhitong and {Zhang}, Jujia and {Valenti}, Stefano and {Guevel}, David and {Shappee}, Benjamin and {Kochanek}, Christopher S. and {Holoien}, Thomas W. -S. and {Filippenko}, Alexei V. and {Fender}, Rob and {Nyholm}, Anders and {Yaron}, Ofer and {Kasliwal}, Mansi M. and {Sullivan}, Mark and {Blagorodnova}, Nadja and {Walters}, Richard S. and {Lunnan}, Ragnhild and {Khazov}, Danny and {Andreoni}, Igor and {Laher}, Russ R. and {Konidaris}, Nick and {Wozniak}, Przemek and {Bue}, Brian},
        title = "{Energetic eruptions leading to a peculiar hydrogen-rich explosion of a massive star}",
      journal = {\nat},
         year = 2017,
        month = nov,
       volume = {551},
       number = {7679},
        pages = {210-213},
          doi = {10.1038/nature24030},
archivePrefix = {arXiv},
       eprint = {1711.02671},
 primaryClass = {astro-ph.HE},
       adsurl = {https://ui.adsabs.harvard.edu/abs/2017Natur.551..210A}
}

@ARTICLE{2014ApJ...797..118F,
       author = {{Fransson}, Claes and {Ergon}, Mattias and {Challis}, Peter J. and {Chevalier}, Roger A. and {France}, Kevin and {Kirshner}, Robert P. and {Marion}, G.~H. and {Milisavljevic}, Dan and {Smith}, Nathan and {Bufano}, Filomena and {Friedman}, Andrew S. and {Kangas}, Tuomas and {Larsson}, Josefin and {Mattila}, Seppo and {Benetti}, Stefano and {Chornock}, Ryan and {Czekala}, Ian and {Soderberg}, Alicia and {Sollerman}, Jesper},
        title = "{High-density Circumstellar Interaction in the Luminous Type IIn SN 2010jl: The First 1100 Days}",
      journal = {\apj},
         year = 2014,
        month = dec,
       volume = {797},
       number = {2},
          eid = {118},
        pages = {118},
          doi = {10.1088/0004-637X/797/2/118},
archivePrefix = {arXiv},
       eprint = {1312.6617},
 primaryClass = {astro-ph.HE},
       adsurl = {https://ui.adsabs.harvard.edu/abs/2014ApJ...797..118F}
}

@ARTICLE{2016MNRAS.455.2918H,
       author = {{Holoien}, T.~W. -S. and {Kochanek}, C.~S. and {Prieto}, J.~L. and {Stanek}, K.~Z. and {Dong}, Subo and {Shappee}, B.~J. and {Grupe}, D. and {Brown}, J.~S. and {Basu}, U. and {Beacom}, J.~F. and {Bersier}, D. and {Brimacombe}, J. and {Danilet}, A.~B. and {Falco}, E. and {Guo}, Z. and {Jose}, J. and {Herczeg}, G.~J. and {Long}, F. and {Pojmanski}, G. and {Simonian}, G.~V. and {Szczygie{\l}}, D.~M. and {Thompson}, T.~A. and {Thorstensen}, J.~R. and {Wagner}, R.~M. and {Wo{\'z}niak}, P.~R.},
        title = "{Six months of multiwavelength follow-up of the tidal disruption candidate ASASSN-14li and implied TDE rates from ASAS-SN}",
      journal = {\mnras},
         year = 2016,
        month = jan,
       volume = {455},
       number = {3},
        pages = {2918-2935},
          doi = {10.1093/mnras/stv2486},
archivePrefix = {arXiv},
       eprint = {1507.01598},
 primaryClass = {astro-ph.HE},
       adsurl = {https://ui.adsabs.harvard.edu/abs/2016MNRAS.455.2918H}
}

@ARTICLE{2017A&A...605A...6N,
       author = {{Nyholm}, A. and {Sollerman}, J. and {Taddia}, F. and {Fremling}, C. and {Moriya}, T.~J. and {Ofek}, E.~O. and {Gal-Yam}, A. and {De Cia}, A. and {Roy}, R. and {Kasliwal}, M.~M. and {Cao}, Y. and {Nugent}, P.~E. and {Masci}, F.~J.},
        title = "{The bumpy light curve of Type IIn supernova iPTF13z over 3 years}",
      journal = {\aap},
         year = 2017,
        month = aug,
       volume = {605},
          eid = {A6},
        pages = {A6},
          doi = {10.1051/0004-6361/201629906},
archivePrefix = {arXiv},
       eprint = {1703.09679},
 primaryClass = {astro-ph.SR},
       adsurl = {https://ui.adsabs.harvard.edu/abs/2017A&A...605A...6N}
}

@ARTICLE{2011MNRAS.415..773S,
       author = {{Smith}, Nathan and {Li}, Weidong and {Silverman}, Jeffrey M. and {Ganeshalingam}, Mohan and {Filippenko}, Alexei V.},
        title = "{Luminous blue variable eruptions and related transients: diversity of progenitors and outburst properties}",
      journal = {\mnras},
         year = 2011,
        month = jul,
       volume = {415},
       number = {1},
        pages = {773-810},
          doi = {10.1111/j.1365-2966.2011.18763.x},
archivePrefix = {arXiv},
       eprint = {1010.3718},
 primaryClass = {astro-ph.SR},
       adsurl = {https://ui.adsabs.harvard.edu/abs/2011MNRAS.415..773S}
}

@ARTICLE{2018ApJ...863..105W,
       author = {{Woosley}, S.~E.},
        title = "{Models for the Unusual Supernova iPTF14hls}",
      journal = {\apj},
         year = 2018,
        month = aug,
       volume = {863},
       number = {1},
          eid = {105},
        pages = {105},
          doi = {10.3847/1538-4357/aad044},
archivePrefix = {arXiv},
       eprint = {1801.08666},
 primaryClass = {astro-ph.HE},
       adsurl = {https://ui.adsabs.harvard.edu/abs/2018ApJ...863..105W}
}

@ARTICLE{1981PASP...93....5B,
       author = {{Baldwin}, J.~A. and {Phillips}, M.~M. and {Terlevich}, R.},
        title = "{Classification parameters for the emission-line spectra of extragalactic objects.}",
      journal = {\pasp},
         year = 1981,
        month = feb,
       volume = {93},
        pages = {5-19},
          doi = {10.1086/130766},
       adsurl = {https://ui.adsabs.harvard.edu/abs/1981PASP...93....5B}
}

@software{2021zndo....598352N,
       author = {{Newville}, Matt and {Otten}, Renee and {Nelson}, Andrew and {Stensitzki}, Till and {Ingargiola}, Antonino and {Allan}, Dan and {Fox}, Austin and {Carter}, Faustin and {Micha{\l}} and {Osborn}, Ray and {Pustakhod}, Dima and {Weigand}, Sebastian and {lneuhaus} and {Aristov}, Andrey and {Glenn} and {Mark} and {mgunyho} and {Deil}, Christoph and {Hansen}, Allan L.~R. and {Pasquevich}, Gustavo and {Foks}, Leon and {Zobrist}, Nicholas and {Frost}, Oliver and {Stuermer} and {Jaskula}, Jean-Christophe and {Caldwell}, Shane and {Eendebak}, Pieter and {Pompili}, Matteo and {Hedegaard Nielsen}, Jens and {Persaud}, Arun},
        title = "{lmfit/lmfit-py: 1.3.2}",
         year = 2024,
        month = jul,
          eid = {10.5281/zenodo.598352},
          doi = {10.5281/zenodo.598352},
      version = {1.3.2},
    publisher = {Zenodo},
       adsurl = {https://ui.adsabs.harvard.edu/abs/2021zndo....598352N}
}

@INCOLLECTION{2017hsn..book..403S,
       author = {{Smith}, Nathan},
        title = "{Interacting Supernovae: Types IIn and Ibn}",
    booktitle = {Handbook of Supernovae},
         year = 2017,
       editor = {{Alsabti}, Athem W. and {Murdin}, Paul},
        pages = {403},
          doi = {10.1007/978-3-319-21846-5_38},
       adsurl = {https://ui.adsabs.harvard.edu/abs/2017hsn..book..403S}
}

@ARTICLE{2001ApJ...556..121K,
       author = {{Kewley}, L.~J. and {Dopita}, M.~A. and {Sutherland}, R.~S. and {Heisler}, C.~A. and {Trevena}, J.},
        title = "{Theoretical Modeling of Starburst Galaxies}",
      journal = {\apj},
         year = 2001,
        month = jul,
       volume = {556},
       number = {1},
        pages = {121-140},
          doi = {10.1086/321545},
archivePrefix = {arXiv},
       eprint = {astro-ph/0106324},
 primaryClass = {astro-ph},
       adsurl = {https://ui.adsabs.harvard.edu/abs/2001ApJ...556..121K}
}

@ARTICLE{2007MNRAS.382.1415S,
       author = {{Schawinski}, Kevin and {Thomas}, Daniel and {Sarzi}, Marc and {Maraston}, Claudia and {Kaviraj}, Sugata and {Joo}, Seok-Joo and {Yi}, Sukyoung K. and {Silk}, Joseph},
        title = "{Observational evidence for AGN feedback in early-type galaxies}",
      journal = {\mnras},
         year = 2007,
        month = dec,
       volume = {382},
       number = {4},
        pages = {1415-1431},
          doi = {10.1111/j.1365-2966.2007.12487.x},
archivePrefix = {arXiv},
       eprint = {0709.3015},
 primaryClass = {astro-ph},
       adsurl = {https://ui.adsabs.harvard.edu/abs/2007MNRAS.382.1415S}
}

@ARTICLE{2003MNRAS.346.1055K,
       author = {{Kauffmann}, Guinevere and {Heckman}, Timothy M. and {Tremonti}, Christy and {Brinchmann}, Jarle and {Charlot}, St{\'e}phane and {White}, Simon D.~M. and {Ridgway}, Susan E. and {Brinkmann}, Jon and {Fukugita}, Masataka and {Hall}, Patrick B. and {Ivezi{\'c}}, {\v{Z}}eljko and {Richards}, Gordon T. and {Schneider}, Donald P.},
        title = "{The host galaxies of active galactic nuclei}",
      journal = {\mnras},
         year = 2003,
        month = dec,
       volume = {346},
       number = {4},
        pages = {1055-1077},
          doi = {10.1111/j.1365-2966.2003.07154.x},
archivePrefix = {arXiv},
       eprint = {astro-ph/0304239},
 primaryClass = {astro-ph},
       adsurl = {https://ui.adsabs.harvard.edu/abs/2003MNRAS.346.1055K}
}

@ARTICLE{2022A&A...659A..34C,
       author = {{Charalampopoulos}, P. and {Leloudas}, G. and {Malesani}, D.~B. and {Wevers}, T. and {Arcavi}, I. and {Nicholl}, M. and {Pursiainen}, M. and {Lawrence}, A. and {Anderson}, J.~P. and {Benetti}, S. and {Cannizzaro}, G. and {Chen}, T. -W. and {Galbany}, L. and {Gromadzki}, M. and {Guti{\'e}rrez}, C.~P. and {Inserra}, C. and {Jonker}, P.~G. and {M{\"u}ller-Bravo}, T.~E. and {Onori}, F. and {Short}, P. and {Sollerman}, J. and {Young}, D.~R.},
        title = "{A detailed spectroscopic study of tidal disruption events}",
      journal = {\aap},
         year = 2022,
        month = mar,
       volume = {659},
          eid = {A34},
        pages = {A34},
          doi = {10.1051/0004-6361/202142122},
archivePrefix = {arXiv},
       eprint = {2109.00016},
 primaryClass = {astro-ph.HE},
       adsurl = {https://ui.adsabs.harvard.edu/abs/2022A&A...659A..34C}
}

@ARTICLE{2024TNSCR4932....1G,
       author = {{Gagliano}, A.},
        title = "{Transient Classification Report for 2024-12-16}",
      journal = {Transient Name Server Classification Report},
         year = 2024,
        month = dec,
       volume = {2024-4932},
        pages = {1},
       adsurl = {https://ui.adsabs.harvard.edu/abs/2024TNSCR4932....1G}
}

@ARTICLE{2009ApJ...695.1334S,
       author = {{Smith}, Nathan and {Silverman}, Jeffrey M. and {Chornock}, Ryan and {Filippenko}, Alexei V. and {Wang}, Xiaofeng and {Li}, Weidong and {Ganeshalingam}, Mohan and {Foley}, Ryan J. and {Rex}, Jacob and {Steele}, Thea N.},
        title = "{Coronal Lines and Dust Formation in SN 2005ip: Not the Brightest, but the Hottest Type IIn Supernova}",
      journal = {\apj},
         year = 2009,
        month = apr,
       volume = {695},
       number = {2},
        pages = {1334-1350},
          doi = {10.1088/0004-637X/695/2/1334},
archivePrefix = {arXiv},
       eprint = {0809.5079},
 primaryClass = {astro-ph},
       adsurl = {https://ui.adsabs.harvard.edu/abs/2009ApJ...695.1334S}
}

@ARTICLE{2009ApJ...707.1560I,
       author = {{Izotov}, Yuri I. and {Thuan}, Trinh X.},
        title = "{A Type IIn Supernova with Coronal Lines in the Low-Metallicity Compact Dwarf Galaxy J1320+2155}",
      journal = {\apj},
         year = 2009,
        month = dec,
       volume = {707},
       number = {2},
        pages = {1560-1565},
          doi = {10.1088/0004-637X/707/2/1560},
archivePrefix = {arXiv},
       eprint = {0910.5828},
 primaryClass = {astro-ph.CO},
       adsurl = {https://ui.adsabs.harvard.edu/abs/2009ApJ...707.1560I}
}

@ARTICLE{2002ApJ...572..350F,
       author = {{Fransson}, Claes and {Chevalier}, Roger A. and {Filippenko}, Alexei V. and {Leibundgut}, Bruno and {Barth}, Aaron J. and {Fesen}, Robert A. and {Kirshner}, Robert P. and {Leonard}, Douglas C. and {Li}, Weidong and {Lundqvist}, Peter and {Sollerman}, Jesper and {Van Dyk}, Schuyler D.},
        title = "{Optical and Ultraviolet Spectroscopy of SN 1995N: Evidence for Strong Circumstellar Interaction}",
      journal = {\apj},
         year = 2002,
        month = jun,
       volume = {572},
       number = {1},
        pages = {350-370},
          doi = {10.1086/340295},
archivePrefix = {arXiv},
       eprint = {astro-ph/0108149},
 primaryClass = {astro-ph},
       adsurl = {https://ui.adsabs.harvard.edu/abs/2002ApJ...572..350F}
}

@ARTICLE{2019ApJ...883...31F,
       author = {{Frederick}, Sara and {Gezari}, Suvi and {Graham}, Matthew J. and {Cenko}, S. Bradley and {van Velzen}, Sjoert and {Stern}, Daniel and {Blagorodnova}, Nadejda and {Kulkarni}, Shrinivas R. and {Yan}, Lin and {De}, Kishalay and {Fremling}, U. Christoffer and {Hung}, Tiara and {Kara}, Erin and {Shupe}, David L. and {Ward}, Charlotte and {Bellm}, Eric C. and {Dekany}, Richard and {Duev}, Dmitry A. and {Feindt}, Ulrich and {Giomi}, Matteo and {Kupfer}, Thomas and {Laher}, Russ R. and {Masci}, Frank J. and {Miller}, Adam A. and {Neill}, James D. and {Ngeow}, Chow-Choong and {Patterson}, Maria T. and {Porter}, Michael and {Rusholme}, Ben and {Sollerman}, Jesper and {Walters}, Richard},
        title = "{A New Class of Changing-look LINERs}",
      journal = {\apj},
         year = 2019,
        month = sep,
       volume = {883},
       number = {1},
          eid = {31},
        pages = {31},
          doi = {10.3847/1538-4357/ab3a38},
archivePrefix = {arXiv},
       eprint = {1904.10973},
 primaryClass = {astro-ph.HE},
       adsurl = {https://ui.adsabs.harvard.edu/abs/2019ApJ...883...31F}
}

@ARTICLE{2025MNRAS.537.2024W,
       author = {{Wiseman}, P. and {Williams}, R.~D. and {Arcavi}, I. and {Galbany}, L. and {Graham}, M.~J. and {H{\"o}nig}, S. and {Newsome}, M. and {Subrayan}, B. and {Sullivan}, M. and {Wang}, Y. and {Ili{\'c}}, D. and {Nicholl}, M. and {Oates}, S. and {Petrushevska}, T. and {Smith}, K.~W.},
        title = "{A systematically selected sample of luminous, long-duration, ambiguous nuclear transients}",
      journal = {\mnras},
         year = 2025,
        month = feb,
       volume = {537},
       number = {2},
        pages = {2024-2045},
          doi = {10.1093/mnras/staf116},
archivePrefix = {arXiv},
       eprint = {2406.11552},
 primaryClass = {astro-ph.HE},
       adsurl = {https://ui.adsabs.harvard.edu/abs/2025MNRAS.537.2024W}
}

@ARTICLE{2019AJ....157..168D,
       author = {{Dey}, Arjun and {Schlegel}, David J. and {Lang}, Dustin and {Blum}, Robert and {Burleigh}, Kaylan and {Fan}, Xiaohui and {Findlay}, Joseph R. and {Finkbeiner}, Doug and {Herrera}, David and {Juneau}, St{\'e}phanie and {Landriau}, Martin and {Levi}, Michael and {McGreer}, Ian and {Meisner}, Aaron and {Myers}, Adam D. and {Moustakas}, John and {Nugent}, Peter and {Patej}, Anna and {Schlafly}, Edward F. and {Walker}, Alistair R. and {Valdes}, Francisco and {Weaver}, Benjamin A. and {Y{\`e}che}, Christophe and {Zou}, Hu and {Zhou}, Xu and {Abareshi}, Behzad and {Abbott}, T.~M.~C. and {Abolfathi}, Bela and {Aguilera}, C. and {Alam}, Shadab and {Allen}, Lori and {Alvarez}, A. and {Annis}, James and {Ansarinejad}, Behzad and {Aubert}, Marie and {Beechert}, Jacqueline and {Bell}, Eric F. and {BenZvi}, Segev Y. and {Beutler}, Florian and {Bielby}, Richard M. and {Bolton}, Adam S. and {Brice{\~n}o}, C{\'e}sar and {Buckley-Geer}, Elizabeth J. and {Butler}, Karen and {Calamida}, Annalisa and {Carlberg}, Raymond G. and {Carter}, Paul and {Casas}, Ricard and {Castander}, Francisco J. and {Choi}, Yumi and {Comparat}, Johan and {Cukanovaite}, Elena and {Delubac}, Timoth{\'e}e and {DeVries}, Kaitlin and {Dey}, Sharmila and {Dhungana}, Govinda and {Dickinson}, Mark and {Ding}, Zhejie and {Donaldson}, John B. and {Duan}, Yutong and {Duckworth}, Christopher J. and {Eftekharzadeh}, Sarah and {Eisenstein}, Daniel J. and {Etourneau}, Thomas and {Fagrelius}, Parker A. and {Farihi}, Jay and {Fitzpatrick}, Mike and {Font-Ribera}, Andreu and {Fulmer}, Leah and {G{\"a}nsicke}, Boris T. and {Gaztanaga}, Enrique and {George}, Koshy and {Gerdes}, David W. and {Gontcho}, Satya Gontcho A. and {Gorgoni}, Claudio and {Green}, Gregory and {Guy}, Julien and {Harmer}, Diane and {Hernandez}, M. and {Honscheid}, Klaus and {Huang}, Lijuan Wendy and {James}, David J. and {Jannuzi}, Buell T. and {Jiang}, Linhua and {Joyce}, Richard and {Karcher}, Armin and {Karkar}, Sonia and {Kehoe}, Robert and {Kneib}, Jean-Paul and {Kueter-Young}, Andrea and {Lan}, Ting-Wen and {Lauer}, Tod R. and {Le Guillou}, Laurent and {Le Van Suu}, Auguste and {Lee}, Jae Hyeon and {Lesser}, Michael and {Perreault Levasseur}, Laurence and {Li}, Ting S. and {Mann}, Justin L. and {Marshall}, Robert and {Mart{\'\i}nez-V{\'a}zquez}, C.~E. and {Martini}, Paul and {du Mas des Bourboux}, H{\'e}lion and {McManus}, Sean and {Meier}, Tobias Gabriel and {M{\'e}nard}, Brice and {Metcalfe}, Nigel and {Mu{\~n}oz-Guti{\'e}rrez}, Andrea and {Najita}, Joan and {Napier}, Kevin and {Narayan}, Gautham and {Newman}, Jeffrey A. and {Nie}, Jundan and {Nord}, Brian and {Norman}, Dara J. and {Olsen}, Knut A.~G. and {Paat}, Anthony and {Palanque-Delabrouille}, Nathalie and {Peng}, Xiyan and {Poppett}, Claire L. and {Poremba}, Megan R. and {Prakash}, Abhishek and {Rabinowitz}, David and {Raichoor}, Anand and {Rezaie}, Mehdi and {Robertson}, A.~N. and {Roe}, Natalie A. and {Ross}, Ashley J. and {Ross}, Nicholas P. and {Rudnick}, Gregory and {Safonova}, Sasha and {Saha}, Abhijit and {S{\'a}nchez}, F. Javier and {Savary}, Elodie and {Schweiker}, Heidi and {Scott}, Adam and {Seo}, Hee-Jong and {Shan}, Huanyuan and {Silva}, David R. and {Slepian}, Zachary and {Soto}, Christian and {Sprayberry}, David and {Staten}, Ryan and {Stillman}, Coley M. and {Stupak}, Robert J. and {Summers}, David L. and {Sien Tie}, Suk and {Tirado}, H. and {Vargas-Maga{\~n}a}, Mariana and {Vivas}, A. Katherina and {Wechsler}, Risa H. and {Williams}, Doug and {Yang}, Jinyi and {Yang}, Qian and {Yapici}, Tolga and {Zaritsky}, Dennis and {Zenteno}, A. and {Zhang}, Kai and {Zhang}, Tianmeng and {Zhou}, Rongpu and {Zhou}, Zhimin},
        title = "{Overview of the DESI Legacy Imaging Surveys}",
      journal = {\aj},
         year = 2019,
        month = may,
       volume = {157},
       number = {5},
          eid = {168},
        pages = {168},
          doi = {10.3847/1538-3881/ab089d},
archivePrefix = {arXiv},
       eprint = {1804.08657},
 primaryClass = {astro-ph.IM},
       adsurl = {https://ui.adsabs.harvard.edu/abs/2019AJ....157..168D}
}

@ARTICLE{2025arXiv250314745D,
       author = {{DESI Collaboration} and {Abdul-Karim}, M. and {Adame}, A.~G. and {Aguado}, D. and {Aguilar}, J. and {Ahlen}, S. and {Alam}, S. and {Aldering}, G. and {Alexander}, D.~M. and {Alfarsy}, R. and {Allen}, L. and {Allende Prieto}, C. and {Alves}, O. and {Anand}, A. and {Andrade}, U. and {Armengaud}, E. and {Avila}, S. and {Aviles}, A. and {Awan}, H. and {Bailey}, S. and {Baleato Lizancos}, A. and {Ballester}, O. and {Bault}, A. and {Bautista}, J. and {BenZvi}, S. and {Beraldo e Silva}, L. and {Bermejo-Climent}, J.~R. and {Beutler}, F. and {Bianchi}, D. and {Blake}, C. and {Blum}, R. and {Bolton}, A.~S. and {Bonici}, M. and {Brieden}, S. and {Brodzeller}, A. and {Brooks}, D. and {Buckley-Geer}, E. and {Burtin}, E. and {Canning}, R. and {Carnero Rosell}, A. and {Carr}, A. and {Carrilho}, P. and {Casas}, L. and {Castander}, F.~J. and {Cereskaite}, R. and {Cervantes-Cota}, J.~L. and {Chaussidon}, E. and {Chaves-Montero}, J. and {Chen}, S. and {Chen}, X. and {Claybaugh}, T. and {Cole}, S. and {Cooper}, A.~P. and {Cousinou}, M. -C. and {Cuceu}, A. and {Davis}, T.~M. and {Dawson}, K.~S. and {de Belsunce}, R. and {de la Cruz}, R. and {de la Macorra}, A. and {de Mattia}, A. and {Deiosso}, N. and {Della Costa}, J. and {Demina}, R. and {Demirbozan}, U. and {DeRose}, J. and {Dey}, A. and {Dey}, B. and {Ding}, J. and {Ding}, Z. and {Doel}, P. and {Douglass}, K. and {Dowicz}, M. and {Ebina}, H. and {Edelstein}, J. and {Eisenstein}, D.~J. and {Elbers}, W. and {Emas}, N. and {Escoffier}, S. and {Fagrelius}, P. and {Fan}, X. and {Fanning}, K. and {Fawcett}, V.~A. and {Fern\textbackslash'andez-Garc\textbackslash'ia}, E. and {Ferraro}, S. and {Findlay}, N. and {Font-Ribera}, A. and {Forero-Romero}, J.~E. and {Forero-S\textbackslash'anchez}, D. and {Frenk}, C.~S. and {G\textbackslash''ansicke}, B.~T. and {Galbany}, L. and {Garc\textbackslash'ia-Bellido}, J. and {Garcia-Quintero}, C. and {Garrison}, L.~H. and {Gazta\textbackslash\raisebox{-0.5ex}\textasciitildenaga}, E. and {Gil-Mar\textbackslash'in}, H. and {Gnedin}, O.~Y. and {Gontcho}, S. Gontcho A and {Gonzalez-Morales}, A.~X. and {Gonzalez-Perez}, V. and {Gordon}, C. and {Graur}, O. and {Green}, D. and {Gruen}, D. and {Gsponer}, R. and {Guandalin}, C. and {Gutierrez}, G. and {Guy}, J. and {Hahn}, C. and {Han}, J.~J. and {Han}, J. and {He}, S. and {Herrera-Alcantar}, H.~K. and {Honscheid}, K. and {Hou}, J. and {Howlett}, C. and {Huterer}, D. and {Ir\textbackslashv\{s\}i\textbackslashv\{c\}}, V. and {Ishak}, M. and {Jacques}, A. and {Jimenez}, J. and {Jing}, Y.~P. and {Joachimi}, B. and {Joudaki}, S. and {Joyce}, R. and {Jullo}, E. and {Juneau}, S. and {Kara\textbackslashc\{c\}ayl\{\textbackslashi\}}, N.~G. and {Karim}, T. and {Kehoe}, R. and {Kent}, S. and {Khederlarian}, A. and {Kirkby}, D. and {Kisner}, T. and {Kitaura}, F. -S. and {Kizhuprakkat}, N. and {Kong}, H. and {Koposov}, S.~E. and {Kremin}, A. and {Krolewski}, A. and {Lahav}, O. and {Lai}, Y. and {Lamman}, C. and {Lan}, T. -W. and {Landriau}, M. and {Lang}, D. and {Lange}, J.~U. and {Lasker}, J. and {Le Goff}, J.~M. and {Le Guillou}, L. and {Leauthaud}, A. and {Levi}, M.~E. and {Li}, S. and {Li}, T.~S. and {Lodha}, K. and {Lokken}, M. and {Luo}, Y. and {Magneville}, C. and {Manera}, M. and {Manser}, C.~J. and {Margala}, D. and {Martini}, P. and {Maus}, M. and {McCullough}, J. and {McDonald}, P. and {Medina}, G.~E. and {Medina-Varela}, L. and {Meisner}, A. and {Mena-Fern\textbackslash'andez}, J. and {Menegas}, A. and {Mezcua}, M. and {Miquel}, R. and {Montero-Camacho}, P. and {Moon}, J. and {Moustakas}, J. and {Mu\textbackslash\raisebox{-0.5ex}\textasciitildenoz-Guti\textbackslash'errez}, A. and {Mu\textbackslash\raisebox{-0.5ex}\textasciitildenoz-Santos}, D. and {Myers}, A.~D. and {Myles}, J. and {Nadathur}, S. and {Najita}, J. and {Napolitano}, L. and {Newman}, J.~A. and {Nikakhtar}, F. and {Nikutta}, R. and {Niz}, G. and {Noriega}, H.~E. and {Padmanabhan}, N. and {Paillas}, E. and {Palanque-Delabrouille}, N. and {Palmese}, A. and {Pan}, J. and {Pan}, Z. and {Parkinson}, D. and {Peacock}, J. and {Percival}, W.~J. and {P\textbackslash'erez-Fern\textbackslash'andez}, A. and {P\textbackslash'erez-R\textbackslash`afols}, I. and {Peterson}, P.},
        title = "{Data Release 1 of the Dark Energy Spectroscopic Instrument}",
      journal = {arXiv e-prints},
         year = 2025,
        month = mar,
          eid = {arXiv:2503.14745},
        pages = {arXiv:2503.14745},
          doi = {10.48550/arXiv.2503.14745},
archivePrefix = {arXiv},
       eprint = {2503.14745},
 primaryClass = {astro-ph.CO},
       adsurl = {https://ui.adsabs.harvard.edu/abs/2025arXiv250314745D}
}

@BOOK{2006agna.book.....O,
       author = {{Osterbrock}, Donald E. and {Ferland}, Gary J.},
        title = "{Astrophysics of gaseous nebulae and active galactic nuclei}",
         year = 2006,
       adsurl = {https://ui.adsabs.harvard.edu/abs/2006agna.book.....O}
}

@ARTICLE{2024arXiv241017322J,
       author = {{Jones}, D.~O. and {McGill}, P. and {Manning}, T.~A. and {Gagliano}, A. and {Wang}, B. and {Coulter}, D.~A. and {Foley}, R.~J. and {Narayan}, G. and {Villar}, V.~A. and {Braff}, L. and {Engel}, A.~W. and {Farias}, D. and {Lai}, Z. and {Loertscher}, K. and {Kutcka}, J. and {Thorp}, S. and {Vazquez}, J.},
        title = "{Blast: a Web Application for Characterizing the Host Galaxies of Astrophysical Transients}",
      journal = {arXiv e-prints},
         year = 2024,
        month = oct,
          eid = {arXiv:2410.17322},
        pages = {arXiv:2410.17322},
          doi = {10.48550/arXiv.2410.17322},
archivePrefix = {arXiv},
       eprint = {2410.17322},
 primaryClass = {astro-ph.HE},
       adsurl = {https://ui.adsabs.harvard.edu/abs/2024arXiv241017322J}
}

@ARTICLE{2021A&A...647A...1P,
       author = {{Predehl}, P. and {Andritschke}, R. and {Arefiev}, V. and {Babyshkin}, V. and {Batanov}, O. and {Becker}, W. and {B{\"o}hringer}, H. and {Bogomolov}, A. and {Boller}, T. and {Borm}, K. and {Bornemann}, W. and {Br{\"a}uninger}, H. and {Br{\"u}ggen}, M. and {Brunner}, H. and {Brusa}, M. and {Bulbul}, E. and {Buntov}, M. and {Burwitz}, V. and {Burkert}, W. and {Clerc}, N. and {Churazov}, E. and {Coutinho}, D. and {Dauser}, T. and {Dennerl}, K. and {Doroshenko}, V. and {Eder}, J. and {Emberger}, V. and {Eraerds}, T. and {Finoguenov}, A. and {Freyberg}, M. and {Friedrich}, P. and {Friedrich}, S. and {F{\"u}rmetz}, M. and {Georgakakis}, A. and {Gilfanov}, M. and {Granato}, S. and {Grossberger}, C. and {Gueguen}, A. and {Gureev}, P. and {Haberl}, F. and {H{\"a}lker}, O. and {Hartner}, G. and {Hasinger}, G. and {Huber}, H. and {Ji}, L. and {Kienlin}, A. v. and {Kink}, W. and {Korotkov}, F. and {Kreykenbohm}, I. and {Lamer}, G. and {Lomakin}, I. and {Lapshov}, I. and {Liu}, T. and {Maitra}, C. and {Meidinger}, N. and {Menz}, B. and {Merloni}, A. and {Mernik}, T. and {Mican}, B. and {Mohr}, J. and {M{\"u}ller}, S. and {Nandra}, K. and {Nazarov}, V. and {Pacaud}, F. and {Pavlinsky}, M. and {Perinati}, E. and {Pfeffermann}, E. and {Pietschner}, D. and {Ramos-Ceja}, M.~E. and {Rau}, A. and {Reiffers}, J. and {Reiprich}, T.~H. and {Robrade}, J. and {Salvato}, M. and {Sanders}, J. and {Santangelo}, A. and {Sasaki}, M. and {Scheuerle}, H. and {Schmid}, C. and {Schmitt}, J. and {Schwope}, A. and {Shirshakov}, A. and {Steinmetz}, M. and {Stewart}, I. and {Str{\"u}der}, L. and {Sunyaev}, R. and {Tenzer}, C. and {Tiedemann}, L. and {Tr{\"u}mper}, J. and {Voron}, V. and {Weber}, P. and {Wilms}, J. and {Yaroshenko}, V.},
        title = "{The eROSITA X-ray telescope on SRG}",
      journal = {\aap},
         year = 2021,
        month = mar,
       volume = {647},
          eid = {A1},
        pages = {A1},
          doi = {10.1051/0004-6361/202039313},
archivePrefix = {arXiv},
       eprint = {2010.03477},
 primaryClass = {astro-ph.HE},
       adsurl = {https://ui.adsabs.harvard.edu/abs/2021A&A...647A...1P}
}

@ARTICLE{2022ApJS..258...21W,
       author = {{Wang}, Yibo and {Jiang}, Ning and {Wang}, Tinggui and {Yan}, Lin and {Sheng}, Zhenfeng and {Dou}, Liming and {Ding}, Jiani and {Cai}, Zheng and {Sun}, Luming and {Yang}, Chenwei and {Shu}, Xinwen},
        title = "{Mid-infrared Outbursts in Nearby Galaxies (MIRONG). II. Optical Spectroscopic Follow-up}",
      journal = {\apjs},
         year = 2022,
        month = jan,
       volume = {258},
       number = {1},
          eid = {21},
        pages = {21},
          doi = {10.3847/1538-4365/ac33a6},
archivePrefix = {arXiv},
       eprint = {2111.12729},
 primaryClass = {astro-ph.GA},
       adsurl = {https://ui.adsabs.harvard.edu/abs/2022ApJS..258...21W}
}




\appendix

\section{Light curves}
\label{app:lcs}

\begin{ThreePartTable}
\begin{table}
\caption{ZTF and ATLAS photometry}              
\label{table:ZTFandATLASphot}      
\centering                          
\begin{tabular}{l c c c c c c}        
\hline\hline                
\noalign{\vskip 1mm}
MJD    &  Flux       & err\_Flux  & ABMag  & err\_ABMag & Filter & Flag \\    
{[days]} &  [$\mu$Jy]  & [$\mu$Jy]  & [mag.] & [mag.]     &        &        \\
\hline
58205.28 &      0.18 &       5.46 & 20.31 &   $\cdots$ & ZTFg & T \\
58214.25 &     -1.38 &      12.54 & 19.41 &   $\cdots$ & ZTFg & T \\
58216.25 &     -0.26 &       4.27 & 20.58 &   $\cdots$ & ZTFg & T \\
\hline 
\end{tabular}
\begin{tablenotes}
      \small
      \item The F flag in the last column means real detection and the T flag means upper limit. The full table is available in supplementary material.
    \end{tablenotes}
\end{table}
\end{ThreePartTable}
\begin{ThreePartTable}
\begin{table}
\caption{SEDM photometry}              
\label{table:SEDMphot}      
\centering                          
\begin{tabular}{l c c c c c c}        
\hline\hline                
\noalign{\vskip 1mm}
MJD    &  ABmag  &  err\_ABmag &  limiting\_mag &   Flux     &  err\_Flux & Filter \\
{[days]} &  [mag.] & [mag.]      & [mag.]         &  [$\mu$Jy] & [$\mu$Jy]  & \\
\hline
60007.18 & 19.00    &    0.32     &  18.59  &  91.27   &    26.55 &  sdssr\\
60012.32 & 18.98    &    0.07     &  20.15  &  93.25   &     6.31 &  sdssr\\
60012.35 & $\cdots$ &    $\cdots$ &  19.44  & $\cdots$ &    60.96 &  sdssg\\
60012.35 & $\cdots$ &    $\cdots$ &  18.10  & $\cdots$ &   209.65 &  sdssr\\
60071.33 & 18.66    &    0.15     &  19.05  & 124.36   &    17.36 &  sdssr\\
60071.36 & $\cdots$ &    $\cdots$ &  19.53  & $\cdots$ &    55.87 &  sdssg\\
60071.36 & 18.66    &    0.09     &  19.62  & 124.87   &    10.26 &  sdssr\\
60071.36 & 17.88    &    0.05     &  19.39  & 255.08   &    12.75 &  sdssi\\
60263.52 & 18.86    &    0.07     &  20.14  & 103.30   &     6.40 &  sdssr\\
60378.20 & 20.02    &    0.10     &  20.82  &  35.72   &     3.42 &  sdssg\\
60378.21 & 19.77    &    0.06     &  21.08  &  45.06   &     2.68 &  sdssr\\
60378.21 & 18.69    &    0.04     &  20.65  & 120.96   &     3.98 &  sdssi\\
60525.17 & 19.79    &    0.09     &  20.79  &  44.25   &     3.50 &  sdssr\\
60525.17 & 18.83    &    0.05     &  20.41  & 106.88   &     4.97 &  sdssi\\
60529.17 & 19.80    &    0.07     &  21.05  &  43.53   &     2.75 &  sdssr\\
60529.17 & 18.78    &    0.04     &  20.75  & 111.94   &     3.63 &  sdssi\\
60535.16 & $\cdots$ &    $\cdots$ &  20.69  & $\cdots$ &    19.23 &  sdssr\\
60535.16 & 18.77    &    0.04     &  20.65  & 112.63   &     4.00 &  sdssi\\
60730.54 & $\cdots$ &    $\cdots$ &  21.20  & $\cdots$ &    12.04 &  sdssr\\
60730.55 & 19.42    &    0.07     &  20.70  &  61.95   &     3.82 &  sdssi\\
\hline 
\end{tabular}
\end{table}
\end{ThreePartTable}
\begin{ThreePartTable}
\begin{table*}
\caption{WISE photometry}              
\label{table:WISEphot}      
\centering                          
\begin{tabular}{l c c c c c c c c c}        
\hline\hline                
\noalign{\vskip 1mm}
MJD    &  Flux(W1)   &  err\_Flux(W1) &  Flux(W2)   & err\_Flux(W2) &  W1\_Vega &  err\_W1\_Vega &  W2\_Vega &  err\_W2\_Vega & W1\_Vega$-$W2\_Vega\\
{[days]} &   [$\mu$Jy] & [$\mu$Jy]      &  [$\mu$Jy]  & [$\mu$Jy]     &  [mag.]   &  [mag.]        &  [mag.]   & [mag.]         & [mag.] \\
\hline
57320.00 &      31.10 &       15.00 &      49.40 &       25.00 & 17.09 & $\cdots$ & 15.90 & $\cdots$ & $\cdots$\\
57484.00 &      39.90 &       15.00 &      56.50 &       25.00 & 17.09 & $\cdots$ & 15.90 & $\cdots$ & $\cdots$\\
57687.00 &      58.30 &       15.00 &      77.10 &       25.00 & 16.81 & $\cdots$ & 15.87 & $\cdots$ & $\cdots$\\
57845.00 &      47.20 &       15.00 &      79.90 &       25.00 & 17.04 & $\cdots$ & 15.83 & $\cdots$ & $\cdots$\\
58051.00 &       0.40 &       15.00 &      45.00 &       25.00 & 17.09 & $\cdots$ & 15.90 & $\cdots$ & $\cdots$\\
58208.00 &     -10.90 &       15.00 &      42.50 &       25.00 & 17.09 & $\cdots$ & 15.90 & $\cdots$ & $\cdots$\\
58415.00 &      14.90 &       15.00 &      32.30 &       25.00 & 17.09 & $\cdots$ & 15.90 & $\cdots$ & $\cdots$\\
58573.00 &      -3.40 &       15.00 &      92.70 &       25.00 & 17.09 & $\cdots$ & 15.67 & $\cdots$ & $\cdots$\\
58782.00 &       2.80 &       15.00 &       3.70 &       25.00 & 17.09 & $\cdots$ & 15.90 & $\cdots$ & $\cdots$\\
58937.00 &      30.50 &       15.00 &       8.00 &       25.00 & 17.09 & $\cdots$ & 15.90 & $\cdots$ & $\cdots$\\
59147.00 &      -6.00 &       15.00 &     -18.80 &       25.00 & 17.09 & $\cdots$ & 15.90 & $\cdots$ & $\cdots$\\
59304.00 &     -11.50 &       15.00 &      37.30 &       25.00 & 17.09 & $\cdots$ & 15.90 & $\cdots$ & $\cdots$\\
59511.00 &      -8.70 &       15.00 &      25.60 &       25.00 & 17.09 & $\cdots$ & 15.90 & $\cdots$ & $\cdots$\\
59668.00 &     -16.40 &       15.00 &       7.10 &       25.00 & 17.09 & $\cdots$ & 15.90 & $\cdots$ & $\cdots$\\
59878.00 &    1732.90 &       15.00 &    1700.30 &       25.00 & 13.13 & 0.01     & 12.51 &   0.02   & 0.62\\
60032.00 &    2359.40 &       15.00 &    2821.20 &       25.00 & 12.79 & 0.01     & 11.96 &   0.01   & 0.83\\
60242.00 &    2325.30 &       15.00 &    3129.70 &       25.00 & 12.81 & 0.01     & 11.85 &   0.01   & 0.96\\
60399.00 &    2128.70 &       15.00 &    3215.40 &       25.00 & 12.91 & 0.01     & 11.82 &   0.01   & 1.09\\
\hline 
\end{tabular}
\begin{tablenotes}
      \small
      \item Magnitudes without an associated errorbar denote upper limits. AB magnitudes are obtained as W1\_AB $=$ W1\_Vega $+$ 2.699 and W2\_AB $=$ W2\_Vega $+$ 3.339, respectively (see \url{https://wise2.ipac.caltech.edu/docs/release/allsky/expsup/sec4_4h.html#conv2flux}).
    \end{tablenotes}
\end{table*}
\end{ThreePartTable}

\begin{ThreePartTable}
\begin{table}
\caption{Pseudo bolometric light curve}              
\label{table:pbolLC}      
\centering                          
\begin{tabular}{c c c}        
\hline\hline                
\noalign{\vskip 1mm}
Phase &   Log$_{10}$(Lbol) &  err\_Log$_{10}$(Lbol) \\
{[days]} & [erg/s]          & [erg/s] \\
\hline
-34.48 & 42.48 &      0.06 \\
-34.39 & 42.49 &      0.06 \\
-34.29 & 42.49 &      0.06 \\
\hline 
\end{tabular}
\begin{tablenotes}
      \small
      \item The full table is available in supplementary material.
    \end{tablenotes}
\end{table}
\end{ThreePartTable}

\section{Spectral log and measurements}

\begin{ThreePartTable}
\begin{table}
\caption{Spectral log}              
\label{table:speclog}      
\centering                          
\begin{tabular}{l c c c l}        
\hline\hline                
\noalign{\vskip 1mm}
Target    &  Date    &   MJD   &    phase  &  Instrument\\
          &         & [days]   & [days]   &            \\
\hline
AT~2022rze  & 2022-11-16 & 59899.49 &  58.24   & SEDM      \\
AT~2022rze  & 2022-12-01 & 59914.23 &  71.87   & SPRAT     \\
AT~2022rze  & 2023-03-04 & 60007.18 & 157.79   & SEDM      \\
AT~2022rze  & 2023-03-09 & 60012.32 & 162.54   & SEDM      \\
AT~2022rze  & 2023-03-20 & 60023.47 & 172.84   & LRIS      \\
AT~2022rze  & 2023-04-16 & 60050.45 & 197.78   & LRIS      \\
AT~2022rze  & 2023-05-07 & 60071.33 & 217.09   & SEDM      \\
AT~2022rze  & 2023-05-21 & 60085.31 & 230.01   & LRIS      \\
AT~2022rze  & 2023-06-15 & 60110.25 & 253.07   & LRIS      \\
AT~2022rze  & 2023-06-15 & 60110.31 & 253.12   & LRIS      \\
AT~2022rze  & 2023-11-15 & 60263.52 & 394.74   & SEDM      \\
AT~2022rze  & 2023-11-30 & 60278.16 & 408.28   & ALFOSC    \\
AT~2022rze  & 2023-12-15 & 60293.20 & 422.18   & ALFOSC    \\
AT~2022rze  & 2024-01-31 & 60340.04 & 465.48   & ALFOSC    \\
AT~2022rze  & 2024-03-26 & 60395.88 & 517.09   & ALFOSC    \\
AT~2022rze  & 2024-06-15 & 60476.23 & 591.37   & DBSP      \\
AT~2022rze  & 2024-06-29 & 60490.27 & 604.35   & LRIS      \\
Galaxy B    & 2022-03-15 & 59653.00 & -169.61  & DESI      \\
Galaxy A    & 2023-12-15 & 60293.20 & 422.18   & ALFOSC    \\
Galaxy A    & 2024-01-31 & 60340.04 & 465.48   & ALFOSC    \\
\hline 
\end{tabular}
\end{table}
\end{ThreePartTable}
\begin{ThreePartTable}
\begin{table}
\caption{Balmer decrement and velocity}              
\label{table:BalmerDecandVel}      
\centering                          
\begin{tabular}{l c c c c c}        
\hline\hline                
\noalign{\vskip 1mm}
Target     & Phase    & H$\alpha$/H$\beta$ & err\_H$\alpha$/H$\beta$ & Vel(H$\alpha$) & err\_Vel(H$\alpha$) \\
           & {[days]} &                    &                         &  [km/s]        &   [km/s]            \\
\hline
AT~2022rze &  58.24  & $\cdots$     & $\cdots$  & $\cdots$    & $\cdots$                                   \\
AT~2022rze &  71.87  & 4.4          & 0.6	& 897.5       & 248.9                                      \\
AT~2022rze & 157.79  & $\cdots$     & $\cdots$	& $\cdots$    & $\cdots$                                   \\
AT~2022rze & 162.54  & $\cdots$     & $\cdots$	& $\cdots$    & $\cdots$                                   \\
AT~2022rze & 172.84  & 5.3          & 0.1	& 1200.2      & 7.8                                        \\
AT~2022rze & 197.78  & 4.9          & 0.1	& 1471.6      & 30.8                                       \\
AT~2022rze & 217.09  & $\cdots$     & $\cdots$	& $\cdots$    & $\cdots$                                   \\
AT~2022rze & 230.01  & 4.9          & 0.1	& 1547.6      & 1.0                                        \\
AT~2022rze & 253.07  & 5.0          & 0.1	& 1596.9      & 65.3                                       \\
AT~2022rze & 253.12  & $\cdots$     & $\cdots$	& $\cdots$    & $\cdots$                                   \\
AT~2022rze & 394.74  & $\cdots$     & $\cdots$	& $\cdots$    & $\cdots$                                   \\
AT~2022rze & 408.28  & 5.9          & 0.2	& 1993.4      & 1.0                                        \\
AT~2022rze & 422.18  & 4.0          & 0.1	& 1693.9      & 51.4                                       \\
AT~2022rze & 465.48  & 4.9          & 0.1	& 2023.7      & 139.8                                      \\
AT~2022rze & 517.09  & 6.0          & 0.3	& 2028.8      & 64.5                                       \\
AT~2022rze & 591.37  & 8.3          & 0.2	& 1524.6      & 78.4                                       \\
AT~2022rze & 604.35  & 8.0          & 0.1	& 1604.2      & 36.9                                       \\
Galaxy B   & -169.61 & 10.4         & 1.1       & $\cdots$    & $\cdots$                                   \\
Galaxy A   & 422.18  & 3.5          & 0.8       & $\cdots$    & $\cdots$                                   \\
Galaxy A   & 465.48  & 2.6          & 0.7       & $\cdots$    & $\cdots$                                   \\
\hline 
\end{tabular}
\end{table}
\end{ThreePartTable}


\bsp	
\label{lastpage}
\end{document}